\documentclass[12pt]{article}
\usepackage{}
\usepackage{}
\usepackage{mathrsfs}

\usepackage{amsfonts}
\usepackage{amsmath,amssymb,bm,graphicx,latexsym,graphics,epsfig,subfigure,color,amsfonts}
\usepackage{amsthm}%this package is used to set a theorem with no tag
\usepackage{bbm}
\usepackage{xcolor}
\usepackage{epstopdf}
\usepackage{setspace}
\usepackage{indentfirst}
\usepackage{booktabs}
\usepackage{float}
\usepackage[utf8]{inputenc}
\usepackage[title]{appendix}
\usepackage{easyReview}
\usepackage{mathrsfs}
\usepackage{multirow}
\usepackage{diagbox}
\usepackage{cases}
\usepackage[colorlinks=true,linkcolor=blue,urlcolor=blue,citecolor=blue]{hyperref}
\usepackage{geometry}
\usepackage{algorithm,algorithmic}

\usepackage{pifont}
\usepackage{bbding}

\usepackage[authoryear]{natbib}
\usepackage{hyperref}
\usepackage{enumitem}
\hypersetup{bookmarksnumbered, bookmarksopen,
colorlinks,citecolor=blue,linkcolor=blue,CJKbookmarks}

\newtheorem*{Lem*}{Lemma}
\newtheorem{The}{Theorem}

\newtheorem{Pro}{Proposition}[section]
\newtheorem{Lem}{Lemma}[section]
\newtheorem{Cor}{Corollary}
\newtheorem{Def}{Definition}
\newtheorem{Exa}{Example}
\newtheorem{Rem}{Remark}

\newcommand{\ignore}[1]{}

\newtheorem{assump}{Assumption}

\renewcommand\theassump{\arabic{assump}}

\def\E{{\mathbb{E}}}

\def\I{{\mathbb{I}}}

\def\gA{{\mathcal{A}}}
\def\gD{{\mathcal{D}}}
\def\gL{{\mathcal{L}}}
\def\gX{{\mathcal{X}}}

\def\gR{{\mathcal{R}}}

\def\gU{{\mathcal{U}}}
\def\gT{{\mathcal{T}}}
\def\gC{{\mathcal{C}}}
\def\gH{{\mathcal{H}}}

\def\gP{{\mathcal{P}}}
\def\gN{{\mathcal{N}}}

\def\FDR{\operatorname{FDR}}

\newcommand{\Eqmark}[2]{\stackrel{(#1)}{#2}}

\usepackage{xr}    
\usepackage{ hyperref}

\title{\Large Unified Conformalized Multiple Testing with Full Data Efficiency}
\date{}

\author{
Yuyang Huo$^{1}$\thanks{Equal contribution.},
Xiaoyang Wu$^{1}$\footnotemark[1],
Changliang Zou$^1$ and 
Haojie Ren$^2$\thanks{Corresponding author: \href{Email:email-id.com}{haojieren@sjtu.edu.cn}}
\\
{\small $^1$School of Statistics and Data Science, Nankai University}
\vspace{-1em}
\\
{\small $^2$School of Mathematical Sciences, Shanghai Jiao Tong University}
}

\begin{document}

\maketitle

\begin{abstract}

Conformalized multiple testing offers a model-free way to control predictive uncertainty in decision-making. Existing methods typically use only part of the available data to build score functions tailored to specific settings. We propose a unified framework that puts data utilisation at the centre: it uses all available data—null, alternative, and unlabelled—to construct scores and calibrate p-values through a full permutation strategy. This unified use of all available data significantly improves power by enhancing non-conformity score quality and maximising calibration set size while rigorously controlling the false discovery rate. Crucially, our framework provides a systematic design principle for conformal testing and enables automatic selection of the best conformal procedure among candidates without extra data splitting. Extensive numerical experiments demonstrate that our enhanced methods deliver superior efficiency and adaptability across diverse scenarios. 

\end{abstract}

\newpage

\section{Introduction}\label{sec:intro}

In recent years, conformalized multiple testing has attracted much attention in the statistics and machine learning communities. It provides uncertainty quantification to identify test samples with unobserved labels of interest, with applications such as outlier detection \citep{bates2023testing} and sample selection \citep{jin2023selection}.
%\textcolor{red}{It provides uncertainty quantification for identifying test samples whose unobserved labels are of interest using black-box predictive models, as in outlier detection \citep{bates2023testing} and sample selection \citep{jin2023selection}.} %It provides uncertainty quantification for identifying multiple individuals with unobserved labels of interest, when implementing a black-box model, such as in the scenarios of outlier detection \citep{bates2023testing} and sample selection \citep{jin2023selection}.
{For example, detecting benchmark samples that may have appeared in the training corpora of large language models is essential for reliable evaluation \citep{dekoninck2024constat}; }
%\textcolor{blue}{For example, identifying test samples that may have been used in large language model training, with statistical guarantees, is essential for reliable evaluation \citep{dekoninck2024constat}; }
%For example, detecting reliable data for large language models training with statistical guarantees is essential for accurately evaluating benchmark performance \citep{dekoninck2024constat}; 
in drug discovery, deep learning models are often used to screen large candidate pools for potential drugs, where error-rate control is desired \citep{dara2022machine}.

Suppose we observe independent and identically distributed (i.i.d.) data pairs $(X, Y)$, where $X\in\mathcal{X}$ is a covariate and $Y\in\{0,1\}$ is the binary response. Generally, conformalized multiple testing involves three types of datasets: a negative (null) labelled dataset $\gD_{0}=\{(X_{i},Y_i)\}_{i=1}^{n_0}$ with all $Y_i=0$, a positive (non-null/alternative) labelled dataset $\gD_{1}=\{(X_{i},Y_i)\}_{i=n_0+1}^{n_0+n_1}$ with all $Y_i=1$ (this dataset is optional) and a test/unlabelled dataset $\gD_{u}=\{X_{j}\}_{j=n+1}^{n+m}$ with unobserved responses, where $n=n_0+n_1$. Let the null labelled samples be indexed by $\gL_0$, the non-null labelled samples by $\gL_1$, and the unlabelled samples by $\gU$. %Let the null labelled samples be indexed by $\gL_0$, the non-null labelled samples by $\gL_1$, and the unlabelled samples by $\gU$.
Each test point $X_j$, $j\in\gU$ is then associated with a hypothesis:
\begin{equation}\label{eq:main_test}
    H_{0,j}: Y_j=0\quad \text{vs.}\quad H_{1,j}: Y_j=1.
\end{equation}
A rejection set $\gR$ is then determined from $\gU$ such that the false discovery rate (FDR) \citep{Benjamini1995ControllingTF} is controlled at a given level $\alpha\in (0,1)$:

$$ \FDR=\E\left[\frac{\sum_{j\in\gU}\I\{j\in\gR,Y_j=0\}}{1\vee|\gR|}\right]\leq\alpha.$$

{A conformalized testing procedure usually proceeds by constructing a non-conformity score, comparing test scores with properly calibrated null scores to form conformal p-values, and applying a multiple testing rule such as BH \citep{Benjamini1995ControllingTF} or conditional calibration \citep{lei2020cc}. The key principle is to maintain {\it exchangeability} among null scores. This exchangeability guarantees that conformal p-values under the null hypothesis are super-uniform and exhibit a well-structured dependence, which together enable finite-sample FDR control in a distribution-free manner. 

A growing line of work has developed with this principle in different ways. As pioneers of conformalized multiple testing, \cite{bates2023testing} built the split conformal p-value, using part of null data $\gD_0$ for score construction and the remaining part for calibration. \cite{adadetect} further improved upon \cite{bates2023testing} by
incorporating the test data $\gD_u$ into score construction.  \cite{liang2024integrative} used both null and non-null labelled samples to enhance the score, followed by conditional calibration to handle the resulting dependence. These methods differ not only in the form of the score function, but also in their data utilisation schemes. In particular, they vary in which
parts of $(\gD_0,\gD_1,\gD_u)$ are used to construct the score, and in how much of
$\gD_0$ is retained for calibration. Such choices are directly tied to statistical
power: a more informative score can better separate nulls from non-nulls, while a
larger calibration set can produce conformal p-values with higher resolution. }

{However, the structural connections among these methods are not
fully transparent, because each method is developed and justified
under its own setting and data utilisation scheme. 
This naturally raises a central question: can we develop a unified framework that
encompasses these existing methods? More importantly, can such a framework enable a more flexible and principled use of the available data for both score construction and calibration?}

\subsection{Preview of contributions}

In this paper, we propose a unified framework that systematically incorporates existing conformalized multiple testing approaches, with a focus on how they utilise available data. The key is to view conformalized multiple testing through the
lens of permutation testing: for each $j\in\gU$, we construct the individualised score function $S^{(j)}$, and compute a new form of conformal p-value  by
\begin{equation}\label{eq:perm_pvalue}
         p_j=\frac{1}{|\Omega_j|}\sum_{\sigma\in\Omega_j}\I\{S^{(j)}(X_j)\leq S^{(j)}_{\sigma}(X_{\sigma(j)})\},
     \end{equation}
    where $\Omega_j$ is the set of all permutations of $\mathcal{L}_0\cup \mathcal{L}_1\cup\gU$ with every index outside of $\gC\cup\{j\}$ being fixed, {$\gC\subseteq \mathcal{L}_0$ is the calibration index set, and $S_{\sigma}^{(j)}$ is the score function constructed on the datasets permuted by $\sigma$.
    This p-value construction is the core of our framework: it treats score construction, data reuse, and calibration within a single exchangeability-preserving scheme.
Then, combined with a carefully designed conditional calibration
step, these p-values yield finite-sample, distribution-free FDR control.}

Our unified view leads to the following main contributions.

\begin{itemize}
 \item[(i)] {\textbf{A unified framework for conformalized multiple testing.}
    We introduce a unified framework based on individualised score functions and full
    permutation p-values in \eqref{eq:perm_pvalue}. When the score 
    functions satisfy suitable symmetry properties, the full permutation procedure
    reduces to simpler and more computationally efficient forms. Consequently, most existing methods can be recovered as
    special cases of our framework under different symmetry structures.} %\citep{bates2023testing,adadetect,liang2024integrative}.}
    \item[(ii)] \textbf{Enhanced approaches by fully utilising data information}: {  The unified framework suggests a direct way to improve power: use all informative data to construct the score and retain the entire null labelled sample for calibration whenever exchangeability permits.  Following this principle, we develop enhanced procedures that jointly use $\gD_0$, $\gD_1$ and $\gD_u$ for score construction while keeping $\gD_0$ for calibration. }%Current methods typically use only a fraction of the available data for score construction and calibration (p-value computation). We instead use all three datasets—$\gD_0$, $\gD_1$ and $\gD_u$—to build the score and keep the entire $\gD_0$ for calibration. 
    This strategy increases power by both improving the score and enlarging the calibration set.
    \item[(iii)] \textbf{Adaptive selection of testing approaches}: Our framework also enables a data-driven selection strategy, automatically picking the most powerful approach from several candidates. Reusing the same data for selecting approaches and conducting tests would normally inflate the FDR; we avoid this by embedding the selection step into the score construction. The resulting procedure keeps finite-sample FDR control without extra data splitting.
\end{itemize}

\subsection{Related work}

 Conformal inference originally exploits the exchangeability of data to produce distribution-free prediction intervals \citep{vovk2005algorithmic,lei2018distribution}. Among many others, recent developments include advanced score constructions \citep{CQR,chernozhukov2021distributional}, valid cross-validation schemes \citep{barber2021predictive}, online implementations \citep{gibbs2021adaptive}, %handling distribution shift \citep{barber2023conformal}, 
achieving approximate conditional guarantees \citep{guan2023localized,gibbs2025conformal} and addressing selective issues \citep{bao2023selective,jin2024confidence,gazin2025selecting}.

The same principles have been adapted to test-based tasks. \cite{bates2023testing} first proposed conformal p-values for one-class novelty detection, splitting the null data to retain exchangeability. \cite{adadetect} and \cite{lee2025full} incorporated the unlabelled test points to improve power, while \cite{liang2024integrative} introduced integrative scores that use both null and non-null labelled samples, followed by conditional calibration \citep{lei2020cc} to ensure finite-sample FDR control under dependence induced by data reuse. 
{Along this line, further developments include test-data-driven model selection \citep{zhangautoms}, fairness-adjusted testing \citep{rava2021burden} and post-selection multiple testing \citep{wang2024conformalized}. }
For the sample selection task, \cite{jin2023selection} proposed a variant of conformal p-values, %when labelled and test data are identically distributed. 
which is later extended to the covariate shift setting \citep{jin2026modelfree} and model selection \citep{bai2024optimized}.
Alternatively, {\cite{wu2024optimal} and \cite{nair2025diversifying} developed sample selection schemes that simultaneously control the selection error rate and maximise the diversity of selected samples.} We provide a unified view that encompasses most of the above approaches and enables principled design of new procedures that fully utilise the available data.

\subsection{Organisation of the paper}
The remainder of this paper is organised as follows. Section \ref{sec:framework} introduces the unified framework and discusses several special cases that cover existing methods. Sections \ref{sec:enhanced} and \ref{sec:appro_select} present two key applications: enhancing existing approaches and enabling approach selection, respectively. Section \ref{sec:numer} provides both synthetic and real-data experiments to evaluate the proposed methods. Section \ref{sec:conclu} concludes with future directions. Additional numerical results, methodological extensions, and technical proofs are provided in the Supplementary Material.

Code for reproducing the experimental results in this paper is available at https://github.com/Xiaoyang-Wu/ECOT.

\section{Unified framework for conformalized multiple testing}\label{sec:framework}

%Suppose there exists labelled null data $\gD_{0}$, labelled alternative data $\gD_{1}$ and unlabelled test data $\gD_u$ with index sets $\mathcal{L}_0$, $\mathcal{L}_1$ and $\gU$. We summarize all conformalized multiple testing approaches into the following three steps:
{We now formalise the unified framework previewed in Section~\ref{sec:intro}, referred to as \emph{Enhanced COnformal Testing} (ECOT). Following the general structure of conformalized multiple testing, ECOT consists of three components: score construction, p-value computation, and the final testing procedure.}
%As outlined in Section \ref{sec:intro}, 
%A basic conformalized multiple testing procedure consists of three key steps: score construction, p-value computation, and testing procedure. Building on this structure, we present a unified framework that summarises existing approaches,  %through the lens of permutation testing and data-utilisation, 
%referred to as {\it Enhanced COnformal Testing} (ECOT).
%{\color{red}Maybe we formally state this is the ECOT procedure. Nothe that we do not define what is ECOT}  
 \begin{itemize}
     \item \textbf{Score construction}: For each test sample $j\in\gU$, construct the individualised score function $S^{(j)}$ based on the available data $\gD_0,\gD_1,\gD_u$. A larger value should indicate stronger evidence against the null hypothesis.
     \item \textbf{P-value computation}: Take a subset $\gC\subseteq \mathcal{L}_0$ as the calibration set. For each $j\in\gU$, compute the permutation-based conformal p-value in \eqref{eq:perm_pvalue}. %Figure \ref{fig:illu_perm} illustrates the permutation strategy.
    \item \textbf{Testing procedure}: {To account for the dependence among the conformal
p-values, we apply a conditional calibration correction. %construct a data-dependent testing threshold through auxiliary rejection sets. }
    To be specific, we first define an auxiliary rejection set $\gR_j$ by applying the BH procedure at level $\alpha$ to the modified conformal p-values $\{\tilde{p}_\ell^{(j)}\}_{\ell\in\gU}$, defined as
    \begin{equation}\label{eq:modi_perm_pvalue}
     \tilde{p}_\ell^{(j)}=\frac{1}{|\Omega_j|}\sum_{\sigma\in\Omega_j}\I\{\tilde{S}^{(j)}(X_\ell)\leq S^{(j)}_\sigma(X_{\sigma(j)})\},\quad \ell\neq j\quad\text{and}\quad \tilde{p}_j^{(j)}=0.   
    \end{equation}
    Here we take $\tilde{S}^{(j)}(X_\ell)=\text{Median}\{{S}^{(j)}_\sigma(X_\ell):\sigma\in\Omega_j\}$, which represents the most ‘stable’ score value for the $\ell$-th sample across all permutations. We then perform an initial rejection procedure to obtain $\gR^{\rm init}=\{j\in\gU:p_j\leq {\alpha|\gR_j|}/{m}\}$.} If $|\gR^{\rm init}|\geq|\gR_j|$ for all $j\in\gR^{\rm init}$, output final rejection set $\gR=\gR^{\rm init}$. Otherwise, %apply the pruning procedure: 
    we generate $\varepsilon_j\stackrel{i.i.d.}{\sim}U(0,1)$ and run BH on $\{\varepsilon_j|\gR_j|/|\gR^{\rm init}|\}_{j\in\gR^{\rm init}}$ at level $1$ to obtain the final rejection set $\gR$.
 \end{itemize}

%\begin{figure}[h!]
%    \centering
%    \includegraphics[width=0.9\linewidth]{Figures/Permutation_illustrate.pdf}
 %   \caption{Illustration of the datasets used for constructing original score function $S^{(j)}$ and permuted score function $S^{(j)}_\sigma$. For simplicity, we treat $\gD_0$ and $\gD_1$ as containing covariates only, as their labels are all equal within the dataset.}
 %   \label{fig:illu_perm}
%\end{figure}
{This testing step is inspired by the conditional calibration idea of
\citet{lei2020cc}. Its role is to handle the dependence among conformal p-values induced by data reuse and individualised score construction. 
The construction of $\tilde{S}^{(j)}(X_\ell)$ above makes the modified p-values
$(\tilde{p}_\ell^{(j)}:\ell\neq j)$ invariant to permutations within $C\cup\{j\}$, thereby ensuring their conditional independence from
$p_j$ under the null. The median here is an empirically stable choice; while other
deterministic permutation-invariant summaries (such as mean) are also applicable. %would lead to the same measurability argument.}
%The construction of $\tilde{S}^{(j)}(X_\ell)$ above ensures that the values $(\tilde{p}_\ell^{(j)}:\ell\neq j)$ are conditionally independent of $p_j$ under the null. 
Under suitable symmetry structures of the score function, this general testing step further reduces to the classical BH procedure; see
Section~\ref{sub_sec:joint_sym}.} %The definition of $\tilde{S}^{(j)}(X_\ell)$ may seem abrupt at first glance. As we will see shortly, this construction ensures that the values $(\tilde{p}_\ell^{(j)}:\ell\neq j)$ are conditionally independent of $p_j$ under the null.}

The above ECOT procedure guarantees finite-sample FDR control under an exchangeability assumption in conformal inference \citep{adadetect}. For clarity, we write  $\{X_i : i \in \mathcal{I}\}$ as the unordered set of elements indexed by $\mathcal{I}$, and $(X_i : i \in \mathcal{I})$ as the ordered tuple. Let $\gH_0,\gH_1\subset\gU$ denote the index sets of null and non-null test samples, respectively.

\begin{assump}[Data exchangeability]\label{Assum:data_exch}
$(X_i:i\in \gC\cup\mathcal{H}_0)$ are exchangeable conditional on the remaining data $\big((X_i,Y_i):i\in \mathcal{L}_1\cup\gH_1\cup(\mathcal{L}_0\setminus\gC)\big)$.
\end{assump}

\begin{The}\label{thm:FDR_control}
    Suppose Assumption \ref{Assum:data_exch} holds. Then
    \begin{itemize}
        \item[(i)] The conformal p-value constructed in \eqref{eq:perm_pvalue} is super-uniform, i.e.
        \[
        \Pr(p_j\leq t\mid Y_j=0,\Psi_j)\leq t\quad\text{for any }t\in[0,1], \quad\text{where}
        \]
 \[\Psi_j=\Big(\big\{X_k:k\in\gC\cup\{j\}\big\}, \big(X_k:k\in  \mathcal{L}_1\cup(\gU\setminus\{j\})\cup(\mathcal{L}_0\setminus\gC)\big),(Y_k:k\in\gL_1\cup\gL_0)\Big),\]
        which contains an unordered set of covariates in $\gC\cup\{j\}$, the remaining covariates and responses in the labelled datasets.
        \item[(ii)] The modified conformal p-values $\{\tilde{p}_{\ell}^{(j)}\}_{\ell\in\gU}$ defined in \eqref{eq:modi_perm_pvalue} are measurable with respect to $\Psi_j$ given $Y_j=0$.
        \item[(iii)] The final rejection set $\gR$ of our procedure satisfies $\FDR\leq \alpha\E[|\mathcal{H}_0|/|\gU|]\leq\alpha$.
    \end{itemize}
\end{The}

The core insight of our unified procedure ECOT is to leverage data exchangeability through permutation testing. {Theorem \ref{thm:FDR_control} formalises this idea in three steps.} Theorem \ref{thm:FDR_control}-(i) is a consequence of the permutation-test principle, obtained by the careful computation of p-values in \eqref{eq:perm_pvalue}.
Theorem \ref{thm:FDR_control}-(ii) further exploits the exchangeable structure after permutation by efficiently reusing the permutation results to identify quantities that remain invariant under $\sigma\in\Omega_j$. 
%{\color{blue}Theorem \ref{thm:FDR_control}-(ii) further exploits the exchangeable structure induced by permutation. In particular, it efficiently reuses the permutation results to identify quantities that remain invariant under $\sigma\in\Omega_j$.} 
This serves as a building block for handling dependence in multiple testing. Finally, incorporating the conditional calibration procedure in the testing step provides a distribution-free and model-agnostic FDR guarantee.  

\begin{Rem}
%\noindent Remark 1{\it    
As shown in Theorem \ref{thm:FDR_control}-(iii), having access to the null proportion $|\mathcal{H}_0|/|\gU|$ allows for stricter FDR control \citep{storey2004strong}. We discuss two standard ways to estimate this proportion and rescale the conformal p-values to improve power; see Section A.1 of the Supplementary Material.
\end{Rem}

In fact, conformal inference is closely connected to permutation tests \citep{angelopoulos2024theoretical, barber2025unifying}. The adoption of a full permutation strategy simplifies the analysis of conformalized multiple testing in two important ways: first, it imposes no restrictions on the score function, permitting flexible designs; second, it allows the calibration set to be arbitrarily chosen within $\gL_0$ as long as Assumption \ref{Assum:data_exch} holds, making it possible to use the entire null labelled sample for calibration. %$\mathcal{L}_0$ to be used as the calibration set.

%\begin{Rem}
%Other forms of modified p-values are possible beyond the definition in \eqref{eq:modi_perm_pvalue}. The only requirement is permutation invariance over $\sigma\in\Omega_j$. %, allowing great flexibility in their construction. 
%For instance, adding 1 to both the numerator and denominator in \eqref{eq:modi_perm_pvalue} can help improve the stability of $\gR_j$ and has been considered in \citet{liang2024integrative}. 
%\end{Rem}

Performing a full permutation is computationally intensive. Moreover, standard Monte Carlo permutation schemes are not directly applicable here, because additional sampling randomness breaks the measurability property in Theorem~\ref{thm:FDR_control}-(ii). Fortunately, when the score function satisfies certain properties, ECOT can be simplified to more efficient implementations, which is the case for many existing methods.

\subsection{Special case: calibration-symmetric score function}\label{sub_sec:cali_sym}
We first consider a scenario where the $j$-th score function $S^{(j)}$ is invariant under every permutation $\sigma\in\Omega_j$.

    \begin{Def}[Calibration-symmetric score function]\label{Assum:cali_sym}
    The collection of score functions $\{S^{(j)}\}_{j\in\gU}$ is calibration-symmetric, if for each $j\in\gU$, $S^{(j)}$ is constructed symmetrically with respect to $\{X_i:i\in\gC\cup\{j\}\}$, i.e., $S^{(j)}_{\sigma}(x)=S^{(j)}(x),\forall \sigma\in\Omega_j,x\in\gX$.
\end{Def}

By the definition of calibration-symmetry, the p-value in \eqref{eq:perm_pvalue} simplifies to

\begin{align}\label{eq:reduced_p_cali_sym}
  p_j&=\frac{1}{(|\gC|+1)!}\sum_{i\in\gC\cup\{j\}}\underbrace{\sum_{\sigma\in\Omega_j,\sigma(j)=i}\I\{S^{(j)}(X_j)\leq S^{(j)}_\sigma(X_{\sigma(j)})\}}_{=|\gC|!\I\{S^{(j)}(X_j)\leq S^{(j)}(X_{i})\} \text{ by symmetry}}\nonumber\\
  &=
\frac{1}{|\gC|+1}\sum_{i\in\gC\cup\{j\}}\I\{S^{(j)}(X_j)\leq S^{(j)}(X_{i})\}.  
\end{align}
A similar reduction applies to the modified p-values in \eqref{eq:modi_perm_pvalue}:
\begin{equation}\label{eq:reduced_p_modified_cali_sym}
     \tilde{p}_\ell^{(j)}=\frac{1}{|\gC|+1}\sum_{i\in\gC\cup\{j\}}\I\{{S}^{(j)}(X_\ell)\leq S^{(j)}(X_{i})\},\quad \ell\neq j\quad\text{and}\quad \tilde{p}_j^{(j)}=0.  
\end{equation}
Thus, there is no need to reconstruct the score functions after permutations, and the number of scores required for p-value computation is reduced from $(|\gC|+1)!$ to $|\gC|+1$, significantly lowering the computational complexity while maintaining FDR control.

\begin{Pro}\label{pro:cali_sym}
    Suppose Assumption \ref{Assum:data_exch} holds and the score functions $\{S^{(j)}\}_{j\in\gU}$ are calibration-symmetric.
    The final rejection set $\gR$ obtained from the ECOT procedure is equivalent to that obtained by replacing the full permutation-based conformal p-values in \eqref{eq:perm_pvalue} and \eqref{eq:modi_perm_pvalue} with their reduced forms in \eqref{eq:reduced_p_cali_sym} and \eqref{eq:reduced_p_modified_cali_sym}, respectively.
\end{Pro}

We next provide some examples covered by ECOT under calibration-symmetry.

\begin{Exa}[Integrative conformal p-value]\label{Exa:integ}
\citet{liang2024integrative} considered the setting where both $\gD_0$ and $\gD_1$ are available. 
%\citet{liang2024integrative} enhanced the quality of score functions by assuming access to both $\gD_0$ and $\gD_1$. 
After splitting both datasets into training and calibration sets, they train separate one-class classifiers $s_0$ and $s_1$ on the training sets of $\gD_0$ and $\gD_1$, respectively. For each test point $j\in\gU$, test-specific score functions $S^{(j)}$ are constructed by integrating $s_0$ and $s_1$. Due to the complicated dependence structure of integrative p-values, a conditional calibration procedure \citep{lei2020cc} is used as the testing rule. The score construction is symmetric with respect to the calibration samples and the $j$-th test point, and hence integrative conformal testing is covered by ECOT under calibration-symmetry.%\textcolor{red}{This method uses both $\gD_0$ and $\gD_1$ to improve power, but it still relies on data splitting and does not fully use $\gD_u$ to learn the score.} 
\end{Exa}
%The first example is the integrative conformal p-value in Example \ref{Exa:integ} \citep{liang2024integrative}. In that method, the score function $S^{(j)}$ is carefully designed to be symmetric with respect to $\gD_c\cup\{X_j\}$. 
%Therefore, this approach falls within the unified ECOT framework. We next present another example.

%\begin{Exa}[Integrative conformal p-value \citep{liang2024integrative}]\label{Exa:integ}
%They split both data $\gD_0=\gD_{t}\cup\gD_{c},\gD_1=\gD_{1,t}\cup\gD_{1,c}$. One-class classifiers $S_0$ and $S_1$ are trained separately on $\gD_{t}$ and $\gD_{1,t}$. For each $j\in\gU$, the initial $p$-values $\widehat{u}_0$ and $\widehat{u}_1$ are computed using the corresponding score functions and calibration datasets as $\gD_{c}\cup\{X_j\}$ and $\gD_{1,c}$. The final p-value is constructed with score $\widehat{u}_0/\widehat{u}_1$ and calibration data $\gD_{c}\cup\{X_j\}$. Therefore, the score function is symmetric to $\gD_c\cup\{X_j\}$. And a subsequent conditional calibration procedure will be applied to these p-values to obtain rejection set. Therefore, their approach is covered by our unified procedure.
%\end{Exa}

\begin{Exa}[Localised conformal p-value]
    The localised conformal prediction interval \citep{guan2023localized,hore2023conformal} can be inverted into a p-value, studied by \cite{wu2024conditional} in conditional testing problems.
    First, split $\gD_0=\gD_{t}\cup\gD_{c}$ and train the score function $S(\cdot)$ on $\gD_{t}$. Then, take the score function $S^{(j)}$ as the kernel estimator:
    \begin{equation*}
        S^{(j)}(x)=\frac{\sum_{i\in\gC\cup\{j\}}H(X_i,x)\I\{S(x)\geq S(X_i)\}}{\sum_{i\in\gC\cup\{j\}}H(X_i,x)},
    \end{equation*}
where $H: \mathcal{X} \times \mathcal{X} \rightarrow \mathbb{R}$ is a kernel function that captures the similarity  between two covariates. The final p-value is constructed by the score $S^{(j)}$ and the calibration data $\gD_{c}$. %, and the rejection set is also given by the conditional calibration procedure. 
Because $S^{(j)}$ is symmetric with respect to $\gD_c\cup\{X_j\}$, the multiple testing procedure with localised conformal p-values \citep{wu2024conditional} is covered by our ECOT framework.
\end{Exa}

\subsection{Special case: joint-symmetric score function}\label{sub_sec:joint_sym}

%We consider another special case, the joint-symmetric score function, which is essentially similar to the score function considered by \citet{adadetect}. 

We next consider another special case, the joint-symmetric score function, under which the conditional calibration step is no longer needed and the ECOT procedure reduces to ordinary BH applied to conformal p-values.

\begin{Def}[Joint-symmetric score functions]\label{Assum:joint_sym}
The series of score functions $\{S^{(j)}\}_{j\in\gU}$ are joint-symmetric, if for each $j\in\gU$, the score function $S^{(j)}$ is identical with $S^{(j)}\equiv S$, and $S$ is constructed symmetrically with respect to $\{X_i:i\in\gC\cup\mathcal{H}_0\}$.
\end{Def}

In practice, a sufficient condition is that $S$ is constructed symmetrically with respect to all samples in $\gC\cup\gU$, which automatically implies symmetry with respect to $\gC\cup\gH_0$. Clearly, joint-symmetric score functions are also calibration-symmetric. The following proposition establishes the equivalence between applying the BH and conditional calibration procedures when the score function is joint-symmetric.

%A key advantage of this setting is that the third step of conditional calibration can be simplified to directly applying the BH procedure, facilitating an easier implementation.

%Furthermore, if score functions are identical for all $j$, i.e. $S^{(j)}(\cdot)\equiv S(\cdot)$ and $S(\cdot)$ are symmetric with respect to data in $\gC\cup\mathcal{H}_0$, our procedure reduces to the case of \citet{adadetect} for joint-exchangeable score functions.

\begin{Pro}\label{pro:joint_sym}
    If Assumption \ref{Assum:data_exch} holds and the score functions $\{S^{(j)}\}_{j\in\gU}$ are joint-symmetric, then  the final rejection set $\gR$ output by ECOT is equal to the set output by the BH procedure applied to the conformal p-values constructed by 
\begin{equation}\label{eq:marg_cp}
    p_j=\frac{\sum_{i\in\gC\cup\{j\}}\I\{S(X_j)\leq S(X_i)\}}{|\gC|+1},\quad j\in\gU.
\end{equation}
\end{Pro}

This result covers many existing methods based on the BH procedure. %Examples \ref{Exa:split_cp} and \ref{Exa:ada} fall into this category. 

\begin{Exa}[Basic conformal p-value]\label{Exa:split_cp}
%As pioneers of conformalized multiple testing, 
\citet{bates2023testing} considered a novelty detection setting without non-null samples ($\gD_1$). %To ensure exchangeability, 
They split the null data $\gD_0$ into a training set $\gD_t$ and a calibration set $\gD_c$. A one-class classifier is trained on $\gD_t$ to produce a score function $S$. Conformal p-values are computed as in \eqref{eq:marg_cp} 
and processed by the BH procedure.  Since the score function is fixed after training and does not depend on the calibration and test data, it satisfies joint-symmetry.
\end{Exa}
Moreover, if $\gD_1$ is available, basic conformal p-values described in Example \ref{Exa:split_cp} can be naturally extended by using a binary classifier as the score function. This also satisfies joint-symmetry and is covered by our framework.

\begin{Exa}[AdaDetect]\label{Exa:ada}
    \citet{adadetect} %improved upon \citet{bates2023testing} by 
    incorporated the test data $\gD_u$ into score construction. Similarly, they split $\gD_0$ into $\gD_t$ and $\gD_c$. Instead of a one-class classifier, a binary classifier is trained to distinguish between $\gD_t$ and the combined dataset $\gD_c \cup \gD_u$ %, using its output
    as the score function $S$. The p-value computation and testing procedure remain the same as in Example \ref{Exa:split_cp}. The construction is symmetric with respect to calibration samples and null test samples, and hence satisfies joint-symmetry. 
\end{Exa}

\begin{Exa}[Full conformal novelty detection]\label{Exa:fullND}
\cite{lee2025full} proposed a novelty detection strategy that trains a classifier based on $\{X_j\}_{j\in\gL_0\cup\gU}$ as the score function, which satisfies joint-symmetry. Although it is implemented with the e-BH procedure \citep{wang2022false}, \cite{lee2025full} have shown that in this case, the e-BH procedure is equivalent to applying the BH procedure on conformal p-values. Therefore, this method is also encompassed within our framework.
\end{Exa}

Beyond joint-symmetry, there exists another class of score functions that permits the implementation of the BH procedure.  This class generalises the oracle Jackknife conformal prediction method \citep{barber2021predictive}. Further details are provided in Section A.5 of the Supplementary Material.

%\begin{Rem}
%   Our framework can also be extended to encompass the procedure in \citet{jin2023selection}, where a different type of  p-value is constructed using a subset of both labelled null and non-null data %$\mathcal{D}_0\cup \mathcal{D}_1$ 
%   instead of $\mathcal{D}_0$. %They construct a different type of conformal p-value %under the assumption that data in $\mathcal{D}_0\cup \mathcal{D}_1$ is identically distributed as those in $\gU$. 
%  Details are provided in Section A.3 of the Supplementary Material.
%\end{Rem}
%\vspace{-0.5em}

%\begin{Rem}
%Our unified perspective can also be extended to the label-integrated conformal p-values of \citet{jin2023selection}, where calibration may involve both labelled null and non-null samples, rather than only $D_0$. This extension is discussed in Section~A.3 of the Supplementary Material.
%\end{Rem}

\subsection{General design strategy for conformalized multiple testing}
Beyond unifying existing conformal testing approaches, ECOT guides the design of new conformalized multiple testing procedures, especially tailored to specific requirements.
%Our framework not only unifies existing conformal testing approaches but also provides a general and practical strategy for designing new conformalized multiple testing procedures, especially tailored to specific requirements. 

The design strategy consists of two key steps. First, one starts by building initial score functions that directly address the task-specific goal--for example, using all available data to train predictive models. These scores can be arbitrarily complex. %While ECOT already guarantees FDR control using these initial scores, 
However, ECOT with these initial scores typically requires about $m \times (|\gC|+1)!$ model-training runs due to full permutations, which is often computationally infeasible. To reduce computational cost (when computational budget is limited), 
we introduce the second step: slightly adjusting the initial score construction to enforce calibration- or joint-symmetry at the cost of mild contamination in the training procedure. This adjustment enables computationally efficient implementations as in Proposition \ref{pro:cali_sym} or \ref{pro:joint_sym}. %, yielding procedures that remain valid.
%and requirement-specific. 
Below is an illustrative example.

\begin{Exa}[Enhanced AdaDetect]
    The original AdaDetect in Example 4 constructs valid p-values via data splitting. %, but this leads to information loss of null-labelled data. 
    %Suppose instead that we require full use of $\gD_0$ for constructing scores, 
    If we instead aim to use the entire $\gD_0$ for score construction, 
    a classifier that distinguishes between $\gD_0$ and $\gD_u$ can be used as the initial score function, and the unified ECOT is then applied. To reduce computational cost, we can train a classifier on $\gD_0\cup\{X_j\}$ and $\gD_u\setminus\{X_j\}$ instead, ensuring calibration-symmetry. Then by setting $\gC = \gL_0$, the procedure in Section \ref{sub_sec:cali_sym} applies. This enhanced version, also explored by \citet{lee2025full} from a cross-validation perspective, demonstrates the validity of our design strategy in practice.
\end{Exa}

In the following sections, based on the above design strategy, we present two key directions of our framework: (1) developing new enhanced methods that fully leverage $\gD_0,\gD_1,\gD_u$ simultaneously, and (2) developing an adaptive procedure that selects the best conformal testing approach while accounting for the post-selection issue.

\section{Enhanced approaches by fully utilising data}\label{sec:enhanced} %information

\begin{figure}[htbp]
    \centering
    \includegraphics[width=0.8\linewidth]{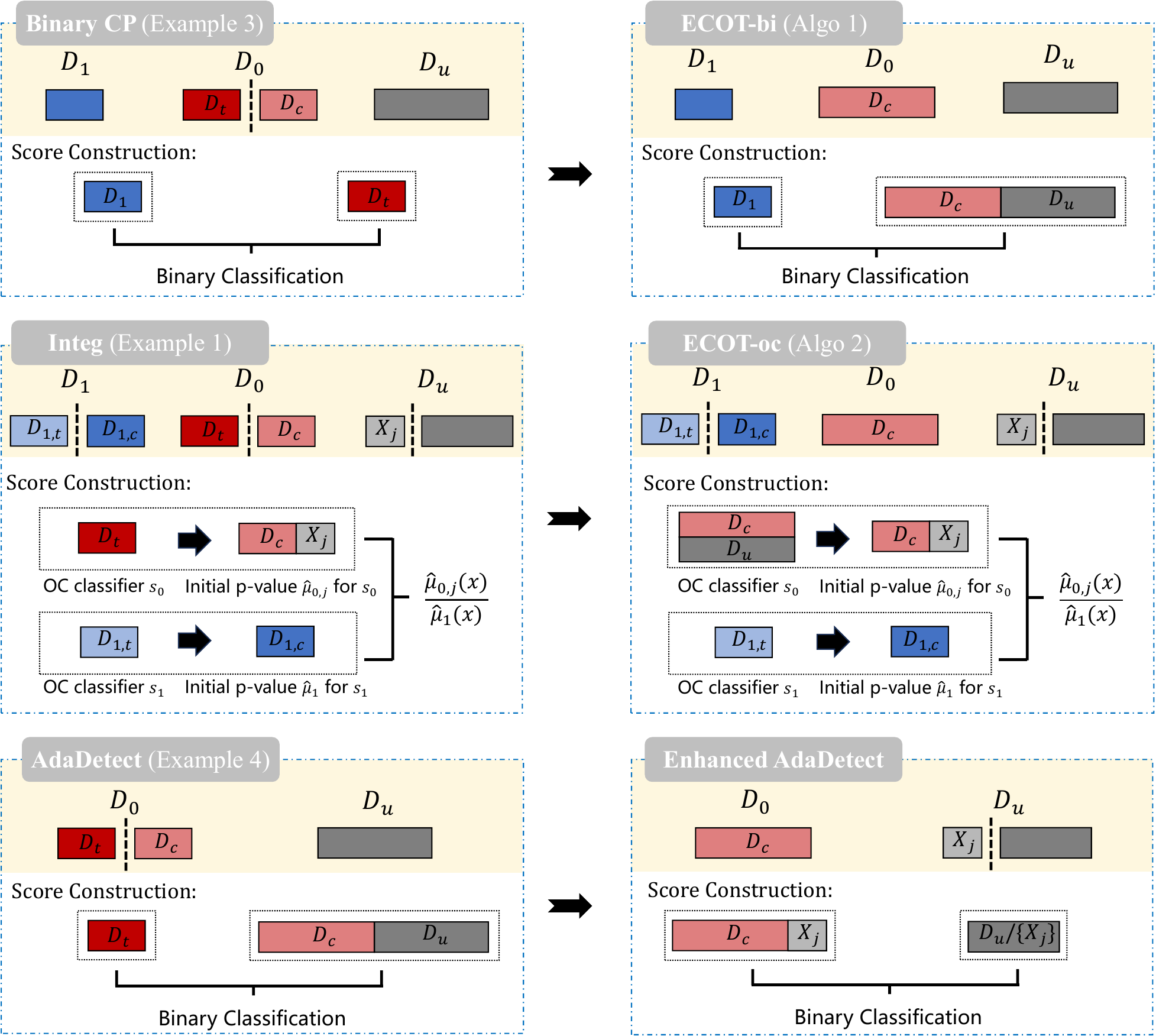}
    \caption{Illustration of the existing approaches and our enhanced approaches with respect to the score construction procedures.}
    \label{fig:illu}
\end{figure}

In this section, we present two improved approaches based on our unified ECOT framework, each addressing data-utilisation limitations in existing methods. Figure \ref{fig:illu} provides an overview, illustrating the procedures for score construction in both the original and enhanced methods. In Section \ref{subsec:ECOT_bi}, we introduce ECOT-bi, an improved version of the method based on conformal p-values with binary classifiers \citep{jin2023selection}. %, which fully uses the information in $\gD_1$, $\gD_0$ and $\gD_u$. 
Section \ref{subsec:one_class_prefer} presents ECOT-oc, an enhancement of the integrative conformal p-value approach \citep{liang2024integrative}, tailored for settings favouring one-class classification. %Finally, Section \ref{subsec:improved_ada} provides an improved method for scenarios with only null-labelled data, building upon and addressing the limitations of AdaDetect \citep{adadetect}.

\subsection{New approach ECOT-bi: full use of       information with $\gD_1$}\label{subsec:ECOT_bi}

Following our framework, when labelled non-null data are available, a natural idea is to jointly  use $\gD_0,\gD_1,\gD_u$ for constructing the score function, which is expected to improve power. As shown in \citet{adadetect}, under the assumption of independent data points, the optimal score function for conformalized multiple testing is any monotone transformation of the density ratio:

$$r(x)=\frac{(1-\pi) f_1(x)}{\pi f_0(x)+(1-\pi) f_1(x)},$$ where $\pi=\Pr(Y=0)$ is the null proportion, and $f_0(x)$ and $f_1(x)$ are the conditional density functions of $X$ given $Y=0$ and $Y=1$, respectively. 

To approximate this optimal score, existing approaches typically use the output of a binary classifier. In the setting of AdaDetect (Example \ref{Exa:ada}), as $\gD_1$ is not available,
$\gD_0$ is split into a training set $\gD_t$ and a calibration set $\gD_c$. A binary classifier is trained by treating $\gD_t$ as one class and $\gD_c \cup \gD_u$ as the other. In contrast,  \cite{jin2023selection} considered training a classifier to distinguish between $\gD_t$ and $\gD_1$, but their method ignores $\gD_u$, losing valuable distributional information. Moreover, both approaches require splitting $\gD_0$, using only a subset for calibration, which may diminish statistical power, particularly when the sample size of $\gD_0$ is limited. %We next show how the proposed design strategy guide us to develop a procedure that employs a binary classifier to distinguish between $\{X_j\}_{j\in\gL_1}$ and $\{X_j\}_{j\in\gL_0\cup\gU}$ as the score function, thereby fully utilising all available data information. 

%However, both methods fail to fully utilise the available data. AdaDetect does not incorporate labelled non-null data, relying entirely on the test data to represent the non-null distribution. Since the signal proportion in $\gD_u$ is typically small, the distributions of $\gD_t$ and $\gD_c \cup \gD_u$ may be quite similar in finite samples, making them hard to separate and potentially degrading the performance of the resulting score function. On the other hand, the approach in \citet{jin2023selection} uses labelled non-null data in $\gD_1$ but ignores the unlabelled test data $\gD_u$, thus missing valuable distributional information. Moreover, both approaches require splitting $\gD_0$, using only a subset for calibration, which may diminish statistical power, particularly when the sample size of $\gD_0$ is limited.

We next show how the proposed design strategy leads to an enhanced procedure. Specifically, we train a binary classifier to distinguish the labelled non-null samples $\{X_j\}_{j\in\gL_1}$ from the pooled samples $\{X_j\}_{j\in\gL_0\cup\gU}$, and use its output as the score function, thereby making full use of all available data. 
Since $\gD_u$ consists mainly of null samples, the mixture distribution of $\{X_j\}_{j\in\gL_0\cup\gU}$ roughly approximates the null distribution, while $\gD_1$ directly characterises the non-null distribution. Therefore, this setup provides a clearer separation between the two distributions and facilitates a more accurate approximation of the density ratio. Moreover, 
incorporating $\gD_1$ avoids the need for data splitting on $\gD_0$. 
%, leading to more efficient data usage. 
Taking $\gC = \gL_0$ as the calibration set, the score function still satisfies joint-symmetry. Consequently, the procedure given in Section 2.2 can be applied, resulting in {\it Enhanced COnformal Testing with binary classification} (abbreviated as ECOT-bi); see Algorithm \ref{alg:main}.
%we can directly use $\gD_0$ as calibration set, compute conformal p-values, and apply the BH procedure to obtain the final rejection set, which is detailed in Algorithm \ref{alg:main}.

\begin{Cor}
    Suppose Assumption \ref{Assum:data_exch} holds. The rejection set $\gR$ output by Algorithm \ref{alg:main} ensures ${\rm FDR}\leq \alpha\E[|\mathcal{H}_0|/|\gU|]\leq\alpha$.
\end{Cor}

Finally, we show that the proposed score function possesses certain optimality properties under the criterion defined by \citet{adadetect}, when obtained by minimising a standard population loss over all measurable functions. This follows directly by establishing its connection to the density ratio $r(x)$. Let $\bar f$ denote the population density of a randomly drawn observation from the pooled sample $\{X_j:j\in\gL_0\cup\gU\}$.  We use $\ell(Y,S(X))$ to denote a binary classification loss that measures the discrepancy between a response $Y\in\{0,1\}$ and a probability-type score $S(x)\in [0,1]$.

\begin{Pro}\label{pro:optimal}
Define the population score as
$S^*=\arg\min_{S}\E_{X\sim \bar{f}}\ell(0,S(X))+\gamma\E_{X\sim f_1}\ell(1,S(X))$, 
where $\gamma>0$ is a trade-off parameter, and the minimisation is taken over all measurable functions. Assume that there exists
a dominating measure $\nu$ such that $f_0$ and $f_1$ are positive $\nu$-almost everywhere. If $\ell(\cdot,\cdot)$ is the 0\textendash1 loss or cross-entropy, then $S^*$ is an optimal score function in the sense: among all procedures of the form of rejected set $\gR(S)=\{j \in \gU\::\: S(X_j)\geq c(\alpha)\}$ where $c(\alpha)\in (0,1)$ is chosen such that ${\rm mFDR}:=\E[|\gR(S)\cap\gH_0|]/\E[|\gR(S)|]=\alpha$, the procedure based on $S^*$ achieves the largest expected number of rejections $\E[|\gR(S^*)|]$.
\end{Pro}

% \textcolor{blue}{
% \begin{Pro}\label{pro:optimal}
% Define the score function as the minimiser of a population loss $\ell(\cdot, \cdot)$ over all measurable functions:
% %$S^*=\arg\min_{S } \E_{X\sim \{X_j\}_{j\in\gL_0\cup\gU}}\ell(1,S(X))+\lambda \E_{X\sim \{X_j\}_{j\in\gL_1}}\ell(-1,S(X)),$
% %where $\lambda > 0$ is a trade-off parameter.
% {\color{red}$S^*=\arg\min_{S}\E_{X\sim \bar{f}}\ell(1,S(X))+\gamma\E_{X\sim f_1}\ell(-1,S(X)),$
% where $\bar{f}$ denotes the pooled covariate density represented by the samples in $\gL_0\cup \gU$, and $\gamma>0$ is a trade-off parameter. Assume that there exists
% a dominating measure $v$ such that $f_0$ and $f_1$ are positive $v$-almost everywhere}. If $\ell(\cdot,\cdot)$ is the 0\textendash1 loss, hinge loss or cross entropy, then the corresponding $S^*$ is an optimal score function in the sense that among all procedures that reject
% hypotheses in the form of $\gR(S)=\{j \in \gU\::\: S(X_j)\geq c(\alpha)\}$ where $c(\alpha)\in (0,1)$ is chosen such that ${\rm mFDR}:=\E[\gR\cap\gH_0]/\E[\gR]=\alpha$, the procedure based on $S^*$ achieves the largest expected number of rejections $\E[|\gR(S^*)|]$.
% \end{Pro}
% }

\begin{algorithm}[htbp]
  \caption{Enhanced COnformal Testing with binary classification (ECOT-bi)}\label{alg:main}
  \begin{algorithmic}[1]
  \REQUIRE Labelled data $\gD_0,\gD_1$ and test data $\gD_u$; FDR target level $\alpha\in(0,1)$; Binary classification algorithm $\gA$.
  \vspace{0.05in} 
  \STATE {\bf Score construction}: fit a binary classification model $S(\cdot)=\gA(\{X_j\}_{j\in\gL_1},\{X_j\}_{j\in\gL_0\cup\gU})$ as the score function;
  \STATE {\bf P-value computation}: take $\gD_0$ as the calibration set, compute p-values
  \begin{equation*}
      p_j=\frac{1}{|\mathcal{L}_0|+1}\sum_{i\in \mathcal{L}_0\cup\{j\}}\I\left\{S(X_j)\leq S(X_i)\right\},\;\; j\in\gU;
  \end{equation*}
  \STATE {\bf Testing procedure}: apply the BH procedure to $\{p_j\}_{j\in\gU}$ at level $\alpha$, obtain the rejection set $\gR$;
  \vspace{0.05in}
  \ENSURE Rejection set $\gR$.
  \end{algorithmic}
\end{algorithm}

\subsection{One-class classification preference}\label{subsec:one_class_prefer}

%In some scenarios where it is difficult to distinguish null and non-null samples with binary classifiers, the approximation could be unreliable. This includes simulation settings in \cite{bates2023testing,liang2024integrative} and \cite{lee2025full}. 
In some scenarios, binary classifiers cannot accurately approximate the density ratio $r(x)$ and yield inefficient testing procedures. This issue arises in simulation settings studied by \citet{bates2023testing}, \citet{liang2024integrative} and \citet{lee2025full}, where %
the distributions of null and non-null samples differ mainly in variances.
%are outliers with higher variance than null samples. 
In such cases, one-class classifiers often outperform binary classifiers (or are more robust to such cases). The integrative conformal p-value (Example \ref{Exa:integ}) addresses this by training two one-class classifiers separately on null and non-null data to better capture the structure. We show that an enhanced version can be achieved within our framework.

The original integrative method has two main limitations: it excludes $\gD_u$  from the score construction and requires data splitting on $\gD_0$. %, which reduces the effective calibration sample size. 
Our remedy is as follows. First, we do not conduct splitting on $\gD_0$ and use it as the calibration set to replace its splitting counterpart in \cite{liang2024integrative}. Second, we directly train a one-class classifier $s_{0}$ on $\{X_j\}_{j\in\gL_0\cup\gU}$, since $\gD_u$ is believed to consist mostly of null samples. % Second, after training, the first-step p-value function $\widehat{u}_0$ is constructed by taking $\gD_0\cup\{j\}$ as the calibration set, avoiding splitting $\gD_0$. 
%By defining $s_1$ and $\widehat{u}_1$ as in \cite{liang2024integrative}, 
The enhanced version of the integrative conformal p-value,
{\it Enhanced COnformal Testing with one-class classification} (ECOT-oc), is presented in Algorithm \ref{alg:integ}.

As is clear from the procedure, the final score functions $\{S^{(j)}\}_{j\in\gU}$ do not satisfy joint-symmetry but still satisfy calibration-symmetry. Together with the exchangeability condition, Algorithm \ref{alg:integ} ensures FDR control in finite samples.

\begin{algorithm}[htbp]
  \caption{Enhanced Conformal Testing with one-class classification (ECOT-oc)}\label{alg:integ}
  \begin{algorithmic}[1]
  \REQUIRE Labelled data $\gD_0,\gD_1$ and test data $\gD_u$; FDR target level $\alpha\in(0,1)$; One-class classification algorithm $\gA$.
  \vspace{0.05in} 
  \STATE {\bf Score construction}: Split $\gD_1=\gD_{1,t}\cup\gD_{1,c}$ with $\mathcal{L}_1=\gT_1\cup\gC_1$. Fit one-class classifiers $s_0(\cdot)=\gA(\{X_j\}_{j\in\gL_0\cup\gU}),s_1(\cdot)=\gA(\{X_j\}_{j\in\gT_{1}})$. For each $j$, compute first-step p-value functions
  \begin{equation*}
      \widehat{u}_{0,j}(x)=\frac{\sum_{i\in\mathcal{L}_0\cup\{j\}}\I\{s_0(x)\leq s_0(X_i)\}}{n_0+1},\;\;\widehat{u}_{1}(x)=\frac{\sum_{i\in\gC_1}\I\{s_1(x)\leq s_1(X_i)\}+1}{|\gC_1|+1}.
  \end{equation*}
  Take $S^{(j)}(x)=\widehat{u}_{0,j}(x)/\widehat{u}_1(x)$ as the score function for the $j$-th sample;
  \STATE {\bf P-value computation}: take $\gD_0$ as the calibration set, and compute p-values
  \begin{equation*}
      p_j=\frac{1}{|\mathcal{L}_0|+1}\sum_{i\in \mathcal{L}_0\cup\{j\}}\I\left\{S^{(j)}(X_j)\leq S^{(j)}(X_i)\right\},\;\; j\in\gU;
  \end{equation*}
  \STATE {\bf Testing procedure}: apply the testing procedure as described in the third step of the ECOT framework on $\{p_j\}_{j\in\gU}$ at level $\alpha$, and obtain the rejection set $\gR$;
  \vspace{0.05in}
  \ENSURE Rejection set $\gR$.
  \end{algorithmic}
\end{algorithm}

\begin{Cor}
    Suppose Assumption \ref{Assum:data_exch} holds. The rejection set $\gR$ output by Algorithm \ref{alg:integ} ensures ${\rm FDR}\leq \alpha\E[|\mathcal{H}_0|/|\gU|]\leq\alpha$.
\end{Cor}

As a side note, using $\gD_u$ for score training may introduce contamination due to the mixed nature of the unlabelled data. We propose a purification step to address this issue, which can improve power when the learning algorithm is sensitive to such contamination. This applies to ECOT-bi, ECOT-oc and any other method using $\gD_u$. See Section A.2 of the Supplementary Material.

%Benefiting from the full utilisation of data, numerical experiments in Section \ref{sec:numer} show that this enhanced version consistently outperforms the original integrative conformal method in power across all scenarios. 

\section{Adaptive selection of conformal testing approaches}\label{sec:appro_select}

As discussed in Section %\ref{sec:enhanced},
\ref{subsec:one_class_prefer}, 
neither binary nor one-class classifier approaches uniformly dominate one another; the optimal choice depends on the specific scenario. This naturally motivates the development of an adaptive strategy that selects the most suitable approach for any situation. %, while still ensuring valid FDR control. 

Suppose we have $K$ candidate conformal testing approaches, such as ECOT-bi and ECOT-oc. A natural idea is to select the one that yields the most rejections, which is a good proxy of power given that the FDR is controlled below the nominal level.
{Denote $R_k$ as the number of hypotheses rejected by the $k$-th approach}. %Let $R_k\big({\gC_k,\gU}\big)$ denote the number of hypotheses rejected by the $k$-th approach, where $\gC_k$ is the corresponding calibration set and $\gU$ is the test set.
The selected approach can be $k^*=\arg\max_{k\in [K]}R_k.$ 
Different from minimising the size of conformal prediction sets \citep{yang2024selection,liang2024conformal}, this metric is testing-oriented and directly aligns with the objective of maximising %statistical
power.

After approach selection, one may directly apply the $k^*$-th approach to obtain the final rejection set. However, reusing the data for both approach selection and testing introduces the ``double dipping'' issue known in post-selection inference \citep{taylor2015statistical,gao2024selective}. As shown empirically in existing literature \citep{zhangautoms}, using the selected approach without proper correction would lead to an inflated FDR, especially for small-sample settings. We therefore develop an  approach-selection strategy, which utilises our framework to ensure finite-sample FDR control.

%This introduces two important issues: how to evaluate the performance of different conformal testing approaches under a fair comparison, and how to account for the potential ``post-selection'' problem, introduced by reusing the data for approach selection. Luckily, as our framework unifies conformal testing approaches together, both issues can be well addressed. 

%As our procedure unifies existing approaches together, it is direct for this task under our framework.
 
%Note that for different approaches unified under our framework, the steps of p-value formation and testing procedure are the same if using the same calibration set. To select the best conformal testing approach, it suffices to choosing the most suitable score function.

\subsection{Approach-selection strategy}

We begin by taking a deeper look at the post-selection issue through the lens of score construction.
Denote $S^{(j),k}$ as the score function corresponding to the $k$-th conformal testing approach and the $j$-th test sample. The score function for the selected approach is $S^{(j),k^*}$. In our framework, directly employing the $k^*$-th approach is equivalent to performing a conformal testing procedure using the p-values as follows:

\begin{equation}\label{eq:naive_cp_approach_selection}
     p'_j=\frac{1}{|\Omega^{k^*}_j|}\sum_{\sigma\in\Omega^{k^*}_j}\I\{S^{(j),k^*}(X_j)\leq S_{\sigma}^{(j),k^*}(X_{\sigma(j)})\},
\end{equation}
where $\Omega^{k^*}_j$ is defined as the sets of permutations on $\gL_0\cup\gL_1\cup\gU$ that fixes indices outside of $\gC_{k^*}\cup\{j\}$. However, $k^*$ is data-dependent. The permuted score function accounts only for data dependence from the basic score construction and ignores that from the approach selection step. As a result, the observed null score $S^{(j),k^*}(X_j)$ and its permuted counterpart $S_{\sigma}^{(j),k^*}(X_{\sigma(j)})$ may not be exchangeable.
%As a result, $S_{\sigma}^{(j),k^*}(X_{\sigma(j)})$ may not be exchangeable with the original score $S^{(j),k^*}(X_j)$. 
Hence, the p-values in \eqref{eq:naive_cp_approach_selection} are no longer super-uniform under the null, rendering downstream inference procedures potentially invalid. 

To address this, we treat $S^{(j),k^*}$ as a ``new'' score function, where the procedure of data-driven approach-selection is considered as an integral step in score construction. This interpretation allows us to apply our unified ECOT framework to produce valid post-selection conformal p-values. More specifically, for the $k$-th approach, we express the rejection number as $R_k=R\left((S^{(j),k}:j\in\gU),(X_i:i\in\gC_k\cup\gU)\right)$,  which makes explicit its dependence on the score functions, the test covariates, and the calibration covariates. We then define the final calibration set $\gC=\bigcup_{k\in[K]}\gC_k$ as the union of the calibration sets for all candidate approaches. 
%More specifically, we express the rejection number of $k$-th approach $R_k=R\left(\{(S^{(j),k},X_j):{j\in\gU}\},\{X_i:i\in\gC_k\}\right)$ as a function of the score–data pairs in the test set and the calibration points and define the final calibration set $\gC=\bigcup_{k\in [K]}\gC_k$ as the union set of calibration sets of all candidate approaches. 
Next, recall $\Omega_j$ as the set of permutations on  $\mathcal{L}_0\cup \mathcal{L}_1\cup\gU$ that fixes indices outside of $\gC\cup\{j\}$.  For each $\sigma\in\Omega_j$, we first permute the data by $\sigma$, and then compute the rejection number for each approach
based on the permuted data to obtain the best approach, denoted as 
\[
k^*_\sigma=\arg\max_{k\in [K]}R\left(\big(S^{(j),k}_\sigma:j\in\gU\big),\big(X_{\sigma(i)}:i\in\gC_k\cup\gU\big)\right).
\]
The p-value can be computed accordingly by

\begin{equation}\label{eq:cp_approach_selection}
    p_j=\frac{1}{|\Omega_j|}\sum_{\sigma\in\Omega_j}\I\{S^{(j),k^*}(X_j)\leq S_{\sigma}^{(j),k^*_\sigma}(X_{\sigma(j)})\}.
\end{equation}
Compared to the naive p-values in \eqref{eq:naive_cp_approach_selection}, the formulation in \eqref{eq:cp_approach_selection} preserves super-uniformity by accounting for data dependence from both score construction and approach selection.  %thereby restoring exchangeability. 
The new p-values in \eqref{eq:cp_approach_selection} are then adapted to our unified framework, yielding finite-sample FDR control. The complete procedure is summarised in Section B.2 of the Supplementary Material.%Algorithm~\ref{alg:ECOT-as}.

\begin{Cor}
    Suppose Assumption \ref{Assum:data_exch} holds. The rejection set output by our ECOT based on the score functions \{$S^{(j),k^*}\}_{j\in\gU}$  ensures ${\rm FDR}\leq \alpha\E[|\mathcal{H}_0|/|\gU|]\leq\alpha$.
\end{Cor}

\subsection{Adjusted strategy for practical implementation}\label{sec:adj_strat}

One practical challenge of the proposed solution is again the computational cost associated with full permutation procedures. To mitigate this issue, we adopt our design strategy with an adjustment step. For simplicity, suppose for each $k\in[K]$, $\{S^{(j),k}\}_{j\in\gU}$ are calibration-symmetric and each approach uses a common calibration set $\gC$, i.e. $\gC_k=\gC$. In this case, we can adjust our procedure such that the selected score functions also satisfy calibration-symmetry, thereby reducing the number of model re-fittings required. That is, for each $j\in\gU$ and each approach $k$, we define an adjusted rejection number ${R}_k^{(j)}$, which is obtained by running one testing procedure (such as BH) at level $\alpha$ to modified p-values $\{\tilde{p}_\ell^{(j),k}:\ell\in\gU\}$, where

\begin{equation}\label{eq:adap_appro_sel_modi_p_value}
      \tilde{p}_\ell^{(j),k}=\frac{1}{|\gC|+1}\sum_{i\in\gC\cup\{j\}}\I\{S^{(j),k}(X_\ell)\leq S^{(j),k}(X_{i})\},\quad \ell\neq j\quad \text{and}\quad\tilde{p}_j^{(j),k}=0. 
\end{equation}
This ${R}_k^{(j)}$ serves as a proxy of the performance of each method on sample $j$, %and can be seen as a function of { $R_k^{(j)}=R\left(S^{(j),k},\{X_\ell:\ell\in\gU\setminus\{j\}\},\{X_i:i\in \gC\bigcup\{j\}\}\right)$}.
and can be seen as a function of $S^{(j),k}$, $\{X_
\ell:\ell\in\gU\setminus\{j\}\}$ and $\{X_i:i\in\gC\cup\{j\}\}$.
The best method is selected as:
\begin{equation}\label{eq:adap_appro_sel_k_star}
 k^*_j=\arg\max_{k\in [K]} {R}^{(j)}_k.   
\end{equation}

The score function of the selected approach $S^{(j),k^*_j}$ satisfies calibration symmetry {by the careful construction of $R_k^{(j)}$}. Accordingly, the computation can be simplified as in Section \ref{sub_sec:cali_sym}. We summarise it in Algorithm \ref{alg:ECOT-ras}. %An alternative adjustment strategy ensuring joint symmetry is provided in Section B.2 of the Supplementary Material. %as a special case of selecting among conformal testing approaches, 
%Moreover, our algorithm also encompasses the selection of candidate score functions within a single approach as a special case. This score selection problem has also been studied by \citet{bai2024optimized} in a similar form, though from a different perspective. 

\begin{algorithm}[htbp]
  \caption{Enhanced Conformal Testing - adjusted adaptive approach selection}\label{alg:ECOT-ras}
  \begin{algorithmic}[1]
  \REQUIRE Labelled data $\gD_0,\gD_1$ and test data $\gD_u$; FDR target level $\alpha\in(0,1)$; $K$ candidate conformal testing approaches, each $k\in [K]$ has 
  score functions $\{S^{(j),k}\}_{j\in\gU}$ and the same calibration set $\gD_c$.
  \STATE {\bf Score construction}: for $j\in\gU$ and each $k\in [K]$, compute the adjusted evaluation criterion $R^{(j)}_{k}$, which is obtained by running the BH procedure at level $\alpha$ to modified p-values $\{\tilde{p}_\ell^{(j),k}:\ell\in\gU\}$ as in \eqref{eq:adap_appro_sel_modi_p_value}.
Then the best approach for sample $j$ is defined as in 
\eqref{eq:adap_appro_sel_k_star}, and the $j$-th score function is $S^{(j),k^*_j}$;
  \STATE {\bf P-value computation}: take $\gD_c$ as the calibration set, compute p-values
  \begin{equation*}
      p_j=\frac{1}{|\mathcal{C}|+1}\sum_{i\in \mathcal{C}\cup\{j\}}\I\left\{S^{(j),k^*_j}(X_j)\leq S^{(j),k^*_j}(X_i)\right\},\;\; j\in\gU;
  \end{equation*}
  \STATE {\bf Testing procedure}: apply the conditional calibration procedure over $\{p_j\}_{j\in\gU}$ at level $\alpha$. The rejection set $\gR_j$ is obtained by applying the BH procedure over $\{\tilde{p}_\ell^{(j),k^*_j}\}_{\ell\in\gU}$;
  \ENSURE Rejection set $\gR$.
  \end{algorithmic}
\end{algorithm}

\begin{Cor}\label{cor:ECOT-ras}
    Suppose Assumption \ref{Assum:data_exch} holds and the candidate score functions $\{S^{(j),k}\}_{j\in\gU}$ are calibration-symmetric for each $k\in[K]$. The rejection set $\gR$ output by Algorithm \ref{alg:ECOT-ras} ensures ${\rm FDR}\leq \alpha\E[|\mathcal{H}_0|/|\gU|]\leq\alpha$.
\end{Cor}

Moreover, our algorithm also encompasses the selection of candidate score functions as a special case within a single approach. A related score-selection problem was studied by \citet{bai2024optimized}, who developed the OptCS framework based on the label-integrated conformal p-values of \citet{jin2023selection}. Their calibration set involves both labelled null and non-null samples, and their score functions are allowed to depend on both the covariate and the label. This requires a stronger exchangeability condition between the labelled and test data, making the procedure potentially sensitive to shifts in the non-null proportion between the two datasets. %[{\it deleted?}] 
Nevertheless, the label-integrated p-values still admit a full-permutation interpretation, allowing us to develop a {\it similar variant covered by our ECOT} framework; see details in Section~A.3 of the Supplementary Material. %This perspective also clarifies how the model-selection procedure in OptCS can be viewed as a reduced implementation, analogous to the reduction in Proposition~\ref{pro:cali_sym}.
%This perspective also clarifies how the model-selection procedure in OptCS can be viewed as a {\color{brown}reduced implementation [what's the meaning?]} under {\color{brown}the extended ECOT framework. 
Further comparisons with OptCS, including robustness to non-null proportion shifts and power are given in Section \ref{sec:rd-optcs} and Sections~C and D in the Supplementary Material.

\begin{Rem}
   In \citet{adadetect} and \citet{liang2024integrative}, specific model or score selection strategies are proposed for their respective settings. Within our framework, those strategies can also be unified by treating model selection as a special case of approach selection.    Our adaptive selection sheds light on alternative choices. Further details are provided in Section B.4 of the Supplementary Material.
\end{Rem}

\section{Numerical studies}\label{sec:numer}

In this section, we illustrate the empirical performance of our proposed methods, and compare them with existing methods covered by our framework. Throughout, we consider the multiple testing problem in \eqref{eq:main_test} without further specification. All experiments are conducted over 500 replicates for reliability.

% The methods are considered:  %the following methods:
% \vspace{-0.8em}
% \begin{itemize}[itemsep=0pt, parsep=0pt]
%     \item {\bf ECOT-bi}: proposed Algorithm \ref{alg:main} with a binary classification score;
%     \item {\bf ECOT-oc}: proposed Algorithm \ref{alg:integ} with a one-class classification score;
%     \item {\bf ECOT-as}: proposed Algorithm \ref{alg:ECOT-ras} with adaptive approach selection;
%     \item {\bf Integ}:  integrative conformal p-value method in \cite{liang2024integrative}
%     \item {\bf AdaDetect}: adaptive novelty detection method in \cite{adadetect}
%     \item {\bf FullND}: full novelty detection method in \cite{lee2025full}
%     \item {\bf CP-bi}: conformal p-value with binary classification score in Section A.1 of \cite{jin2023selection}
%     \item {\bf CP-oc}: conformal p-value with one-class classification score in \cite{bates2023testing}
% \end{itemize}

We compare the following methods:  {\bf ECOT-bi} proposed in Algorithm \ref{alg:main}, {\bf ECOT-oc} proposed in Algorithm \ref{alg:integ}, and {\bf ECOT-as} proposed in Algorithm \ref{alg:ECOT-ras}; {\bf Integ} \citep{liang2024integrative}; {\bf AdaDetect} \citep{adadetect}; {\bf FullND} \citep{lee2025full}; {\bf CP-bi} in Section A.1 of \cite{jin2023selection}; and {\bf CP-oc} \citep{bates2023testing}. 
For ECOT-as, we consider three candidate methods: ECOT-bi, ECOT-oc, and FullND, as they all take the whole $\gD_0$ as calibration set and are methodologically more efficient than the remaining alternatives as discussed in Section \ref{sec:appro_select}. Since the results of applying the BH and conditional calibration procedure are quite close, we report only the BH results (which is also adopted in \citet{liang2024integrative}); a detailed comparison is provided in Section D.3 of the Supplementary Material. Binary and one-class classifiers are implemented with random forest and isolation forest, respectively. %\citep{breiman2001random}, and isolation forest \citep{liu2008isolation}, respectively.
In our experiments, we further refine the modified p-values in \eqref{eq:adap_appro_sel_modi_p_value} to improve power while preserving FDR control. Details are provided in Section B.1 of the Supplementary Material.

\subsection{Simulation examples}

We conduct simulated experiments to evaluate the performance of different methods under two scenarios reflecting distinct preferences in score construction.  These experiments are designed to validate the earlier discussion
that binary-classification and one-class-classification scores may be preferable in different settings. %thereby validating our earlier discussion. 
Additional results are provided in Section D of the Supplementary Material.

%Here, we present simulated experiments to evaluate the performance of different methods under two scenarios, and to validate our earlier discussions. The two scenarios are designed to reflect distinct preferences in score construction, which align with our earlier discussion. Additional simulation results can be found in Appendix \ref{sec:add_simu}.

\subsubsection{Binary classification preference}\label{sec:bi_prefer}
We first consider the simulation setting in \cite{adadetect}, where binary-classification scores are preferable to one-class scores. To be specific, we consider the following data generation:
$X\mid Y=0\sim\gN({\bm 0},I_d), \quad X\mid Y=1\sim \gN({\bm \mu},I_d),$
where ${\bf\mu}$ is a vector with its first 5 coordinates being $\sqrt{a\log(d)}$ and the rest being 0, with larger $a>0$ indicating stronger signals. Unless otherwise stated, we fix dimension $d=50$, nominal level $\alpha=0.1$, labelled sample size $n$ with ratio $n_0:n_1=4:1$, and test sample size $m=1000$. For each replicate, the test data $\gD_u$ contain $1-\pi$ fraction of non-null samples, which represents the signal ratio. %When not varying, the parameters are set to $n=n_0+n_1=500, \pi=0.95$ and $a=1$.

\begin{figure}[h]
    \centering
    \includegraphics[width=0.75\linewidth]{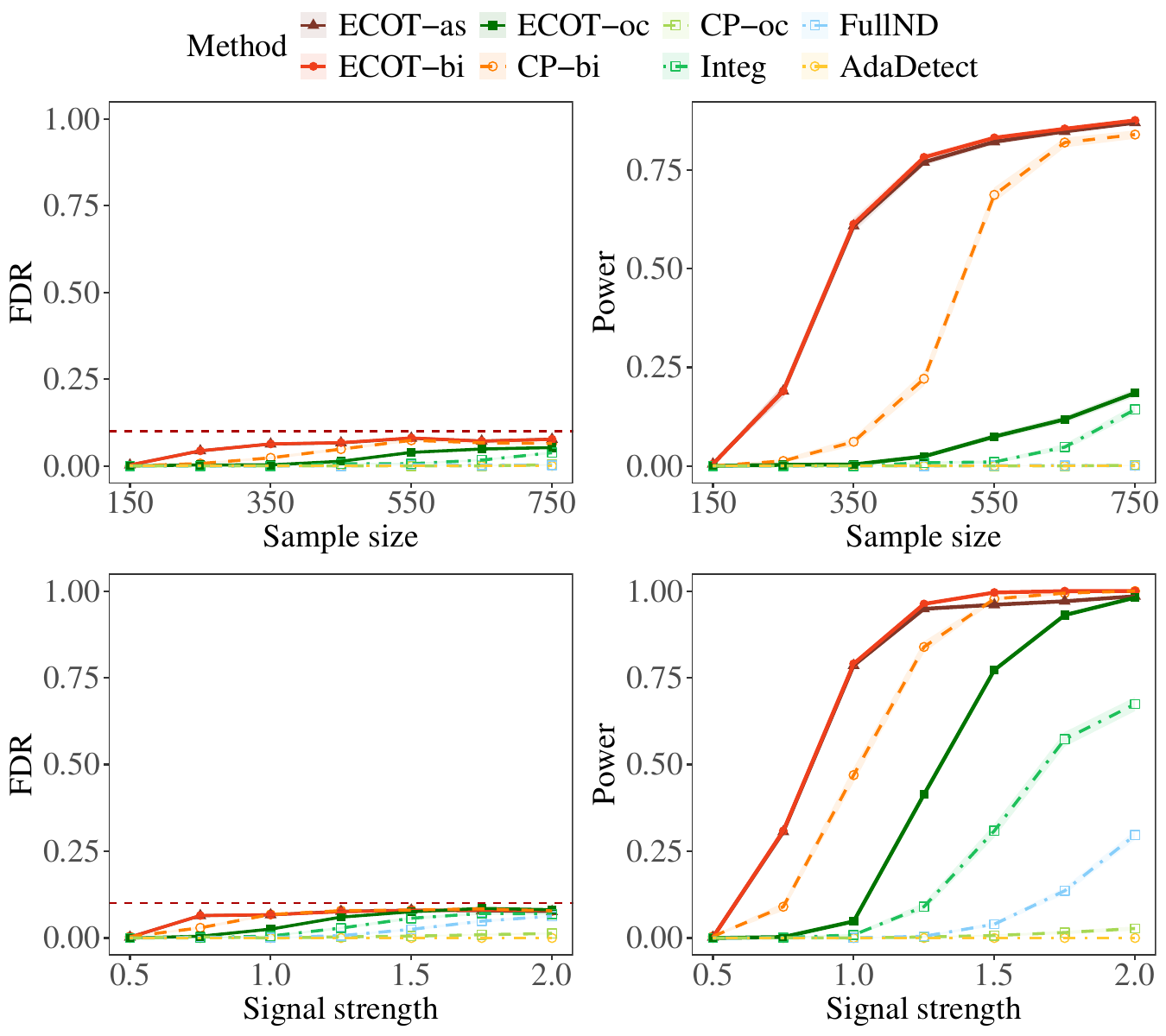}
    \caption{FDR and power of different methods with varying labelled sample size $n$ (top row) under $a=1$ and signal strength $a$ (bottom row) under $n=500$ when {$\pi=0.95$}.}
    \label{fig:bi}
\end{figure}

\begin{figure}[htbp]
    \centering
    \includegraphics[width=0.8\linewidth]{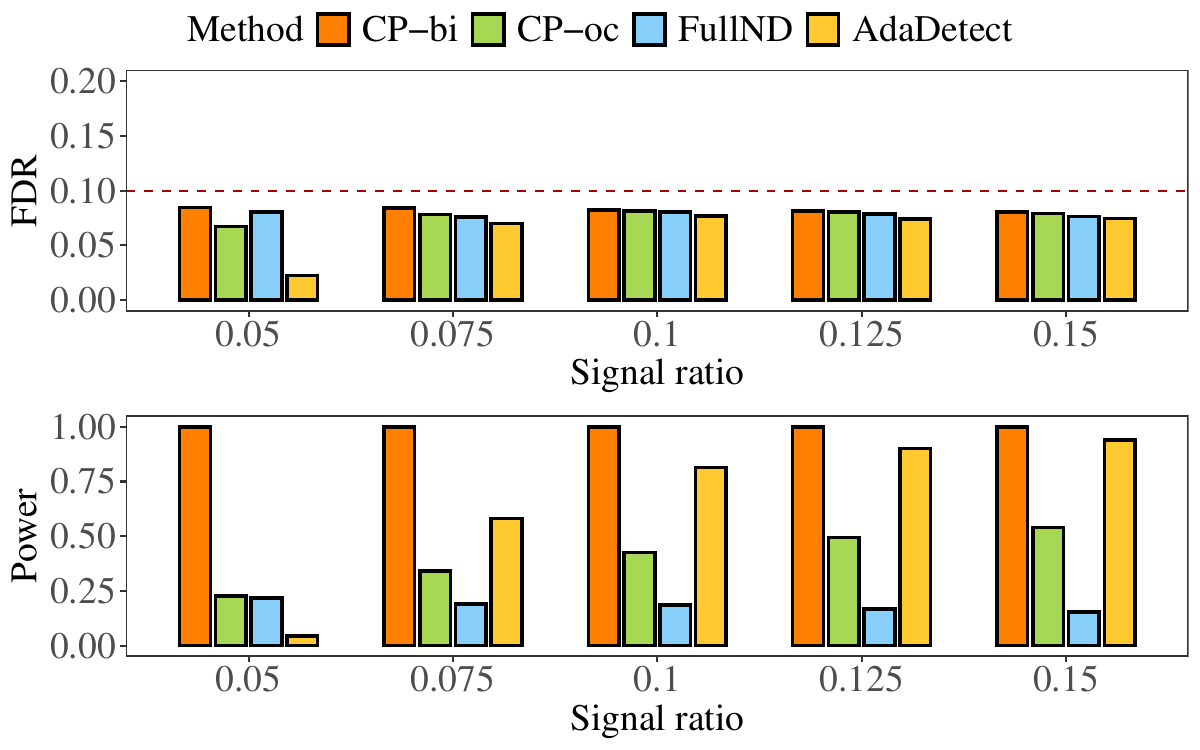}
    \caption{FDR and power of three methods not utilising $\gD_1$ and CP-bi when {varying signal ratio $1-\pi$}. The parameters are fixed at $n=2000,m=1000$ and $a=1.5$. The red dashed line denotes the target FDR level $\alpha=0.1$. }
    \label{fig:sig_ratio}
\end{figure}

Figure \ref{fig:bi} shows the results in FDR and power across different methods as $n$ and $a$ vary. ECOT-bi and ECOT-as achieve the highest power among all methods as expected in the current scenario. While ECOT-oc is relatively less powerful mainly due to the inefficiency of one-class classifiers here, it performs well among those approaches based on one-class classifiers thanks to its full utilisation of data. Among the remaining baselines, CP-bi is the most powerful, but its power is limited by the sample splitting step, which reduces the calibration set size, rendering it ineffective with small sample sizes. 
Notably, five methods leveraging labelled non-null samples from $\gD_1$ (e.g., ECOT series, Integ, CP-bi) demonstrate marked power gains compared to those (e.g., CP-oc, FullND, AdaDetect) that do not. To highlight this contrast, Figure \ref{fig:sig_ratio} compares the performance of the latter three methods with CP-bi as a representative $\gD_1$-based method. With a large sample size $n$ and increasing signal ratios, all methods exhibit observable power, but CP-bi consistently achieves power close to 1 and outperforms the others significantly. This underscores the advantage of incorporating $\gD_1$, even in the simplest way.
Among the three $\gD_1$-free methods, AdaDetect leads due to its binary classification score. As the signal ratio increases, both AdaDetect and CP-oc show rising power. In contrast, FullND declines as more non-null samples in $\gD_u$ degrade the performance of the one-class classifier trained on $\gD_0 \cup \gD_u$.

\subsubsection{One-class classification preference}\label{sec:oc_prefer}

Next, we consider the simulation setting in \cite{bates2023testing} and \cite{lee2025full}, where one-class classification scores are more efficient. To be specific, a fixed set $\mathcal{W}$ of $d$ independent samples is drawn from ${\rm U}[-3,3]^{d}$. %and fix the set throughout all experiments. 
Data are then generated as
    $X = \sqrt{1+a\cdot\I\{Y=1\}}V+W,$
where $V\sim\gN({\bf 0}, I_d)$ and $W$ is sampled from $\mathcal{W}$ independently. Unless otherwise specified, parameters follow the previous scenario. %Here we take $d=1000$ to show performance in high dimensions.

\begin{figure}[htbp]
    \centering
    \includegraphics[width=0.8\linewidth]{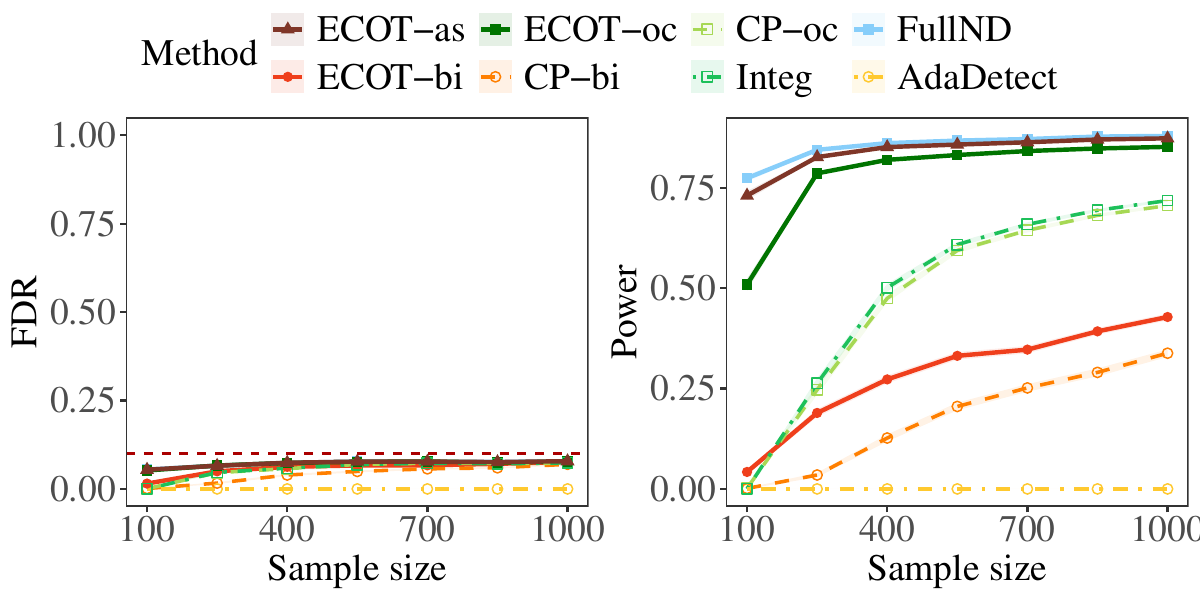}
    \caption{FDR and power of different methods with varying labelled sample size $n$ when $a=3.5$ and $\pi=0.8$.}
    \label{fig:oc}
\end{figure}

Figure \ref{fig:oc} depicts FDR and power across different methods as labelled sample size $n$ varies. In the current scenario, one-class classifiers outperform binary ones, which further leads to a different power ranking: FullND, ECOT-as, and ECOT-oc emerge as the most powerful approaches. Notably, as an enhanced version, ECOT-oc significantly improves power over the integrative conformal method Integ. Additionally, ECOT-oc and FullND achieve nearly identical power, with a similar relationship between CP-oc and Integ, as the information contained in $\gD_0$ is sufficient to detect all distinguishable non-null samples, rendering $\gD_1$ less useful. When the size of $\gD_1$ is very small, incorporating its noisy information could even be detrimental and degrade the power. As illustrated in Figures \ref{fig:bi} and \ref{fig:oc}, even if ECOT-oc and ECOT-bi may not always be top performers individually, ECOT-as can track the best-performing method through the adaptive selection step.

\begin{figure}[htbp]
    \centering
    \includegraphics[width=0.8\linewidth]{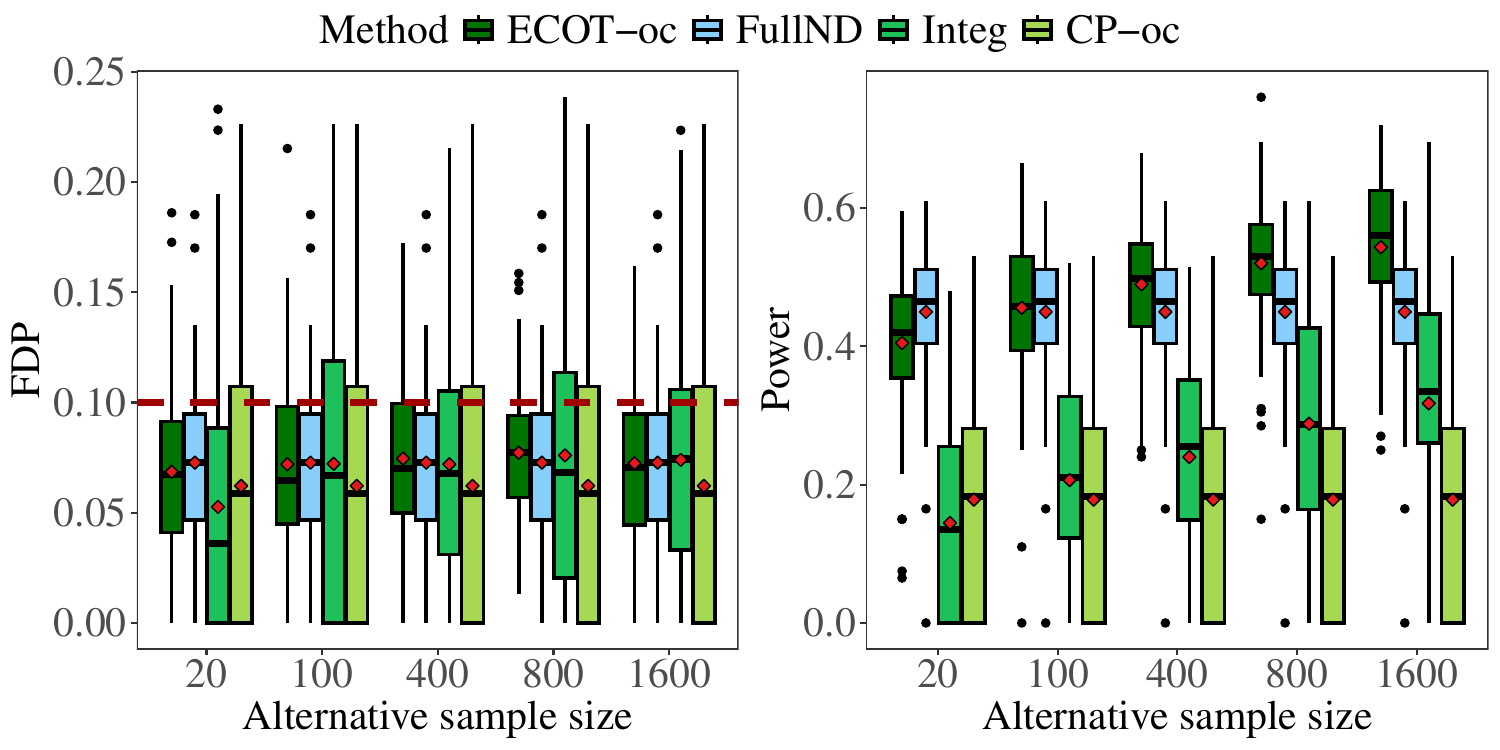}
    \caption{FDP and power of ECOT-oc and FullND with varying non-null sample size and $d=1000$. The red dashed line denotes the target FDR level $\alpha=0.1$.}
    \label{fig:oc_box}
\end{figure}

To further investigate the benefit of labelled non-null data $\gD_1$, we consider a more challenging high-dimensional setting with $d = 1000$. Figure \ref{fig:oc_box} illustrates the performance of four related methods as $n_1$ increases. Here, relying solely on $\gD_0$ is no longer sufficient. As $n_1$ grows, the enhanced information from $\gD_1$ becomes more reliable, and the power of ECOT-oc and Integ gradually surpasses that of FullND and CP-oc, respectively.

\subsection{Real data evaluation}
\subsubsection{Main experiments on approach selection}
In this section, we validate the performance of our proposed methods on several outlier detection datasets. Table \ref{tab:datasets} provides a brief summary of the datasets used. For each dataset, labelled data $\gD_0$, $\gD_1$, as well as test data $\gD_u$, are sampled independently from the inlier and outlier parts. Treating inliers as null samples and outliers as non-null samples, the goal is to detect outliers in $\gD_u$, which corresponds to the multiple testing problem in \eqref{eq:main_test}. For each replicate, we sample $n_0=400$ inliers as $\gD_0$ and $n_1=100$ outliers as $\gD_1$. The test data $\gD_u$ are constructed by sampling $950$ inliers and $50$ outliers, so that $m=1000$ and the signal ratio is $1-\pi = 0.05$.

\begin{table}[htbp]
\caption{Summary of different datasets considered in the experiments.}\label{tab:datasets}
\centering
\begin{tabular}{ccccccc}
\hline
         & Credit Card & Satellite & Shuttle & CovType & Mammography \\ \hline
\# Features &  30  &  36  &  9  &  10 & 6 \\
\# Inliers  &  284,315  &  5,702  &  45,586  &  283,301 & 10,922 \\
\# Outliers  &  492  &  703  &  12,414  &  2,747 & 260 \\ \hline
\end{tabular}
\end{table}

Table \ref{tab:realdata} shows the performance of all methods across different datasets. Our  three ECOT methods consistently achieve higher power than existing baselines. Specifically, ECOT-oc performs best on the Satellite dataset, while ECOT-bi dominates on the other datasets. ECOT-as consistently achieves power close to the best among all methods, highlighting the benefit of adaptively selecting approaches. Except for CP-bi, the remaining baselines exhibit little to no power, likely due to limited labelled data. AdaDetect is also severely influenced by a low signal ratio in the test dataset. These results suggest that binary classifiers are generally more effective than one-class methods on most real-world datasets. %, since methods based on one-class classifiers are not as powerful as those based on binary classifiers.

\begin{table}[htbp]
\centering
\caption{FDR and power results for compared benchmarks across five different real datasets. The nominal FDR level is $\alpha=0.1$. The highest two values of power for each dataset are shown in bold.}\label{tab:realdata}
\vspace{0.5em}
\resizebox{\textwidth}{!}{%
\renewcommand{\arraystretch}{1} 
\begin{tabular}{cccccccccc}
\hline
        &       & ECOT-as & ECOT-bi & ECOT-oc & CP-bi & CP-oc & AdaDetect & Integ & FullND \\ \hline
\multirow{2}{*}{Credit Card} & FDR   & 0.067    & 0.067    & 0.039    & 0.044  & 0.012  & 0.000      & 0.000  & 0.012   \\
        & Power & \textbf{0.710}    & \textbf{0.729}    & 0.355    & 0.210  & 0.033   & 0.000      & 0.000   & 0.017    \\
        \\
\multirow{2}{*}{Satellite}   & FDR   & 0.088    & 0.082    & 0.087    & 0.070  & 0.050  & 0.000   & 0.005  & 0.047   \\
        & Power & \textbf{0.865}    & 0.862    & \textbf{0.902}    & 0.572  & 0.230  & 0.000    & 0.006   & 0.160   \\
        \\
\multirow{2}{*}{Shuttle} & FDR   &  0.073    &  0.073    &  0.023    &  0.083     &  0.000    &  0.000    &  0.000     &  0.038      \\
        & Power &  \textbf{0.988}    &  \textbf{0.988}    &  0.046    &  0.923     &  0.000     &  0.000      &  0.000     &  0.065      \\
        \\
\multirow{2}{*}{CovType} & FDR   &  0.079    &  0.079    &  0.000       &  0.083     &  0.000     &  0.000    &  0.000    &  0.000    \\
        & Power &  \textbf{0.895}       &  \textbf{0.895}       &  0.000       &  0.586     &  0.000     &  0.000      &  0.000     &  0.000      \\
        \\
        \multirow{2}{*}{Mammography} & FDR   &  0.062    &  0.062    &  0.003       &  0.029     &  0.010     &  0.000    &  0.000    &  0.000    \\
        & Power &  \textbf{0.170}       &  \textbf{0.170}       &  0.004       &  0.060     &  0.010     &  0.000      &  0.000     &  0.000      \\\hline
        
\end{tabular}
}
\end{table}

{
\subsubsection{An illustration of model selection task}\label{sec:rd-optcs}
As discussed in Section \ref{sec:adj_strat}, our approach selection task encompasses the model selection task within a single approach as a special case. Here we compare the model selection counterpart of ECOT-as to the two model selection methods (OptCS-MSel and OptCS-Full-MSel) proposed by \citet{bai2024optimized}. For the latter two methods to be applicable, we use the same signal ratio $1-\pi=0.05$ for both labelled and unlabelled data with $n_0=380,n_1=20,m=1000$. To compare the complete versions of the two frameworks, we further apply the purification step we proposed in Section A.2 of the Supplementary Material to ECOT-as. We consider six candidate models, as described in Section D.2 of the Supplementary Material.

\begin{table}[htbp]
\centering
\caption{FDR, power and computation time (seconds) for compared benchmarks across five different real datasets. The nominal FDR level is $\alpha=0.1$. The highest value of power for each dataset is shown in bold. }\label{tab:RE-rd-maintext}
\vspace{0.2em}
\footnotesize
\begin{tabular}{ccccc}
\hline
        &       & ECOT-as & OptCS-MSel & OptCS-Full-MSel \\ \hline
\multirow{3}{*}{Credit Card} & FDR   & 0.067    & 0.061    & 0.059  \\
        & Power & \textbf{0.769}    & 0.207   & 0.437 \\
        & Time & 30.0 & 19.1 & 2465.3        \\
        \\
\multirow{3}{*}{Satellite}   & FDR   & 0.077    & 0.094    & 0.078  \\
        & Power & \textbf{0.889}    &  0.443   & 0.741  \\
        & Time & 25.6 & 12.2 & 1346.4        \\
        \\
\multirow{3}{*}{Shuttle} & FDR   &  0.069    &  0.063    &  0.080      \\
        & Power &  \textbf{0.815}    &  0.297    &  0.452\\
        & Time & 23.6 & 14.4 & 670.2
        \\
        \\
\multirow{3}{*}{CovType} & FDR   &  0.095 &  0.068       &  0.085     \\
        & Power &  \textbf{0.867}     &  0.506       &  0.590     \\
        & Time & 23.6 & 14.2 & 685.6        \\
        \\
        \multirow{3}{*}{Mammography} & FDR   &  0.050    &  0.018       &  0.035      \\
        & Power &  \textbf{0.157}       &  0.035       &  0.061   \\
        & Time & 23.5 & 17.0 & 758.6        \\
        \hline
        
\end{tabular}

\end{table}

In Table \ref{tab:RE-rd-maintext}, we observe that our ECOT method consistently achieves the highest power. The full conformal version of OptCS is relatively more powerful than the split version, which can be attributed to the small labelled sample size used here. From a computational perspective, ECOT and the split implementation of OptCS have similar runtimes within 1 minute. As a full conformal implementation, OptCS-Full-MSel incurs substantially higher computational cost, requiring at least 10 minutes for a single run. 
}

\section{Concluding Remarks}\label{sec:conclu}

%In this article, we propose a unified framework for conformalized multiple testing in view of how the available data are utilised, which encompasses most existing methods as special cases. Building upon this framework, we introduce two novel procedures, ECOT-bi and ECOT-oc, that fully exploit the available data, leading to enhanced power compared to existing approaches. Furthermore, our framework enables the development of an algorithm that selects the optimal approach to maximize power, while still ensuring finite-sample FDR control.

We conclude by outlining several directions for future research. First, our framework relies on data exchangeability, a standard condition in conformal inference. An important extension would be to relax this assumption to accommodate non-exchangeable data, such as sequences generated by two-state hidden Markov models \citep{zhao2024false}. Second, the unified procedure ECOT requires full permutation operations to compute valid p-values, which can be computationally intensive and requires careful adjustment to ensure smooth execution. Exploring techniques to accelerate permutation-based inference would be highly beneficial for tackling this problem. Third, while our approach is designed for offline analysis, there is a growing demand for online decision-making, as reflected in recent work on online multiple testing \citep{javanmard2018online}. Extending our framework to the online conformalized setting presents an exciting avenue for further study.

%\section*{Acknowledgments}
%The authors thank the Editor, Associate Editor and two reviewers for their valuable comments and suggestions, which have greatly improved the manuscript.

%\section*{Supplementary material}
%The Supplementary Material contains additional numerical results, methodological extensions, and technical proofs.

\bibliographystyle{abbrvnat}
\bibliography{ref}

\newpage
\appendix

\renewcommand{\theThe}{\Alph{section}.\arabic{The}}
\setcounter{The}{0}

\renewcommand{\theCor}{\Alph{section}.\arabic{Cor}}
\setcounter{Cor}{0}

\renewcommand{\thePro}{\Alph{section}.\arabic{Pro}}
\setcounter{Pro}{0}

\renewcommand{\theequation}{\Alph{section}.\arabic{equation}}
\setcounter{equation}{0}

\renewcommand{\theassump}{\Alph{section}.\arabic{assump}}
\setcounter{assump}{0}

\renewcommand{\thefigure}{\Alph{section}.\arabic{figure}} 
\renewcommand{\thetable}{\Alph{section}.\arabic{table}}   
\renewcommand{\thealgorithm}{\Alph{section}.\arabic{algorithm}}  
\setcounter{figure}{0}    
\setcounter{table}{0}
\setcounter{algorithm}{0}

\section{Extensions for the framework ECOT}

\subsection{Incorporation of null proportion}\label{appen_sec:null_prop}
We can incorporate the null proportion $|\mathcal{H}_0|/|\gU|$ to make our unified procedure more powerful. As $\gH_0$ is unknown in practice, we introduce two methods for estimating the null proportion. 

\paragraph{Storey-type estimator via auxiliary score}
The first approach adapts Storey's estimator \citep{storey2004strong} to our conformalized setting using an additional joint-symmetric score function $S^{\rm join}$ that satisfies Definition \ref{Assum:joint_sym}. Importantly, this auxiliary score function is only used to estimate the null proportion and does not constrain the main score function $S^{(j)}$ for testing, which can be constructed freely. In practice, $S^{\rm join}$ can be obtained via an outlier detection model trained on $\gC \cup \gU$.

Then for each $j\in\gU$, we can define the auxiliary p-values as 
\begin{equation}\label{eq:p_aux_join}
 \tilde{p}^{(j),{\rm join}}_\ell=\frac{1}{|\gC|+1}\sum_{i\in\gC\cup\{j\}}\I\{S^{\rm join}(X_\ell)\leq S^{\rm join}(X_i)\},\quad \ell\in\gU\setminus\{j\}.   
\end{equation}
Based on $\{\tilde{p}^{(j),{\rm join}}_\ell\}_{\ell\in\gU\setminus\{j\}}$, we estimate the null proportion by
\begin{equation}\label{eq:pi_storey}
    \hat{\pi}_j=\frac{1+\sum_{\ell\in\gU\setminus\{j\}}\I\{\tilde{p}^{(j),{\rm join}}_\ell\geq\lambda\}}{m(1-\lambda)}
\end{equation}
for a fixed $\lambda\in(0,1)$ \citep{storey2004strong}. Then, applying conditional calibration over $\{\hat{\pi}_jp_j\}_{j\in\gU}$ at level $\alpha$, the procedure continues to control the FDR.

\begin{Pro}\label{pro:storey_est}
    Suppose Assumption \ref{Assum:data_exch} holds. If the null proportion estimator $\hat{\pi}_j$ takes the form in \eqref{eq:pi_storey} with $\tilde{p}^{(j),{\rm join}}_\ell$ in \eqref{eq:p_aux_join} and $S^{\rm join}$ satisfies Definition \ref{Assum:joint_sym}, then our ECOT procedure applied to $\{\hat{\pi}_jp_j\}_{j\in\gU}$ instead of $\{p_j\}_{j\in\gU}$ at level $\alpha$ still controls FDR at $\alpha$.
\end{Pro}
The results can be extended to Storey’s estimator with an adaptively chosen $\lambda$, treated as a specific stopping time. For further details, see \citet{gao2025adaptive} and \citet{lee2025full}.

\paragraph{Label-assisted null proportion estimator}

The second approach applies in scenarios where the labelled data are drawn from the same distribution as the test data, i.e. Assumption \ref{Assum:data_full_exch} in Section \ref{appex_sec:jin_pvalue}. In this case, the empirical null proportion $|\gL_0|/(|\gL_0| + |\gL_1|)$ approximates the true null proportion $|\gH_0|/|\gU|$ \citep{jin2023selection}. Specifically, we select subsets $\gC_0 \subset \gL_0$ and $\gC_1 \subset \gL_1$, and estimate the null proportion via
\begin{equation}\label{eq_appen:label_asis_null_prop_est}
\hat{\pi} = \frac{1 + |\gC_0|}{1 + |\gC_0| + |\gC_1|}.
\end{equation}

Setting the calibration set $\gC = \gC_0$, we apply conditional calibration to $\{\hat{\pi} p_j\}_{j \in \gU}$ at level $\alpha$. Under suitable assumptions on the data and score functions, the procedure continues to control FDR.

\begin{Pro}\label{pro:cali_est}
Let $\gC_0\subset\gL_0$ and $\gC_1\subset\gL_1$.
If Assumption \ref{Assum:data_full_exch} holds and the score function $S^{(j)}$ is symmetric to data in $\gC\cup\gC_1\cup\{j\}$, then our ECOT procedure applied to $\{\hat{\pi}p_j\}_{j\in\gU}$ with $\hat{\pi}$ taken in \eqref{eq_appen:label_asis_null_prop_est} at level $\alpha$ still controls FDR at $\alpha$.
\end{Pro}

\subsection{Test dataset purification}\label{sec:pure}
Some conformal testing approaches leverage the test data during model training to capture information about the alternative hypothesis or improve efficiency. These include AdaDetect \citep{adadetect}, FullND \citep{lee2025full}, ECOT-bi and ECOT-oc as stated in Section \ref{sec:numer}. The AdaDetect method uses the test dataset $\gD_u$ to represent the alternative distribution, whereas the other methods treat it as representative of the null distribution. Although training on such mixtures can yield an effective classifier asymptotically, the performance of the fitted model may deteriorate depending on the underlying data distribution.

Here we take our ECOT-bi method as an example, which employs $\{X_j\}_{j\in\gL_1}$ and $\{X_j\}_{j\in\gL_0\cup\gU}$ to fit a binary classifier as the predictive model. While combining $\gD_0$ and $\gD_u$ preserves the exchangeability between $\gD_0$ and the null samples in $\gD_u$, it also introduces alternative samples from $\gD_u$ into a dataset treated as null in the training phase. If the proportion of alternative samples in $\gD_u$ is large, the influence of this contamination could be significant. To mitigate this issue, we introduce a purification step to remove most of the alternative samples from $\gD_u$ and improve the quality of the fitted model.

Specifically, we first train a probabilistic classifier and construct an initial score function $S$ as in the ECOT-bi procedure. Although this score is affected by contamination, it can still approximately distinguish null and alternative samples through the magnitude of its values. We then rank the scores $\{S(X_j)\}_{j \in \gL_0 \cup \gU}$ and remove the samples corresponding to the largest $\lceil \eta m \rceil$ scores, where $0 < \eta < 1$. Since samples from $H_1$ tend to have larger scores, they are more likely to be removed in this step. Formally, let $S_{(1)}\leq S_{(2)}\leq\cdots\leq S_{(n_0+m)}$ be the order statistics of $\{S(X_j)\}_{j \in \gL_0 \cup \gU}$. We define the purified index set and dataset as
\begin{equation}\label{eq:pure}
    \gP=\{j:j\in\gL_0\cup\gU, S(X_j)\leq S_{(n_0+m-\lceil\eta m\rceil)}\},\quad \gD_p=\{X_j:j\in\gP\}.
\end{equation}
After the purification, we fit a classifier to distinguish $\{X_j\}_{j\in\gL_1}$ from $\{X_j\}_{j\in\gP}$ and construct a new score function $S_p$. Since the purification rule is a symmetric function of the scores over $\gL_0\cup\gU$,  this score function is still joint-symmetric. Therefore, the remaining procedure of ECOT-bi can follow directly while the same FDR control
guarantee continues to hold. 

This purification technique can also be applied to other methods that use test data during training, including AdaDetect, FullND, and ECOT-oc. The full procedure with purification for our ECOT methods is summarised in Algorithm~\ref{alg:pure}.

\begin{algorithm}[htbp]
  \caption{Enhanced COnformal Testing with purification (ECOT-pure)}\label{alg:pure}
  \begin{algorithmic}[1]
  \REQUIRE Labelled data $\gD_0,\gD_1$ and test data $\gD_u$; FDR target level $\alpha\in(0,1)$; Score construction algorithm $\gA$.
  \vspace{0.05in} 
  \STATE {\bf First-step score construction}: construct the first-step score function as $S(\cdot)=\gA(\{X_j\}_{j\in\gL_1},\{X_j\}_{j\in\gL_0\cup\gU})$;
  \STATE {\bf Purification}: obtain the purified dataset $\gD_p$ as in \eqref{eq:pure};
  \STATE {\bf Purified score construction}: construct the purified score function as $S_p(\cdot)=\gA(\{X_j\}_{j\in\gL_1},\{X_j\}_{j\in\gP})$;
  \STATE {\bf P-value computation}: take $\gD_0$ as the calibration set, compute p-values
  \begin{equation*}
      p_j=\frac{1}{|\mathcal{L}_0|+1}\sum_{i\in \mathcal{L}_0\cup\{j\}}\I\left\{S_p(X_j)\leq S_p(X_i)\right\},\;\; j\in\gU;
  \end{equation*}
  \STATE {\bf Testing procedure}: apply the BH procedure (for ECOT-bi) or the third step in the unified framework (for ECOT-oc) to $\{p_j\}_{j\in\gU}$ at level $\alpha$, obtain the rejection set $\gR$;
  \vspace{0.05in}
  \ENSURE Rejection set $\gR$.
  \end{algorithmic}
\end{algorithm}

\begin{Cor}
    Suppose Assumption \ref{Assum:data_exch} holds. The rejection set $\gR$ output by Algorithm \ref{alg:pure} ensures ${\rm FDR}\leq \alpha\E[|\mathcal{H}_0|/|\gU|]\leq\alpha$.
\end{Cor}

One may be concerned that this purification step may remove too many samples from the dataset used for training. However, in typical applications we have $|\gL_1| \ll |\gL_0| + |\gU|$. Therefore, reducing the size of $\{X_j\}_{j\in\gL_0\cup\gU}$ will not degrade model efficiency significantly. In experiments, we find that taking $\eta=0.5$ yields good performance in most scenarios. Moreover, the purified versions of ECOT can be naturally integrated into ECOT-as for approach selection. For a more robust implementation, one may also consider multiple values of $\eta$ and perform the selection procedure among them. 

\begin{figure}[h]
            \centering
            \includegraphics[width=0.8\linewidth]{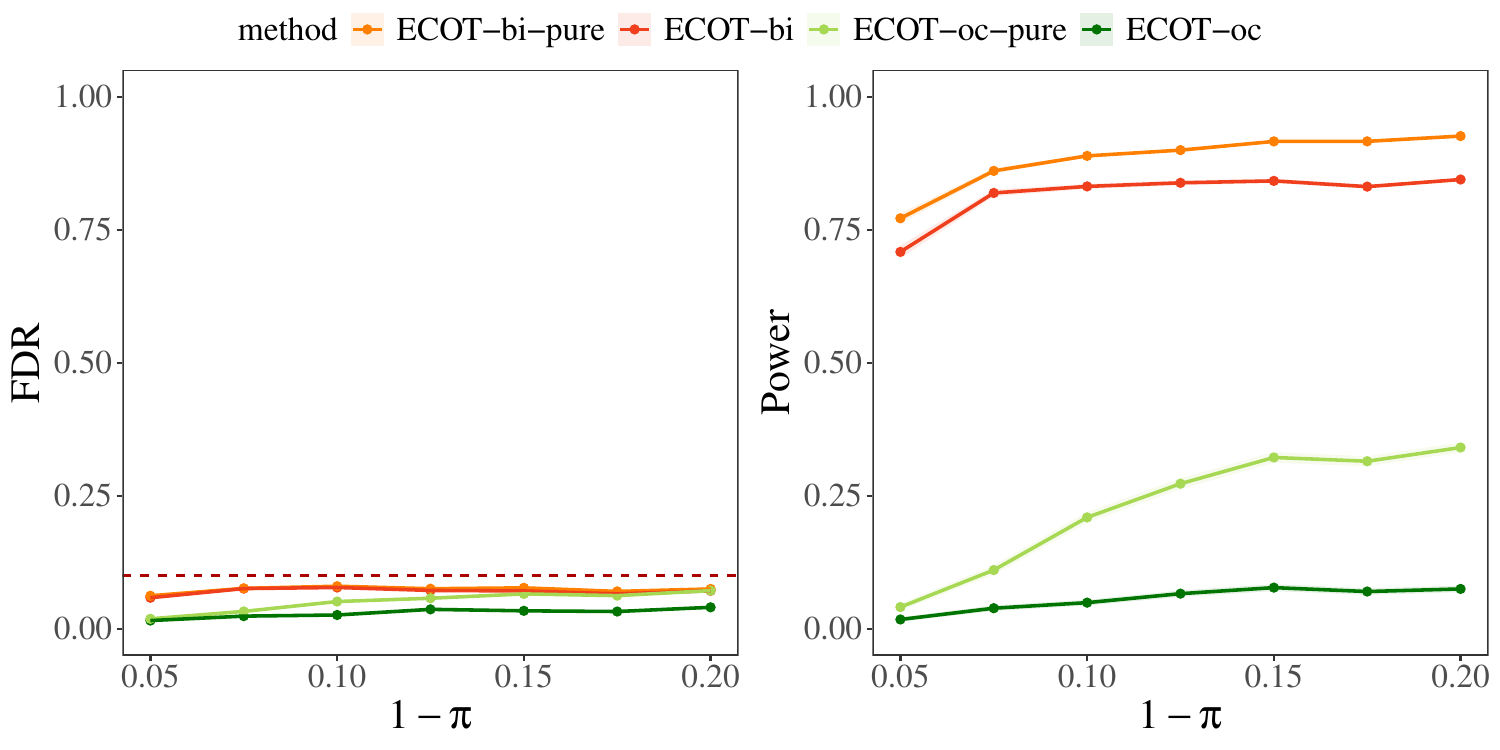}
            \caption{FDR and power of ECOT methods with (-pure) and without the purification step. The red dotted line marks the nominal level $\alpha=0.1$.}
            \label{fig:CLEAN-supp}
        \end{figure}

Here we illustrate the advantage of purification by a simulated example on our binary classification setting (Section \ref{sec:bi_prefer}). Figure \ref{fig:CLEAN-supp} shows that the purified ECOT procedure still controls the FDR below the nominal level. It improves the power of both ECOT-bi and ECOT-oc particularly when the proportion of alternative samples is large.

\subsection{Unified framework based on label-integrated conformal p-values}\label{appex_sec:jin_pvalue}
Under a sample selection scenario, \citet{jin2023selection} proposed an additional form of conformal p-values. Our unified ECOT can be extended to accommodate their setting, offering an alternative perspective on the framework in \citet{bai2024optimized} through the lens of full permutation.

Define the label-integrated score function $S^{(j)}$ to map from $\gX \times \{0,1\}$ to $\mathbb{R}$ (instead of $\gX \mapsto \mathbb{R}$), which is a function of both $X$ and $Y$, and assume it is monotonic in the label, i.e., $S^{(j)}(x,0) \geq S^{(j)}(x,1)$.\footnote{This is the reverse of the original definition in \citet{jin2023selection}, where $S^{(j)}(x,0) \leq S^{(j)}(x,1)$ was used. The revised form aligns closer with our p-value formulation and is adopted here for consistency.} Denote $\gL=\gL_0\cup\gL_1$ as the full labelled dataset. Moreover, we also define a set of pseudo-labels $\{\tilde{Y_k}\}_{k \in \gL \cup \gU}$, where $\tilde{Y}_k = Y_k$ for $k \in \gL$ and $\tilde{Y}_k = 0$ for $k \in \gU$. Let $\gC\subseteq \gL$ be the calibration set, and let $\Omega_j$ be the set of all
permutations of $\gL\cup\gU$ that fix every index outside $\gC\cup\{j\}$. We then consider the following ECOT procedure based on label-integrated p-values.

\begin{itemize}
     \item \textbf{Score construction}: construct individualised score function $S^{(j)}$ for each test sample $j\in\gU$, based on available data $\gD_0,\gD_1,\gD_u$. 
     \item \textbf{P-value computation}: take a subset $\gC\subseteq \mathcal{L}$, and compute label-integrated conformal p-value for the $j$-th sample as
     \begin{equation}\label{eq:perm_pvalue_full}
         p_j=\frac{1}{|\Omega_j|}\sum_{\sigma\in\Omega_j}\I\{{S}^{(j)}(X_j,\tilde{Y}_j)\leq S^{(j)}_{\sigma}(X_{\sigma(j)},\tilde{Y}_{\sigma(j)})\}
     \end{equation}
    where $S_{\sigma}^{(j)}$ is the score function constructed on the dataset permuted by $\sigma$.
    \item \textbf{Testing procedure}: the testing step is the same as the conditional calibration step in the main ECOT framework, with the modified p-values
  \begin{equation}\label{eq:modi_perm_pvalue_full}
     \tilde{p}_\ell^{(j)}=\frac{1}{|\Omega_j|}\sum_{\sigma\in\Omega_j}\I\{\tilde{S}^{(j)}(X_\ell,\tilde{Y}_\ell)\leq S^{(j)}_\sigma(X_{\sigma(j)},\tilde{Y}_{\sigma(j)})\}\quad \ell\neq j\quad\text{and}\quad \tilde{p}_j^{(j)}=0.   
    \end{equation}
    Here $\tilde{S}^{(j)}(X_\ell,\tilde{Y}_\ell)=\text{Median}\{{S}^{(j)}_\sigma(X_\ell,\tilde{Y}_\ell):\sigma\in\Omega_j\}$.
 \end{itemize}

To establish theoretical guarantees in this label-integrated setting, we impose a stronger exchangeability assumption than Assumption~\ref{Assum:data_exch}. While Assumption~\ref{Assum:data_exch} only requires exchangeability of the null data $\big((X_i,0): i\in\gL\cup\gU,\ Y_i=0\big)$, the assumption adopted in \citet{jin2023selection} extends exchangeability to all labelled and unlabelled samples.
\begin{assump}\label{Assum:data_full_exch}
    $\Big((X_i,Y_i):i\in \gL\cup\gU\Big)$ are exchangeable.
\end{assump}

\begin{The}\label{thm:FDR_control_jin}
    Suppose Assumption \ref{Assum:data_full_exch} holds. Then with a little abuse of the notations,
    \begin{itemize}
        \item[(i)] The conformal p-value constructed in \eqref{eq:perm_pvalue_full} satisfies
        $$\Pr(p_j\leq t, Y_j=0\mid\Psi_j)\leq t\quad\text{for any }t\in[0,1],$$
        where $$\Psi_j=\Big( \big((X_k,Y_k):k\in  (\gU\setminus\{j\})\cup(\mathcal{L}\setminus\gC)\big),\big\{(X_k,Y_k):k\in\gC\cup\{j\}\big\}\Big),$$
        which contains an unordered set of covariate-response pairs in $\gC\cup\{j\}$ and the remaining data.
        \item[(ii)] The final rejection set $\gR$ output by the unified procedure satisfies $\FDR\leq \alpha$.
    \end{itemize}
    
\end{The}
The full permutation construction can be reduced when the label-integrated score
functions have suitable symmetry properties, in direct analogy with the reductions in
Sections~\ref{sub_sec:cali_sym} and~\ref{sub_sec:joint_sym}.

\begin{Cor}\label{cor:label_integ_no_weight}
Suppose Assumption \ref{Assum:data_full_exch} holds. Then the following simplifications
hold with the same FDR control.
\begin{itemize}
    \item[(i)] If $\{S^{(j)}\}_{j\in\gU}$ is calibration-symmetric with respect to
$\{(X_i,\tilde Y_i):i\in\gC\cup\{j\}\}$, then the final rejection set $\gR$ obtained based on label-integrated p-values is equivalent to that obtained by replacing the  \eqref{eq:perm_pvalue_full} and \eqref{eq:modi_perm_pvalue_full} with their reduced forms as
    $$  p_j=
\frac{\sum_{i\in\gC\cup\{j\}}\I\{S^{(j)}(X_j,\tilde{Y}_j)\leq S^{(j)}(X_{i},\tilde{Y}_i)\}}{|\gC|+1};\quad \tilde{p}_\ell^{(j)}=\frac{\sum_{i\in\gC\cup\{j\}}\I\{{S}^{(j)}(X_\ell,\tilde{Y}_\ell)\leq S^{(j)}(X_{i},\tilde{Y}_i)\}}{|\gC|+1}.$$
\item[(ii)] If $S^{(j)}\equiv S$ for all $j\in\gU$ and $S$ is constructed
symmetrically with respect to $\{(X_i,\tilde Y_i):i\in\gC\cup\gU\}$, then it is equal to the set output by the BH procedure applied to the conformal p-values constructed by
$$      p_j=\frac{\sum_{i\in\gC\cup\{j\}}\I\{S(X_j,\tilde{Y}_j)\leq S(X_i,\tilde{Y}_i)\}}{|\gC|+1},\quad j\in\gU.
$$
\end{itemize}
\end{Cor}
Thus, the full-permutation view also covers the label-integrated conformal p-values
of \citet{jin2023selection}; the usual simplified implementations arise as symmetry
reductions of the same construction. This perspective also helps clarify
the connection to \citet{bai2024optimized}: their OptCS-MSel procedure for model selection fits naturally into the reduced 
form described in Corollary~\ref{cor:label_integ_no_weight}-(i).

\subsection{Unified framework based on weighted conformal p-values under covariate shift}\label{appex_sec:covariate_shift}
In this subsection, we extend our framework to the covariate-shift setting \citep{tibshirani2019conformal}. 
Specifically, we adopt a weighted full permutation formulation for the label-integrated 
framework of \citet{jin2023selection}. We begin by introducing the data-generating 
assumption that allows for covariate shift between the labelled and test samples.
%Under the sample selection setting of \citet{jin2023selection}, conformal p-values are constructed by integrating the label information into the score function. This important line of work has subsequently been extended to handle covariate shift via weighting \citep{jin2023modelfree} and to allow data-adaptive score selection \citep{bai2024optimized}. In this subsection, we show that our ECOT framework can be extended to accommodate these label-integrated conformal p-values, thereby providing a full-permutation perspective on this family of methods. To make it more general, we consider a covariate shift setting \citep{tibshirani2019conformal}.

%Let the labelled dataset be $\gD_l=\gD_0\cup\gD_1$, with index set $\gL=\gL_0\cup\gL_1$, and let $\gC\subseteq \gL$ be a calibration set drawn from the full labelled data rather than from $\gL_0$ only. To make it more general, we allow for covariate shift between the labelled and test samples and impose the following data generating assumption.

\begin{assump}[Covariate shift]\label{Assum:covariate_shift}
Let $P_X$ and $Q_X$ denote the covariate distributions of the labelled and test samples, respectively, and let $P_{Y\mid X}$ denote the common conditional distribution of the response given the covariate. The data $\{(X_i,Y_i)\}_{i\in\gL\cup\gU}$ are independent, and
\[
X_i\sim P_X\text{ for }i\in\gL,\quad  
X_j\sim Q_X\text{ for }j\in\gU,\quad 
Y_k\mid X_k\sim P_{Y\mid X}\text{ for }k\in\gL\cup\gU.
\]
Moreover, assume that $Q_X$ is absolutely continuous with respect to $P_X$.
\end{assump}
The covariate shift is encoded by the Radon--Nikodym derivative $w(x)=dQ_X/dP_X(x)$, which serves as the weight function in the construction of weighted conformal $p$-values. We then consider the following ECOT procedure based on weighted label-integrated p-values.

\begin{itemize}
     \item \textbf{Score construction}: construct individualised label-integrated score function $S^{(j)}$ for each test sample $j\in\gU$, based on available data $\gD_0,\gD_1,\gD_u$. 
     \item \textbf{P-value computation}: take a subset $\gC\subseteq \mathcal{L}$ as calibration set, and compute weighted label-integrated p-value for the $j$-th sample as \begin{equation}\label{eq:perm_pvalue_full_weighted}
p_j=\frac{\sum_{\sigma\in\Omega_j}w(X_{\sigma(j)})\I\{{S}^{(j)}(X_j,\tilde{Y}_j)\leq S^{(j)}_{\sigma}(X_{\sigma(j)},\tilde{Y}_{\sigma(j)})\}}{\sum_{\sigma\in\Omega_j}w(X_{\sigma(j)})}.
     \end{equation}
    \item \textbf{Testing procedure}: the testing step is the same as the conditional calibration step in the main ECOT framework, with the weighted modified p-values
\begin{equation}\label{eq:modi_perm_pvalue_full_weighted}
\tilde{p}_\ell^{(j)}=\frac{\sum_{\sigma\in\Omega_j}w(X_{\sigma(j)})\I\{\tilde{S}^{(j)}(X_\ell,\tilde{Y}_\ell)\leq S^{(j)}_\sigma(X_{\sigma(j)},\tilde{Y}_{\sigma(j)})\}}{\sum_{\sigma\in\Omega_j}w(X_{\sigma(j)})}\quad \ell\neq j\quad\text{and}\quad \tilde{p}_j^{(j)}=0.   
    \end{equation}
    Here $\tilde{S}^{(j)}(X_\ell,\tilde{Y}_\ell)=\text{Median}\{{S}^{(j)}_\sigma(X_\ell,\tilde{Y}_\ell):\sigma\in\Omega_j\}$.
 \end{itemize}

\begin{The}\label{thm:FDR_control_jin_weighted}
    Suppose Assumption \ref{Assum:covariate_shift} holds. 
    \begin{itemize}
        \item[(i)] The conformal p-value constructed in \eqref{eq:perm_pvalue_full_weighted} satisfies
        $$\Pr(p_j\leq t, Y_j=0\mid\Psi_j)\leq t\quad\text{for any }t\in[0,1],$$
        where $\Psi_j=\Big( \big((X_k,Y_k):k\in  (\gU\setminus\{j\})\cup(\mathcal{L}\setminus\gC)\big),\big\{(X_k,Y_k):k\in\gC\cup\{j\}\big\}\Big).$
        \item[(ii)] The final rejection set $\gR$ output by the unified procedure based on weighted label-integrated p-values satisfies $\FDR\leq \alpha$.
    \end{itemize}
    
\end{The}
The same symmetry reduction also directly applies in the weighted case.

\begin{Cor}\label{cor:label_integ_cali_sym}
   Suppose that the series of label-integrated score functions $\{S^{(j)}\}_{j\in\gU}$ is
calibration-symmetric in the sense that, for each $j\in\gU$, $S^{(j)}$ is constructed
symmetrically with respect to
$\{(X_i,\tilde Y_i): i\in\gC\cup\{j\}\}$.
Then the final rejection set $\gR$ obtained from the weighted label-integrated ECOT procedure is
equivalent to that obtained by replacing \eqref{eq:perm_pvalue_full_weighted} and
\eqref{eq:modi_perm_pvalue_full_weighted} with
    $$  p_j=
\frac{\sum_{i\in\gC\cup\{j\}}w(X_i)\I\{S^{(j)}(X_j,\tilde{Y}_j)\leq S^{(j)}(X_{i},\tilde{Y}_i)\}}{\sum_{i\in\gC\cup\{j\}}w(X_i)};\quad \tilde{p}_\ell^{(j)}=\frac{\sum_{i\in\gC\cup\{j\}}w(X_i)\I\{{S}^{(j)}(X_\ell,\tilde{Y}_\ell)\leq S^{(j)}(X_{i},\tilde{Y}_i)\}}{\sum_{i\in\gC\cup\{j\}}w(X_i)}.$$
\end{Cor}

This corollary recovers the weighted label-integrated conformal p-values of
\citet{jin2026modelfree} from the full-permutation perspective. When $P_X=Q_X$,
we have $w(x)\equiv 1$, and the construction reduces to the unweighted
label-integrated setting in Section~\ref{appex_sec:jin_pvalue}.

The weighting device above can in principle also be adapted to our main
setting when the covariate shift affects only the null samples \citep{lee2025full}. In that case, however, estimating the weight function is substantially more difficult in practice,
because the null test samples are not separately observed. For this reason, we only pursue
the weighted extension in the label-integrated setting considered here.

\subsection{A special class of Jackknife-type score functions}\label{appex_sub_sec:jk_pvalue}

We consider an additional special case of score function, where our procedure also simplifies to the BH method. Specifically, for each $j \in \gU$, the score function is constructed using a leave-one-out strategy.

\begin{Def}[Jackknife-type score function]\label{Assum:jack_type}
The collection of score functions $\{S^{(j)}\}_{j\in\gU}$ is Jackknife-type, if for each $j\in\gU$, the score function $S^{(j)}$ is constructed symmetrically with respect to $\{X_i:i\in\gC\cup\gH_0\setminus\{j\}\}$.
\end{Def}

While Definition \ref{Assum:jack_type} is not implied by Definition \ref{Assum:cali_sym}, it does follow from Definition \ref{Assum:joint_sym}. In practice, a sufficient condition is that
$S^{(j)}$ is constructed symmetrically with respect to
$\{X_i:i\in\gC\cup(\gU\setminus\{j\})\}$ for each $j\in\gU$. Given its special structure, we illustrate how our unified procedure reduces when the score functions satisfy Definition \ref{Assum:jack_type}. The simplified form is as follows:

\begin{itemize}
     \item \textbf{Score construction}: take a subset $\gC\subseteq \mathcal{L}_0$ and construct individualised score function $S^{(j)}$ satisfying Definition \ref{Assum:jack_type} for each test sample $j\in\gU$, based on available data $\gD_0,\gD_1,\gD_u$;
     \item \textbf{P-value computation}: compute p-value for the $j$-th sample by
\begin{equation}\label{eq:jack_pvalue}
         p_j=\frac{1}{|\gC|+1}\sum_{i\in\gC\cup\{j\}}\I\{{S}^{(j)}(X_j)\leq S^{(j)}_{\sigma(i,j)}(X_{i})\}
     \end{equation}
    where $\sigma(i,j)$ denotes the permutation that only swaps the position of $i$ and $j$.
    \item \textbf{Testing procedure}: perform BH procedure over the p-values in \eqref{eq:jack_pvalue}.
    \end{itemize}
Note that $S^{(j)}_{\sigma(i,j)}$ can be interpreted as a score function symmetric to the dataset $\gC \cup \gU \setminus \{i\}$, and thus can be denoted as $S^{(i)}$ for clarity. This procedure coincides with our unified framework, except for a minor adjustment to the modified p-values.
\begin{Pro}\label{pro:jack_type}
    Suppose Assumption \ref{Assum:data_exch} holds and the score functions $\{S^{(j)}\}_{j\in\gU}$ are Jackknife-type. The final rejection set $\gR$ output by the unified procedure ECOT replacing the modified p-values with  $$\tilde{p}_\ell^{(j)}=\frac{1}{|\Omega_j|}\sum_{\sigma\in\Omega_j}\I\{{S}^{(\ell)}(X_\ell)\leq S^{(j)}_\sigma(X_{\sigma(j)})\}\quad \ell\neq j\quad\text{and}\quad \tilde{p}_j^{(j)}=0$$
    is equivalent to the set output by the BH procedure applied to conformal p-values constructed in \eqref{eq:jack_pvalue}.
\end{Pro}
This Jackknife-type score function avoids data splitting by leveraging leave-one-out symmetry, as also considered in \citet{bai2024optimized}. However, it still requires $|\gC| + m$ score construction, whereas our ECOT-bi method only requires a single model fitting.

\section{More discussions on adaptive approach selection}

\subsection{Technique to improve discrimination}\label{sec:loo}

In Section \ref{sec:appro_select}, we propose to select among candidate approaches for each test sample $X_j$ by choosing the one leading to maximum number of rejections. Specifically, we compute modified $p$-values on $\gD_u \setminus \{X_j\}$ and apply the BH procedure to obtain rejection counts. If we define the modified $p$-value in the same way as the testing $p$-value,
\begin{equation}
    \tilde{p}_{l,\rm naive}^{(j)}=\frac{\sum_{i\in\gC\cup\{j\}}\I\{S(X_i)\geq S(X_l)\}+1}{|\gC|+2},
\end{equation}
then for $H_1$ samples $(X_j, Y_j)$, the score $S(X_j)$ tends to be large. As a result, the indicator term $\I\{S(X_j) \geq S(X_l)\}$ is more likely to equal 1, which can substantially reduce the number of rejections. In some cases, this leads to only a single rejection for all candidate approaches (note that at least one rejection always occurs since $\tilde{p}_j^{(j)} \equiv 0$).
        
To resolve this issue and make candidate approaches distinguishable, we remove the contribution of the largest score in ${S(X_i)}_{i \in \gC \cup \{j\}}$ and define the regularised $p$-value as
\begin{equation}\label{eq:reg}
    \tilde{p}_{l,\rm reg}^{(j)}=\frac{\sum_{i\in\gC\cup\{j\}\backslash\{j^*\}}\I\{S(X_i)\geq S(X_l)\}+1}{|\gC|+1},
\end{equation}
    where $j^* = \arg\max_{i \in \gC \cup \{j\}} S(X_i)$. This effectively regularises the modified $p$-values for $H_1$ samples in $\gD_u$, while having minimal impact on $H_0$ samples.        

    However, in our experiments we also find that when the testing task is difficult (e.g., with small sample sizes or weak signals), even using regularised p-values, many samples in $\gD_u$ yield only 1 rejection across all candidate approaches. This does not mean all approaches will have 0 power but is rather a granularity issue of conformal $p$-values. It occurs less frequently for well-performing approaches.
        
    To further mitigate this issue, we first use the regularised p-value to compute rejection counts for candidate approaches. When all approaches have only 1 rejection, we refine the p-value by subtracting 1 on both the denominator and numerator and make it more aggressive
    \begin{equation}\label{eq:mod}
        \tilde{p}_{l,\rm ref}^{(j)}=\frac{\sum_{i\in\gC\cup\{j\}\backslash\{j^*\}}\I\{S(X_i)\geq S(X_l)\}}{|\gC|}.
    \end{equation}
    Without the +1 term, this p-value is no longer lower bounded by $1/(|\gC|+1)$ and allows for finer differentiation among candidate approaches. Empirically, it enables the best-performing approach to achieve the largest rejection count in most cases. Again, we never distinguish $X_j$ from $\gD_c$ in all above operations. Therefore, the FDR control result still applies. In all our experiments, we apply this technique to improve the robustness of ECOT-as.

\subsection{Detailed algorithm of approach selection strategy with full permutation }\label{appen_sub_sec:ECOT-as}
We present the detailed algorithm of approach selection strategy with full permutation in Algorithm \ref{alg:ECOT-as}.

\begin{algorithm}[htbp]
  \caption{Enhanced Conformal Testing - adaptive approach selection}\label{alg:ECOT-as}
  \begin{algorithmic}[1]
  \REQUIRE Labelled data $\gD_0,\gD_1$ and test data $\gD_u$; FDR target level $\alpha\in(0,1)$; $K$ candidate conformal testing approaches, each $k\in [K]$ has 
  score functions $\{S^{(j),k}\}_{j\in\gU}$ and a calibration set $\gC_k$ 
  \STATE {\bf Score construction}: for $j\in\gU$ and each $k\in [K]$, compute the evaluation criterion $R_k=R\left(\big\{(S^{(j),k},X_{j}):j\in\gU\big\},\big\{X_{i}:i\in\gC_k\big\}\right)$, which is obtained by running $k$-th procedure at level $\alpha$ to original p-values $\{{p}_\ell^{k}:\ell\in\gU\}$, where
  $$\frac{1}{|\Omega_j|}\sum_{\sigma\in\Omega_j}\I\{S^{(j),k}(X_j)\leq S^{(j),k}_{\sigma}(X_{\sigma(j)})\}.$$
Then the best approach is defined as
$$k^*=\arg\max_{k\in [K]} R_k.$$
And the $j$-th score function is $S^{(j),k^*}$;  
  \STATE {\bf P-value computation}: take $\gC=\bigcup_{k\in[K]}C_k$ as the calibration set and define $\Omega_j$ as the sets of all {permutations} of 
    $\mathcal{L}_0\cup \mathcal{L}_1\cup\gU$ that fixes indices outside of $\gC\cup\{j\}$. Compute p-values as
  \begin{equation*}
      p_j=\frac{1}{|\Omega_j|}\sum_{\sigma\in\Omega_j}\I\left\{S^{(j),k^*}(X_j)\leq S_{\sigma}^{(j),k^*_{\sigma}}(X_{\sigma(j)})\right\},\;\; j\in\gU.
  \end{equation*}
  Here $$k^*_{\sigma}=\arg\max_{k\in [K]} R_k^{\sigma},$$ and $R_k^\sigma=R\left(\big\{(S^{(j),k}_\sigma,X_{\sigma(j)}):j\in\gU\big\},\big\{X_{\sigma(i)}:i\in\gC_k\big\}\right)$ is the rejection number for $k$-th approach performed on the permuted datasets;
  \STATE {\bf Testing procedure}: apply the conditional calibration procedure over $\{p_j\}_{j\in\gU}$ at level $\alpha$. Specifically, the $\gR_j$ is the rejection set by applying BH procedure over $\{\tilde{p}_\ell^{(j),k^*}\}_{\ell\in\gU}$ as
    $$\tilde{p}_\ell^{(j),k^*}=\frac{1}{|\Omega_j|}\sum_{\sigma\in\Omega_j}\I\{\tilde{S}^{(j),k^*}(X_\ell)\leq S^{(j),k^*_{\sigma}}_\sigma(X_{\sigma(j)})\}\quad \ell\neq j\quad\text{and}\quad \tilde{p}_j^{(j)}=0.$$
    Here $\tilde{S}^{(j),k^*}(X_\ell)=\text{Median}\{{S}^{(j),k^*_\sigma}_\sigma(X_\ell):\sigma\in\Omega_j\}$;
  \ENSURE Rejection set $\gR$.
  \end{algorithmic}
\end{algorithm}

\subsection{Alternative implementation of adjusted approach selection}\label{appen_sub_sec:alter_imple}
 If all score functions satisfy $S^{(j),k}=S^{k}$ and adhere to joint-symmetry, we can also construct the final selected score function $S^{k^*}$ satisfying joint-symmetry, thereby enabling the implementation of the BH procedure. However, evaluating approaches based on rejection numbers introduces asymmetry between $\gC\cup\gU$, as $R_k$ is a function of $\big\{(S^{(j),k},X_{j}):j\in\gU\big\}$ and $\big\{X_{i}:i\in\gC\big\}$. Previous literature \citep{adadetect} addressed this by employing additional data splitting (See the next subsection for more details). 
 To avoid this, we propose an alternative criterion. Let $$M^k=\frac{1}{|\mathcal{L}_0\cup\gU|}\sum_{i\in \mathcal{L}_0\cup\gU}\frac{\sum_{\ell\in \mathcal{L}_1}\mathbb{I}\{S^{k}(X_\ell)\leq S^k(X_i)\}}{|\mathcal{L}_1|}.$$ A smaller $M^k$ indicates $k$-th approach tends to produce smaller p-values for non-nulls. We then select the best approach by $k^*=\arg\min_{k\in [K]}M^k$. The final score function $S^{k^*}$ satisfies the joint symmetry requirement, and the procedure remains simple and effective by directly implementing BH procedure over constructed p-values.

\subsection{Connections to existing model selection strategies in conformalized multiple testing}\label{appen_sub_sec:model_selection}

We review existing model selection strategies in conformalized multiple testing and show how they are encompassed by our unified framework. In addition, the selection strategy proposed in Algorithm \ref{alg:ECOT-ras} offers an alternative model selection approach applicable to prior methods.

Specifically, we assume the availability of $K$ different models, leading to $K$ different sequences of score functions, denoted as $\{S^{(j),k}\}_{j\in\gU}$ for each $k \in [K]$.

\paragraph{Model selection for basic conformal p-values}

\citet{zhangautoms} proposed a direct model selection approach, Auto-MS, for basic conformal p-values \citep{bates2023testing}. In this case, for all $j\in\gU,k\in[K]$, the score function satisfies $S^{(j),k}\equiv S^k$, where each $S^k$ is constructed solely from $\gD_t$ and satisfies Definition \ref{Assum:joint_sym}.

To select the best model, Auto-MS first constructs conformal p-values using \eqref{eq:marg_cp} based on calibration set $\gC$ for each candidate model, then applies the BH procedure over $\gU$ to identify the model $k^*=\arg\max_{k\in[K]}R_k$, which yields the largest number of rejections. However, they then reuse model $k^*$ to reconstruct p-values without any adjustment. In practice, this means the selected score function depends not only on $\gD_t$, but also on $\gD_c$ and $\gD_u$, thus violating joint-symmetry. As a result, Auto-MS offers only asymptotic FDR control, and empirical studies show that it can substantially inflate FDR in finite samples.

Our approach addresses this issue by adjusting the score function after model selection. Below, we present a model-selection-adapted version of Algorithm \ref{alg:ECOT-ras} for basic conformal p-values:

\begin{itemize}
    \item \textbf{Score construction}: for each $k\in[K]$, compute the evaluation criterion $R^{(j)}_{k}=R(S^{(j)},\{X_\ell:\ell\in\gC\cup\{j\}\},\{X_i:i\in\gU\setminus\{j\}\})$, which is the rejection number by applying BH procedure at level $\alpha$ to conformal p-values in $\gU\setminus\{j\}$, and the conformal p-values are constructed by using $S^{k}$ and treating $\gC\cup\{j\}$ as calibration set. The selected score function index is $k^*_j=\arg\max_{k\in[K]}R^{(j)}_{k}$ and the final score function is $S^{k^*_j}$.
   \item \textbf{P-value computation}: take $\gD_c$ as calibration set and $S^{k^*_j}$ as score function to compute conformal p-values in the form as
\begin{equation}\label{eq_appex:model_sel_baisc_p_value}
        p_j=\frac{1}{|\mathcal{C}|+1}\sum_{i\in \mathcal{C}\cup\{j\}}\I\left\{S^{k^*_j}(X_j)\leq S^{k^*_j}(X_i)\right\},\;\; j\in\gU.
   \end{equation}
    \item \textbf{Testing procedure}: run conditional calibration procedure, where the modified p-values are given by
\begin{equation}\label{eq_appex:model_sel_modifed_baisc_p_value}
    \tilde{p}_\ell^{(j)}=\frac{1}{|\mathcal{C}|+1}\sum_{i\in \mathcal{C}\cup\{j\}}\I\left\{S^{k^*_j}(X_\ell)\leq S^{k^*_j}(X_i)\right\},\;\; \ell\neq j\quad \text{and}\quad\tilde{p}_j^{(j)}=0.
   \end{equation}
    
\end{itemize}

\paragraph{Model selection for AdaDetect}

For AdaDetect, \citet{adadetect} proposed a model selection strategy that involves splitting the labelled null data into three subsets $\gD_{t,a},\gD_{t,b},\gD_c$.  We index these as $\gT_a,\gT_b$ and $\gC$, respectively. Their model selection process can be reformulated within our unified framework as follows:

\begin{itemize}
    \item \textbf{Score construction}: for each candidate model $k$, train a score function $S^{k}$ using a binary classifier to distinguish $\{X_j\}_{j\in\gT_{a}}$ from $\{X_j\}_{j\in\gT_{b}\cup\gC\cup\gU}$. Then compute the evaluation criterion $R^{\rm ada}_k=R(S^k,\{X_j\in\gC\cup\gU\},\{X_i:i\in\gT_{b}\})$, defined as the number of rejections obtained by applying the BH procedure at level $\alpha$ to conformal p-values on $\gC\cup\gU$, using $\gT_{b}$ as calibration set. The selected model is
$k^*=\arg\max_{k\in[K]}R^{\rm ada}_k$ and the final score function is $S^{k^*}$.
    \item \textbf{P-value computation}: take $\gD_c$ as calibration set and $S^{k^*}$ as score function to compute conformal p-values in the form as \eqref{eq:marg_cp}.
    \item \textbf{Testing procedure}: run BH procedure over computed conformal p-values.
\end{itemize}

Since model selection is treated as an integral part of score construction, the final score function $S^{k^*}$ still satisfies Definition \ref{Assum:joint_sym}, ensuring that the entire procedure remains within our theoretical framework and retains FDR control.

Although AdaDetect's model selection strategy is straightforward and theoretically justified, the additional data splitting may reduce statistical power. Furthermore, evaluating model quality based on rejections over $\gC\cup\gU$ can deviate from the primary target—rejections over $\gU$ alone.

To address these limitations, our model selection strategy from Algorithm \ref{alg:ECOT-ras} can also be adapted for AdaDetect. It avoids extra data splitting and evaluates model quality more directly based on the target test set. The adapted version is as follows:

\begin{itemize}
    \item \textbf{Score construction}: train candidate score function $S^{k}$ based on binary classification to distinguish $\{X_j\}_{j\in\gT}$ and $\{X_j\}_{j\in\gC\cup\gU}$. For $j$-th test sample, compute the evaluation criterion $R_{k}^{(j)}:=R(S^{(j),k},\{X_\ell\in\gU\setminus\{j\}\},\{X_i:i\in\gC\cup\{j\}\})$. The best score function is $k^*_j=\arg\max_{k\in[K]}R_{k}^{(j)}$ and the final $j$-th score function is $S^{k^*_j}$.
    \item \textbf{P-value computation}: take $\gD_c$ as calibration set and $S^{k^*_j}$ as score function to compute conformal p-values in the form as in \eqref{eq_appex:model_sel_baisc_p_value}
    \item \textbf{Testing procedure}: run conditional calibration procedure with the modified p-values \eqref{eq_appex:model_sel_modifed_baisc_p_value}.
\end{itemize}

\paragraph{Model selection for integrative conformal p-values}

\citet{liang2024integrative} selected the best model for each $j\in\gU$ individually based on predictive performance. We first review how the score function for integrative conformal p-values is constructed. Split both datasets $\gD_0=\gD_{t}\cup\gD_{c},\gD_1=\gD_{1,t}\cup\gD_{1,c}$. One-class classifiers $s_0$ and $s_1$ are trained separately on $\gD_{t}$ and $\gD_{1,t}$. For each $j\in\gU$, the initial $p$-values $\widehat{u}_0$ and $\widehat{u}_1$ are computed using the corresponding score functions and calibration datasets as $\gD_{c}\cup\{X_j\}$ and $\gD_{1,c}$, i.e.
$${u}_{0,j}(X_j)=\frac{\sum_{i\in\gC\cup\{j\}}\I\{s_0(X_j)\leq s_0(X_i)\} }{|\gC|+1},\quad {u}_1(X_j)=\frac{\sum_{i\in \mathcal{L}_{1,c}}\I\{s_1(X_j)\leq s_1(X_i)\} }{|\mathcal{L}_{1,c}|+1}, $$
  where $\mathcal{L}_{1,c}$ is the index set of $\gD_{1,c}$. The final score function is defined as $\widehat{u}_0/\widehat{u}_1$, the ratio of null initial p-value and alternative initial p-value, making it symmetric with respect to $\gD_c\cup\{X_j\}$.

Next, we describe how \citet{liang2024integrative} selected the best model using a criterion tied directly to the predictive performance of the one-class classifiers, assessing how well they separate inliers from outliers. This model selection strategy can also be formulated within our unified framework:

\begin{itemize}
    \item \textbf{Score construction}: for each $j\in\gU$, train a set of candidate one-class classifiers $s_0^k$ and $s_1^k$. The evaluation criterion for the null model is ${\rm MD}^{(j)}(s_0^k)$, %$${\rm MD}^{(j)}(s_0^k)={\rm Median\,Difference}(\{s_0^k(X_i),i\in\gC\cup\{j\}\},\{s_0^k(X_i),i\in\gT_{1,c}\}) $$
    %$${\rm MD}^{(j),k}_0={\rm Median}\{s_0^k(X_i),i\in\gC\cup\{j\}\}-{\rm Median}\{s_0^k(X_i),i\in\gT_{1,c}\} $$
    which is the median difference between the null classifier scores $s_0^k$ evaluated on $\gC\cup\{j\}$ and those evaluated on $\gT_{1,c}$. The best model for $s_0^k$ is $k^{(j),*}_0=\arg\max_{k\in[K]} {\rm MD}^{(j)}(s_0^k)$.
    Similarly, the selected model for $s_1^k$ is $k^{(j),*}_1=\arg\max_{k\in[K]} {\rm MD}^{(j)}(s_1^k)$. Then the final score function $S^{(j)}$ is constructed based on $s_0^{k^{(j),*}_0}$ and $s_1^{k^{(j),*}_1}$.
    \item \textbf{P-value computation}: take $\gD_c$ as calibration set. Compute final p-value by
        $$p_j=\frac{\sum_{i\in\gC\cup\{j\}}\I\{S^{(j)}(X_j)\leq S^{(j)}(X_i)\}}{|\gC|+1}.$$
        \item \textbf{Testing procedure}: perform conditional calibration to the constructed p-values. The modified conformal p-value differs slightly from \eqref{eq:reduced_p_modified_cali_sym} by adding 1 to both the numerator and denominator:
        $$     \tilde{p}_\ell^{(j)}=\frac{1}{|\gC|+2}\left(\sum_{i\in\gC\cup\{j\}}\I\{{S}^{(j)}(X_\ell)\leq S^{(j)}(X_{i})\}+1\right)\quad \ell\neq j\quad\text{and}\quad \tilde{p}_j^{(j)}=0.  $$
\end{itemize}
One key feature of the model selection strategy in \citet{liang2024integrative} is that the evaluation criterion is based on the predictive performance of the one-class classifiers. However, this criterion may not align with the goal of multiple testing—namely, maximising the number of rejections while controlling the FDR—and can potentially lead to suboptimal results. Our approach selection strategy can also be applied to integrative conformal p-values, offering a more direct and targeted method for model evaluation. The main difference is to replace the criterion ${\rm MD}^{(j)}(s_0^k)$ and ${\rm MD}^{(j)}(s_1^k)$ with $R_{k_1,k_2}^{(j)}$, defined as the number of rejections obtained by applying the BH procedure at level $\alpha$ to the modified $p$-values ${\tilde{p}_\ell^{(j),k_1,k_2} : \ell \in \gU}$, where
\begin{equation*}
      \tilde{p}_\ell^{(j),k_1,k_2}=\frac{1}{|\gC|+1}\sum_{i\in\gC\cup\{j\}}\I\{S^{(j),k_1,k_2}(X_\ell)\leq S^{(j),k_1,k_2}(X_i)\},\quad \ell\neq j\quad \text{and}\quad\tilde{p}_j^{(j),k_1,k_2}=0. 
\end{equation*}
Here, $S^{(j),k_1,k_2}$ denotes the score function constructed using the $k_1$-th null classifier and the $k_2$-th non-null classifier.

\section{Comparison with the OptCS framework in \cite{bai2024optimized}}

In this section, we discuss the distinction between our ECOT framework and the OptCS framework proposed in \cite{bai2024optimized}. 

The main difference between the two frameworks lies in their underlying problem settings. The setting in \citet{bai2024optimized} is primarily based on the conformal selection framework of \citet{jin2023selection}. In both works, the conformal p-values are constructed by using a mixture of null and alternatives samples as calibration data

\begin{equation*}
    p_j=\frac{\sum_{i\in\gC}\I\{S^{(j)}(X_i,Y_i)\geq S^{(j)}(X_{n+j},\tilde{Y}_{n+j})\}+1}{|\gC|+1},
\end{equation*}
where the calibration data $\{(X_i,Y_i):i\in\gC\}$ are randomly sampled from the whole labelled data $\gD_l=\gD_0\cup\gD_1$.
For these p-values to be valid, the labelled and unlabelled (test) data must be marginally exchangeable. This contrasts with our assumption that only null samples in the test data are exchangeable with those in the labelled data. The former assumption is more restrictive, as it requires the proportion of alternative ($H_1$) samples in the test data to be the same as that in the labelled data. If this proportion is smaller in the test set, the $p$-values defined in \citet{bai2024optimized} are no longer super-uniform, which can lead to inflated FDR. This is often the case in real-world applications. In contrast, the $p$-values in the ECOT framework are calibrated solely on the null labelled data and therefore admit the signal proportion to be either larger or smaller in the test data.

For the approach selection part, both the ECOT and OptCS frameworks apply the BH procedure to adjusted p-values to determine the optimal candidate with the largest rejection number. However, our scope is broader than the model selection task in \citet{bai2024optimized}. The ECOT framework considers an approach selection task, which can select among different conformal testing approaches. This includes selecting different prediction models within a single approach as a special case. 

Finally, our framework incorporates additional techniques developed in Sections \ref{sec:pure} and \ref{sec:loo}, which further enhance the performance of the methods within the ECOT framework. Beyond their role here, these techniques are also applicable to other conformal testing methods and may therefore be of independent interest.

\section{Additional experiment results}\label{sec:add_simu}
In this section, we provide additional experimental results regarding different varying parameters or settings not considered in the main text.

\subsection{Computational cost}\label{sec:time}
We provide computation time of different methods we mentioned in Section \ref{sec:numer}. For simulated experiments, we use the same setting to those in Section \ref{sec:bi_prefer} and \ref{sec:oc_prefer} with varying labelled sample sizes. For real-data experiments, we use exactly the same setting to that in the main text.

\begin{table}[htbp]
\centering
\caption{Computation time (seconds) of different methods in the first simulation setting.}\label{tab:re-time-set1-supp}
\vspace{0.5em}
\resizebox{\textwidth}{!}{%
\renewcommand{\arraystretch}{1} 
\begin{tabular}{ccccccccc}
\hline
Sample size & ECOT-as & ECOT-bi & ECOT-oc & CP-bi & CP-oc & Integ & FullND & AdaDetect \\
\hline
150 & 25.9 & 0.8 & 0.2 & 0.1 & 0.1 & 0.1 & 0.1 & 1.4 \\
350 & 33.6 & 1.1 & 0.2 & 0.2 & 0.1 & 0.1 & 0.1 & 2.0 \\
550 & 40.7 & 1.4 & 0.2 & 0.3 & 0.1 & 0.1 & 0.1 & 2.3 \\
750 & 47.9 & 1.7 & 0.2 & 0.4 & 0.1 & 0.1 & 0.1 & 2.7 \\
\hline
\end{tabular}}
\end{table}

\begin{table}[htbp]
\centering
\caption{Computation time (seconds) of different methods in the second simulation setting.}\label{tab:re-time-set2-supp}
\vspace{0.5em}
\resizebox{\textwidth}{!}{%
\renewcommand{\arraystretch}{1} 
\begin{tabular}{ccccccccc}
\hline
Sample size & ECOT-as & ECOT-bi & ECOT-oc & CP-bi & CP-oc & Integ & FullND & AdaDetect \\
\hline
100 & 23.5 & 1.0 & 0.2 & 0.1 & 0.1 & 0.1 & 0.1 & 1.1 \\
400 & 35.7 & 2.2 & 0.2 & 0.2 & 0.1 & 0.1 & 0.1 & 1.9 \\
700 & 47.0 & 3.3 & 0.2 & 0.5 & 0.1 & 0.1 & 0.1 & 2.4 \\
1000 & 58.9 & 4.5 & 0.2 & 0.8 & 0.1 & 0.1 & 0.1 & 2.9 \\
\hline
\end{tabular}}
\end{table}

\begin{table}[htbp]
\centering
\caption{Computation time (seconds) of different methods on different real-world datasets.}\label{tab:re-time-rd-supp}
\vspace{0.5em}
\resizebox{\textwidth}{!}{%
\renewcommand{\arraystretch}{1.1} 
\begin{tabular}{ccccccccc}
\hline
 & ECOT-as & ECOT-bi & ECOT-oc & CP-bi & CP-oc & Integ & FullND & AdaDetect \\
\hline
Credit Card & 63.9 & 0.9 & 0.2 & 0.1 & 0.1 & 0.2 & 0.1 & 1.5 \\
Satellite & 63.2 & 0.8 & 0.2 & 0.1 & 0.1 & 0.1 & 0.1 & 1.6 \\
Shuttle & 54.7 & 0.3 & 0.3 & 0.1 & 0.1 & 0.2 & 0.1 & 0.7 \\
CovType & 61.1 & 0.5 & 0.5 & 0.1 & 0.1 & 0.3 & 0.3 & 0.9 \\
Mammography & 62.0 & 0.4 & 0.5 & 0.1 & 0.1 & 0.3 & 0.3 & 0.8 \\
\hline
\end{tabular}}
\end{table}

Tables \ref{tab:re-time-set1-supp}-\ref{tab:re-time-rd-supp} show computation time of a single run of different methods. From the three tables above, we observe that without the approach selection step, all baseline methods complete a single run within 5 seconds. Since our approach selection method requires determining an optimal approach for each $p$-value, it introduces additional computational overhead. Nevertheless, under moderate sample sizes, the total runtime remains within approximately 1 minute, which is acceptable for most practical applications. Moreover, this additional cost scales linearly with respect to both $m$ and $n$. As the procedure naturally decomposes across test samples $j$, it also admits straightforward parallelisation, which can further reduce the computational time by a factor equal to the number of threads used.

\subsection{Comparison with \cite{bai2024optimized} on simulated data}

In this section, we compare ECOT with the OptCS procedures of \cite{bai2024optimized}  in simulated experiments.
Here we implement ECOT-as with the purification step proposed in Section \ref{sec:pure} applied to base models to evaluate the complete ECOT implementation. The experiment is based on our binary classification setting (Section \ref{sec:bi_prefer}). To align with the assumptions in \citet{bai2024optimized}, we generate labelled and unlabelled data with the same proportion of $H_1$ samples. Specifically, we fix $n_0 + n_1 = 300$ and $m = 1000$, and vary the signal proportion $1 - \pi = n_1/(n_0 + n_1)$. We consider 6 candidate models:
\begin{itemize}
    \item 3 random forest models with tree numbers 100, 300 and 500
    \item 3 $\nu$-classification SVM models with $\nu=0.1,0.2$ and 0.4 
\end{itemize}

\begin{figure}[htbp]
    \centering
    \includegraphics[width=0.8\linewidth]{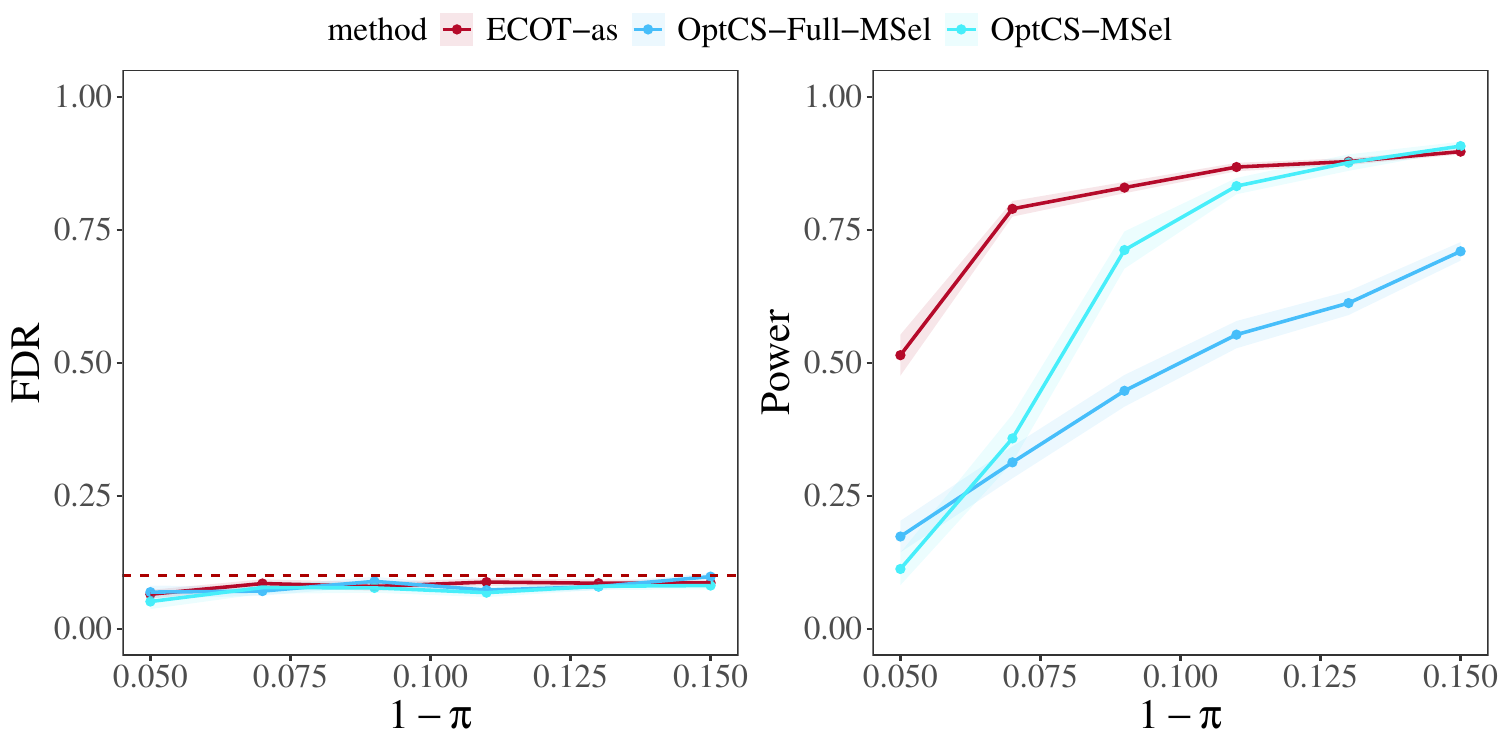}
    \caption{FDR and power comparison of ECOT and OptCS frameworks in the model selection task. The red dotted line marks the nominal level $\alpha=0.1$.}
    \label{fig:FULL}
\end{figure}

As shown in Figure \ref{fig:FULL}, for the model selection task, our ECOT framework exhibits higher power than both OptCS variants. The OptCS-MSel method performs comparably to ECOT when the signal proportion is large, but exhibits low power otherwise. The full conformal variant OptCS-Full-MSel shows the lowest power. This is likely because one of the training datasets $\gD_0 \cup \gD_u \setminus \{X_j\}$ used in OptCS-Full-MSel for each $j \in \gU$ are contaminated by the $H_1$ samples in $\gD_u$.

\subsection{BH procedure and conditional calibration}\label{subsec:BHCC}
We have declared in the main text that for our ECOT-as method, directly applying the BH procedure leads to quite similar performance to that of applying the conditional calibration procedure. Here we illustrate this issue by showing their differences in the binary classification setting in Figure \ref{fig:bi}. We abbreviate these two methods as ECOT-as-BH and ECOT-as-CC.

\begin{figure}[htbp]
    \centering
    \includegraphics[width=0.9\linewidth]{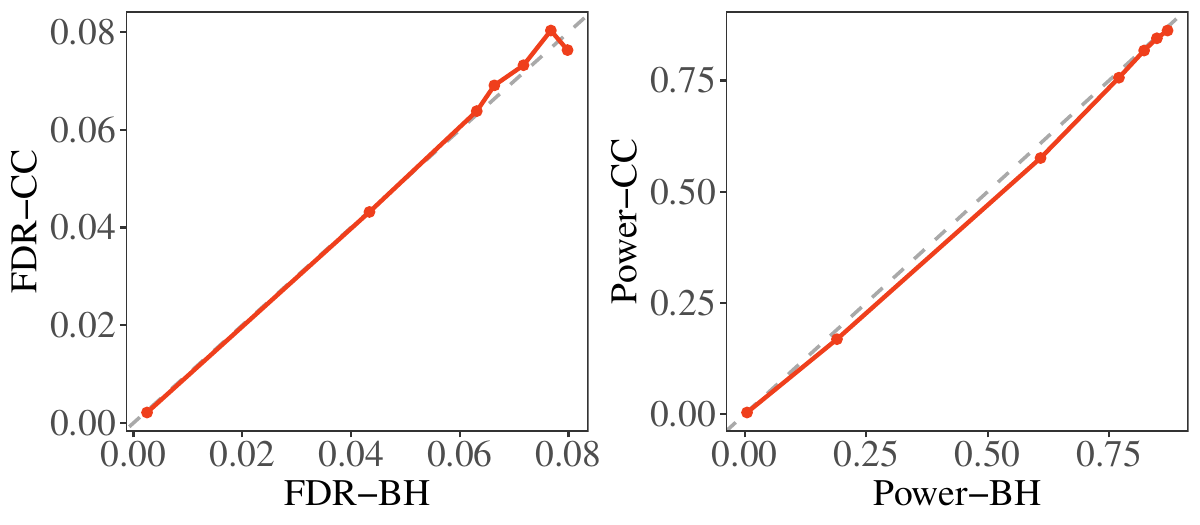}
    \caption{FDR and power of ECOT-as with the BH procedure and conditional calibration (CC) procedure applied.}
    \label{fig:BHCC}
\end{figure}

Figure \ref{fig:BHCC} shows the FDR and power of ECOT-as-BH and ECOT-as-CC across seven different sample sizes, which corresponds to seven points in each figure. It is evident from the figure that the performance gap between the two methods is negligible. This suggests that the conditional calibration step primarily serves as a theoretical refinement and typically does not result in significant power loss.

\subsection{Comparison with the oracle}

In this section, we compare our proposals and other baselines with an oracle competitor to show their gaps and limiting performance. As an oracle benchmark, we consider a procedure that uses the true likelihood ratio as the score function and takes $\gD_0$ as the calibration dataset to compute conformal $p$-values. Applying the BH procedure to these $p$-values yields an optimal testing procedure within the conformal testing framework. This optimality has been established in \citet{adadetect} (Theorem 4.1). In our experiments, we refer to this oracle method as NP.

\begin{figure}[htbp]
\centering
\includegraphics[width=0.5\linewidth]{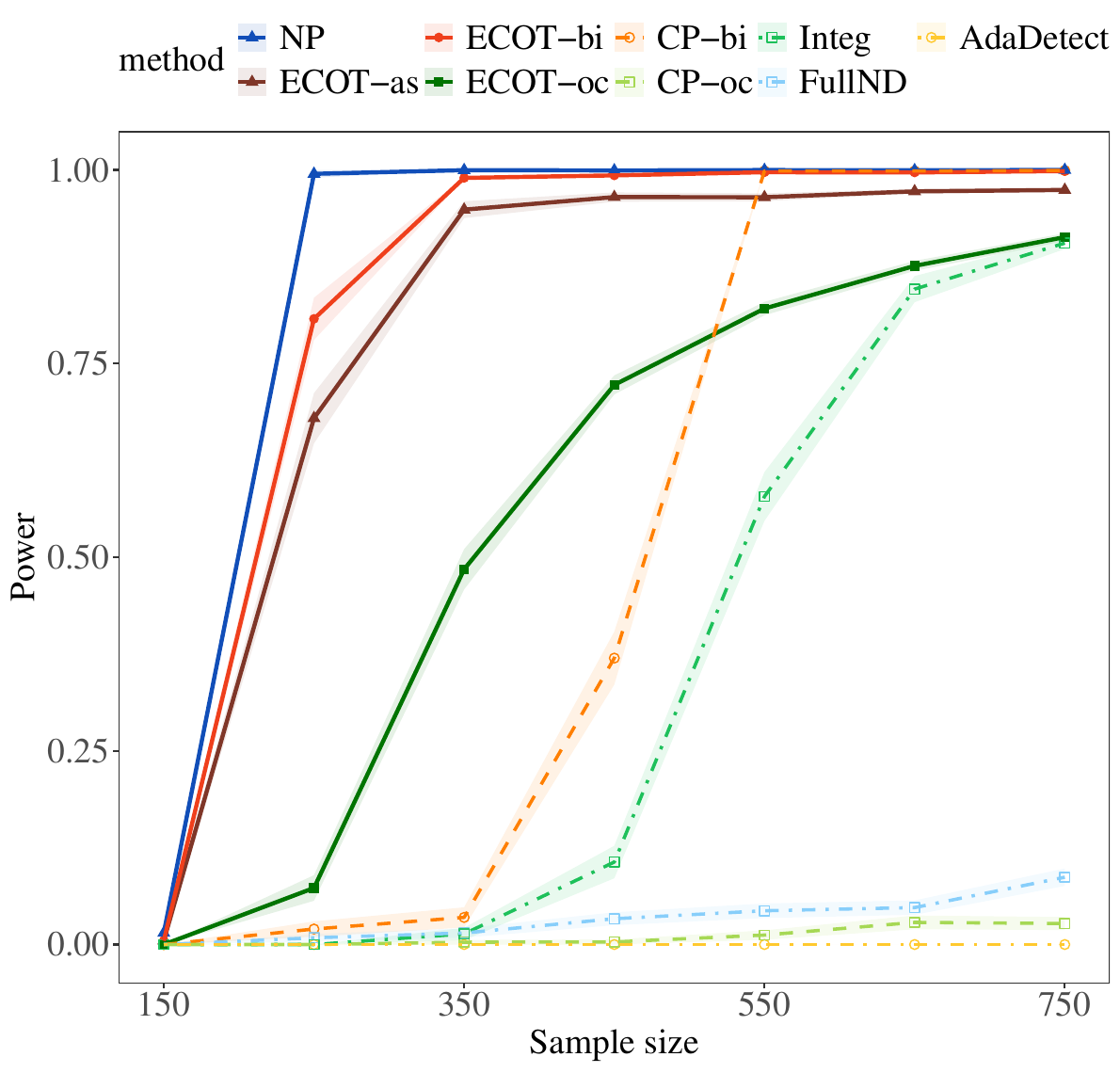}
\caption{Power of different methods compared with the oracle NP method in our binary classification setting (Section \ref{sec:bi_prefer}).}
\label{fig:NP}
\end{figure}

As shown in Figure \ref{fig:NP}, the oracle NP method achieves power 1 with the smallest labelled sample size among all competitors. Among the remaining methods, ECOT-bi performs the closest to NP and also reaches power 1 at a labelled sample size of 350.

\begin{figure}[htbp]
\centering
\includegraphics[width=\linewidth]{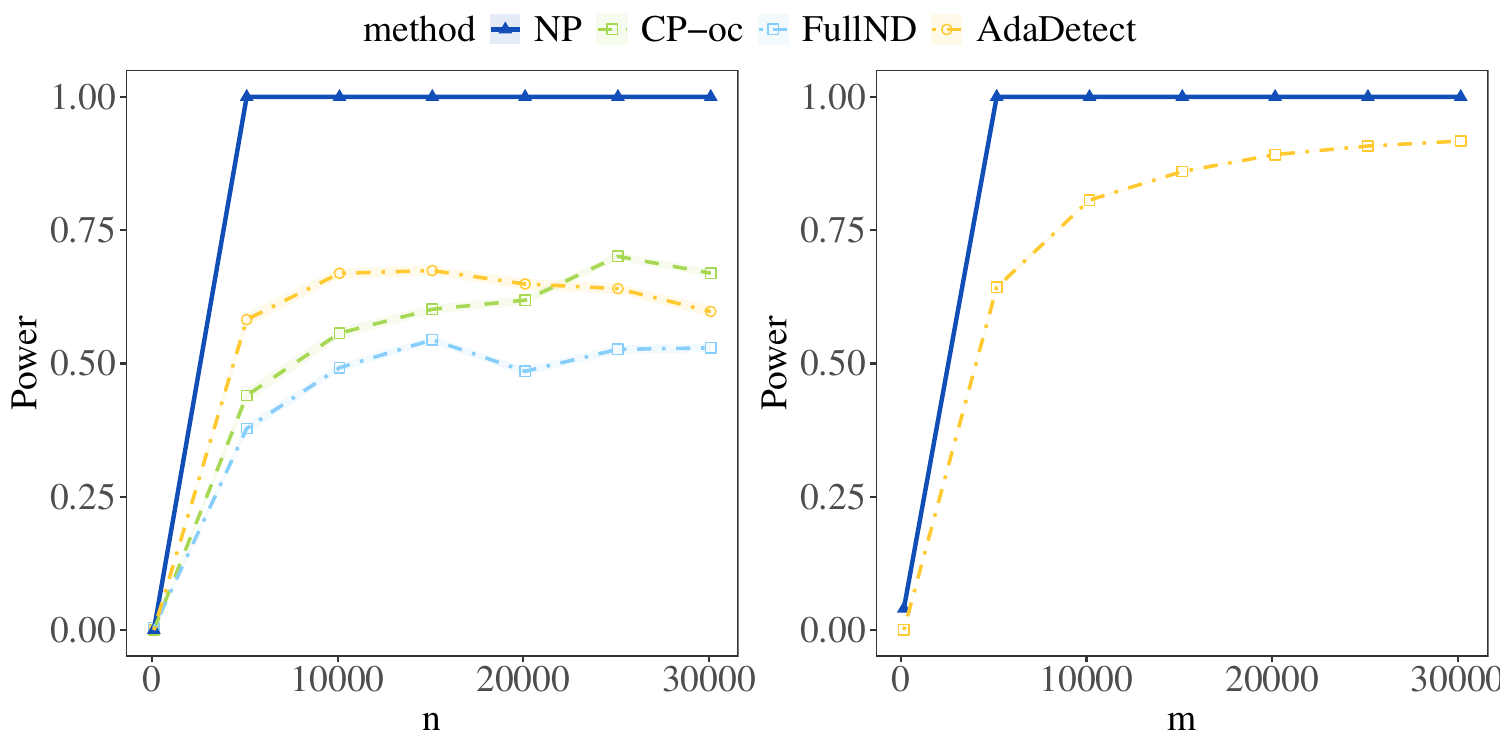}
\caption{Power of NP and three methods not leveraging $\gD_1$ as the labelled and unlabelled sample sizes vary.}
\label{fig:NPCONV}
\end{figure}

For the three methods that rely only on $\gD_0$ and $\gD_u$ (FullND, CP-oc, and AdaDetect), we extend the sample size up to 30,000 to further investigate their limiting behaviour. As shown in the left panel of Figure \ref{fig:NPCONV}, the power of these methods does not converge to 1, even with large sample sizes. For AdaDetect, we note that its score function is expected to converge to the oracle score when the labelled and unlabelled sample sizes grow to infinity simultaneously. To validate this, we fix the ratio $|\gD_0|/|\gD_u| = 1/2$ and vary $|\gD_u| = m$. As shown in the right panel, the power of AdaDetect indeed converges to 1 as $m$ grows beyond 30000.

\subsection{Test sample size $m$}

Here we provide results of varying test sample sizes $m$ under the setting in Section \ref{sec:bi_prefer} with $n=500,a=1$.

\begin{table}[h]
\centering
\caption{Comparison of FDR (top table) and power (bottom table) across different methods and sample sizes $m$. The target FDR level $\alpha=0.1$. The highest two values of power for each $m$ are shown in bold.}
\vspace{0.5em}
\renewcommand{\arraystretch}{1.2}

% ---- FDR Table ----
\begin{tabular}{lccccc}
\toprule
\textbf{Method} & $m=200$ & $m=500$ & $m=1000$ & $m=2000$ & $m=5000$ \\
\midrule
ECOT-as  & 0.077 & 0.075 & 0.070 & 0.067 & 0.067 \\
ECOT-bi  & 0.079 & 0.074 & 0.071 & 0.067 & 0.066 \\
ECOT-oc  & 0.061 & 0.036 & 0.018 & 0.015 & 0.005 \\
CP-bi    & 0.075 & 0.061 & 0.065 & 0.072 & 0.062 \\
CP-oc    & 0.004 & 0.001 & 0.000 & 0.000 & 0.001 \\
AdaDetect   & 0.002 & 0.000 & 0.000 & 0.000 & 0.000 \\
Integ    & 0.020 & 0.007 & 0.006 & 0.004 & 0.006 \\
FullND   & 0.014 & 0.005 & 0.002 & 0.001 & 0.003 \\

\bottomrule
\end{tabular}

\vspace{1em}

% ---- Power Table ----
\begin{tabular}{lccccc}
\toprule
\textbf{Method} & $m=200$ & $m=500$ & $m=1000$ & $m=2000$ & $m=5000$ \\
\midrule
ECOT-as  & \textbf{0.830} & \textbf{0.823} & \textbf{0.790} & \textbf{0.740} & \textbf{0.563} \\
ECOT-bi  & \textbf{0.871} & \textbf{0.843} & \textbf{0.799} & \textbf{0.744} & \textbf{0.563} \\
ECOT-oc  & 0.219 & 0.080 & 0.036 & 0.021 & 0.005 \\
CP-bi    & 0.593 & 0.477 & 0.457 & 0.478 & 0.447 \\
CP-oc    & 0.002 & 0.001 & 0.000 & 0.000 & 0.001 \\
AdaDetect   & 0.000 & 0.000 & 0.000 & 0.000 & 0.000 \\
Integ    & 0.043 & 0.014 & 0.011 & 0.006 & 0.009 \\
FullND   & 0.011 & 0.002 & 0.001 & 0.000 & 0.001 \\

\bottomrule
\end{tabular}
\label{tab:supp-m}
\end{table}

Table \ref{tab:supp-m} shows that all methods exhibit decreasing power when the test sample size $m$ increases. This is more significant for ECOT-bi since the two datasets $\{X_j\}_{j\in\gL_1}$ and $\{X_j\}_{j\in\gL_0\cup\gU}$ become more imbalanced as $m$ grows. For all values of 
$m$, ECOT-bi and ECOT-as are the most powerful among all baselines.

\subsection{Signal ratio under small sample sizes}

In the main text, we presented the performance of one-class classifier-based methods under varying signal ratios, using a large labelled sample size of $n=2000$ to ensure that all methods exhibit observable power. Here, we additionally report the results of ECOT-bi and CP-bi in the first setting when the labelled sample size is reduced to $n=500$.

\begin{figure}[htbp]
    \centering
    \includegraphics[width=0.9\linewidth]{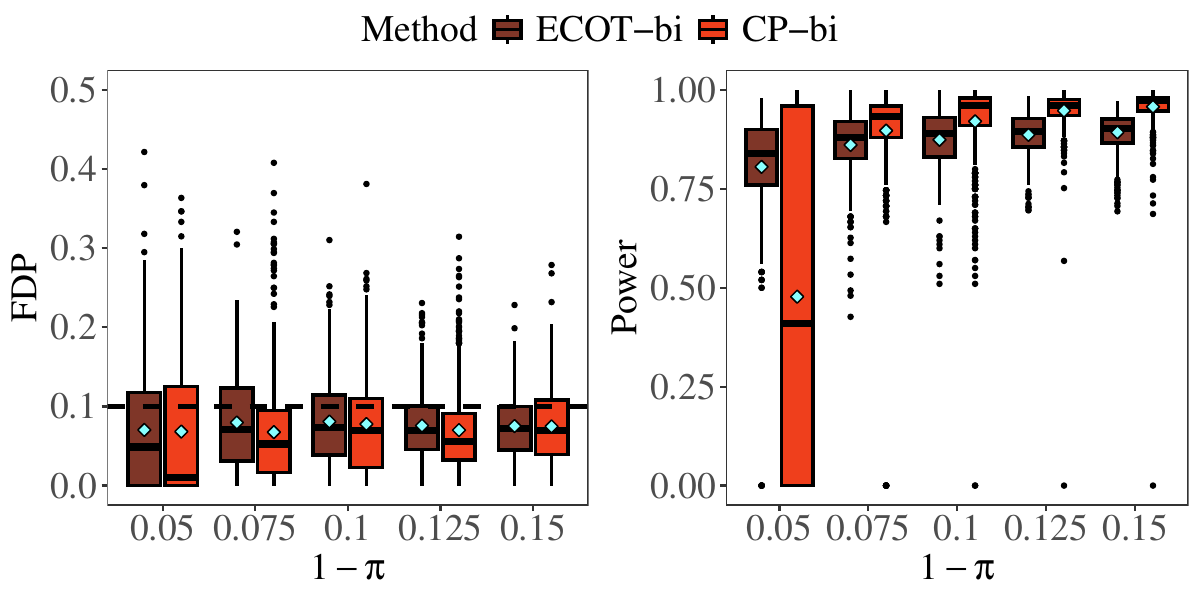}
    \caption{FDP and power of different methods under varying signal ratios. The black dashed line denotes the target FDR level $\alpha=0.1$.}
    \label{fig:ratio_supp}
\end{figure}

From Figure \ref{fig:ratio_supp}, we observe that CP-bi begins to exhibit higher power than ECOT-bi. This is because, although ECOT-bi leverages all parts of the data more efficiently, it has the drawback of using $\{X_j\}_{j\in\gL_0\cup\gU}$ as class 0 in binary classification. While we can theoretically show that this results in a monotonic transformation of the density ratio function in an oracle setting, it introduces a minor negative effect in finite samples. In our setup, the signal strength is already large enough for both methods to be effective, so the efficiency gain from using more data is less impactful. In contrast, the disadvantage of training on a mixture becomes more pronounced as the signal ratio increases. Overall, this issue remains minor, and the power gap is small even when CP-bi performs better.

\subsection{Nominal level $\alpha$}

\begin{figure}[htbp]
    \centering
    \includegraphics[width=0.9\linewidth]{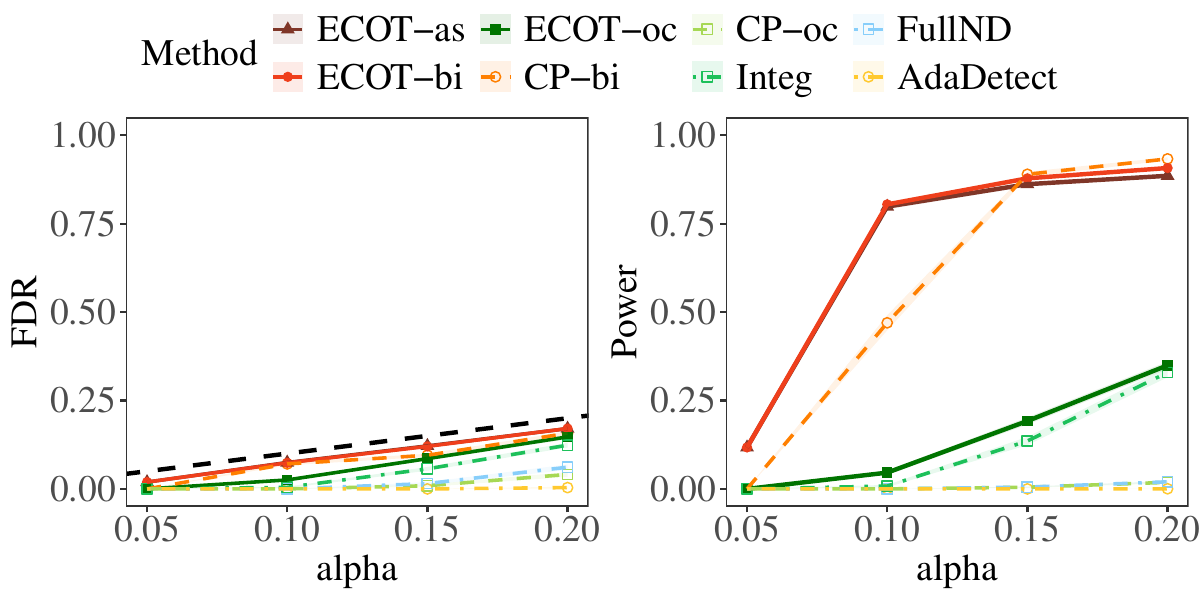}
    \caption{FDR and power of different methods under varying nominal levels. The black dashed line denotes the target FDR level.}
    \label{fig:alpha}
\end{figure}

We also demonstrate the robust performance of our methods under varying nominal levels $\alpha$. Here we consider the setting in Section \ref{sec:bi_prefer} with $n=500$ and $a=1$. Figure \ref{fig:alpha} shows that ECOT-bi and ECOT-as consistently achieve leading power, particularly when the nominal level is small.

\subsection{Training algorithm}

Finally we consider using the support vector machine for score training. We still consider the setting in Section \ref{sec:bi_prefer} with $n=500,a=1$.

\begin{figure}[htbp]
    \centering
    \includegraphics[width=0.9\linewidth]{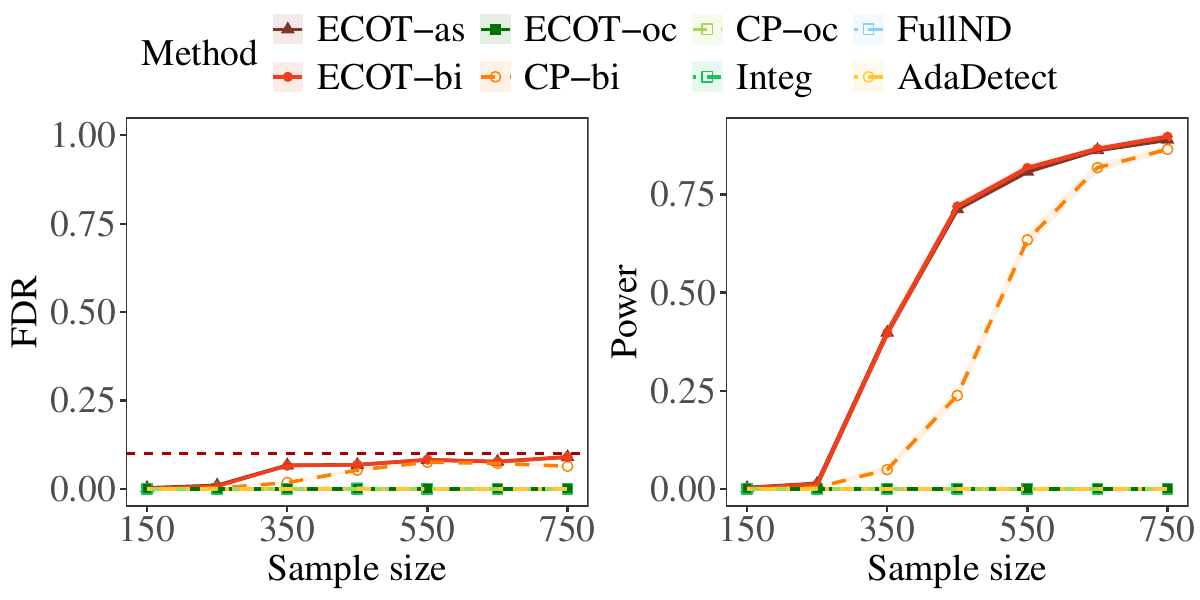}
    \caption{FDR and power of different methods with SVM training algorithms. The red dashed line denotes the target FDR level $\alpha=0.1$.}
    \label{fig:supp_algo}
\end{figure}

Figure \ref{fig:supp_algo} exhibits a pattern similar to the first row of Figure \ref{fig:bi}. The key difference is that both one-class and binary classifier-based methods lose power when trained with SVM algorithms, compared to those trained with random forests in Figure \ref{fig:bi}. This suggests that SVM classifiers are less efficient than random forest classifiers in this data generation scenario. Overall, the consistent patterns across both algorithms demonstrate the robustness of our methods to different training approaches.

\section{Technical details}

\subsection{Auxiliary lemmas}

\begin{Lem}\label{lemma:set_equal}
      Let $\mathcal{A}$ and $\mathcal{B}$ be two sets, $f: \mathcal{A}\mapsto\mathcal{B}$ a fixed function and $\mathcal{E}\subset\mathcal{B}$ a fixed subset. Then, for any bijection 
$g:\mathcal{A}\mapsto\mathcal{A}$,  $$\{f(a)\in\mathcal{E}:a\in\mathcal{A}\}=\{f(g(a))\in\mathcal{E}:a\in\mathcal{A}\}.
    $$
\end{Lem}

\textbf{Proof}:
If $f(a)\in \{f(a)\in\mathcal{E}:a\in\mathcal{A}\}$ with $a\in\mathcal{A}$, we have $a=g(b)$ for some unique $b\in\mathcal{A}$ as $g$ is a bijection. Then $f(a)=f(g(b))\in \{f(g(a))\in\mathcal{E}:a\in\mathcal{A}\}$. Thus $\{f(a)\in\mathcal{E}:a\in\mathcal{A}\}\subset \{f(g(a))\in\mathcal{E}:a\in\mathcal{A}\}$.  

If $f(g(a))\in \{f(g(a))\in\mathcal{E}:a\in\mathcal{A}\}$ with $a\in\mathcal{A}$, we have $g(a)\in\mathcal{A}$, indicating $f(g(a))\in \{f(a)\in\mathcal{E}:a\in\mathcal{A}\}$ and $\{f(g(a))\in\mathcal{E}:a\in\mathcal{A}\}\subset \{f(a)\in\mathcal{E}:a\in\mathcal{A}\}$. Combining together, we have $\{f(a)\in\mathcal{E}:a\in\mathcal{A}\}=\{f(g(a))\in\mathcal{E}:a\in\mathcal{A}\}$.

\subsection{Proof of Theorem \ref{thm:FDR_control}}\label{appen_sub_sec:proof_of_FDR_control}

\paragraph{(i) Validity of conformal p-value}

For notational simplicity, denote the remaining covariates and labelled responses $\Big(\big(X_k:k\in  \mathcal{L}_1\cup(\gU\setminus\{j\})\cup(\mathcal{L}_0\setminus\gC)\big),(Y_k:k\in\gL_1\cup\gL_0)\Big)\Big)$ as $\gD_r$.
Then $\Psi_j=\Big( \gD^r,\big\{X_k:k\in\gC\cup\{j\}\big\}\Big).$
Here $\gC$ is a fixed index set given the responses $(Y_k:k\in\gL_1\cup\gL_0)$.  And $k\in\gC$ implies $Y_k=0$.

Further conditional on $Y_j=0$, by Assumption \ref{Assum:data_exch} that $(X_i:i\in \gC\cup\gU, Y_i=0)$ are exchangeable conditional on remaining data $\gD_r$, we have $(X_i:i\in\gC\cup\{j\})$ are exchangeable too.
Define the sets of all permutations over $\gC\cup\{j\}$ as $\Omega_j$. Therefore, for a sequence of realisations $(x_{k}:k\in\gC\cup\{j\})$ and any permutation $\sigma'\in\Omega_j$,
\begin{align}\label{eq_appen:prob_perm}
 &\Pr\Big((X_k:k\in\gC\cup\{j\})=(x_k:k\in\gC\cup\{j\})\mid \bigcap_{k\in\gC\cup\{j\}}\{Y_k=0\},\Psi_j=\big(\gD^r,\{x_{k}:k\in\gC\cup\{j\}\}\big)\Big)\nonumber\\
 =&\Pr\Big((X_{k}:k\in\gC\cup\{j\})=(x_{\sigma'(k)}:k\in\gC\cup\{j\})\mid \bigcap_{k\in\gC\cup\{j\}}\{Y_k=0\},\Psi_j=\big(\gD^r,\{x_{k}:k\in\gC\cup\{j\}\}\big)\Big) 
\end{align}
The probability equals to $1/|\Omega_j|$ since for each $\sigma'\in\Omega_j$, the above probability is equally taken. 

Denote $ Q_{t}(S_k:{k\in\gA})$ as the $(1-t)$-th quantile in the set $\{S_k:{k\in\gA}\}$. Then we have
\begin{align*}
  &\Pr(p_j\leq t\mid Y_j=0,\Psi_j=\big(\gD^r,\{x_{k}:k\in\gC\cup\{j\}\}\big))\\
  \Eqmark{i}=&\E\left[\I\{S^{(j)}(X_j)\geq Q_{t}(S_\sigma^{(j)}(X_{\sigma(j)}):{\sigma\in\Omega_j})\}\mid \bigcap_{k\in\gC\cup\{j\}}\{Y_k=0\},\Psi_j=\big(\gD^r,\{x_{k}:k\in\gC\cup\{j\}\}\big)\right] \\
\Eqmark{ii}{=}&\sum_{\sigma'\in\Omega_j}\Big[\Pr\Big((X_{k}:k\in\gC\cup\{j\})=(x_{\sigma'(k)}:k\in\gC\cup\{j\})\mid\bigcap_{k\in\gC\cup\{j\}}\{Y_k=0\},\Psi_j=\big(\gD^r,\{x_{k}:k\in\gC\cup\{j\}\}\big)\Big)\\
&\times\I\{S_{\sigma'}^{(j)}(x_{\sigma'(j)})\geq Q_{t}(S_{\sigma\cdot\sigma'}^{(j)}(x_{\sigma\cdot\sigma'(j)}):{\sigma\in\Omega_j})\}\Big]\\
  \Eqmark{iii}{=}&\frac{1}{|\Omega_j|}\sum_{\sigma'\in\Omega_j} \I\{S_{\sigma'}^{(j)}(x_{\sigma'(j)})\geq Q_{t}(S_{\sigma'}^{(j)}(x_{\sigma'(j)}):\sigma'\in\Omega_j)\}\Eqmark{iv}{\leq} t.
\end{align*}
Here $\sigma\cdot\sigma'$ defines the mapping that $\sigma\cdot\sigma'(j)=\sigma(\sigma'(j))$. 
The equality (i) is from the definition of conformal p-value. And the event $\bigcap_{k\in\gC}\{Y_k=0\}$ is contained in $\gD^r$. Equality (ii) comes from the fact that given $\{x_{k}:k\in\gC\cup\{j\}\}$, $Y_j=0$ and $\bigcup_{k\in\gC}\{Y_k=0\}$ and $\gD^r$, the only randomness of $(S^{(j)}_\sigma(x_{\sigma(j)}):\sigma\in\Omega_j)$ is on the order of $\{x_{k}:k\in\gC\cup\{j\}\}$. As for equality (iii), the quantity $1/|\Omega_j|$ is directly from \eqref{eq_appen:prob_perm}. To analyse the summation, it suffices to prove that the sets $\{S_{\sigma'}^{(j)}(x_{\sigma'(j)}):\sigma'\in\Omega_j\}$ is equivalent to $\{S_{\sigma\cdot\sigma'}^{(j)}(x_{\sigma\cdot\sigma'(j)}):\sigma\in\Omega_j\}$. By the bijection property of permutation, if $\sigma'\in\Omega_j$, then $\{\sigma\cdot\sigma':\sigma\in\Omega_j\}=\{\sigma:\sigma\in\Omega_j\}=\Omega_j$. Therefore, noticing that $S_{\sigma'}^{(j)}(x_{\sigma'(j)})$ is a fixed function of $\sigma'$ and by Lemma \ref{lemma:set_equal}, we denote $F(\sigma')=S_{\sigma'}^{(j)}(x_{\sigma'(j)})$ and have
$$\{S_{\sigma\cdot\sigma'}^{(j)}(x_{\sigma\cdot\sigma'(j)}):\sigma\in\Omega_j\}=\{F(\sigma\cdot\sigma'):\sigma\in\Omega_j\}=\{F(\sigma):\sigma\in\Omega_j\}=\{S_{\sigma}^{(j)}(x_{\sigma(j)}):\sigma\in\Omega_j\}, $$
which is equivalent to $\{S_{\sigma'}^{(j)}(x_{\sigma'(j)}):\sigma'\in\Omega_j\}$ by using symbol $\sigma'$ instead of $\sigma$. Finally, equality (iv) holds by the definition of the quantile function $Q_t$. 

\paragraph{(ii) Property of modified p-values}
 
To prove that $\tilde{p}_\ell^{(j)}$ is measurable with respect to $\Psi_j$ given $Y_j=0$, it suffices to prove that the value of $\tilde{p}_\ell^{(j)}$ remains unchanged according to the order of $\{x_{k}:k\in\gC\cup\{j\}\}$ given $\gD^r$,$Y_j=0$ and $\{X_{k}:k\in\gC\cup\{j\}\}=\{x_{k}:k\in\gC\cup\{j\}\}$.

Conditional on the above quantities and suppose $(X_{k}:k\in\gC\cup\{j\})=(x_{\sigma'(k)}:k\in\gC\cup\{j\})$ for a permutation $\sigma\in\Omega_j$,
\begin{align*}
 \tilde{p}_\ell^{(j)}&=\frac{1}{|\Omega_j|}\sum_{\sigma\in\Omega_j}\I\{\tilde{S}_{\sigma'}^{(j)}(x_{\sigma'(\ell)})\leq S^{(j)}_{\sigma\cdot\sigma'}(x_{\sigma\cdot\sigma'(j)})\} \\
 &\Eqmark{i}{=}\frac{1}{|\Omega_j|}\sum_{\sigma\in\Omega_j}\I\{\tilde{S}^{(j)}(x_{\ell})\leq S^{(j)}_{\sigma\cdot\sigma'}(x_{\sigma\cdot\sigma'(j)})\}\\
 &\Eqmark{ii}{=}\frac{1}{|\Omega_j|}\sum_{\sigma\in\Omega_j}\I\{\tilde{S}^{(j)}(x_{\ell})\leq S^{(j)}_{\sigma}(x_{\sigma(j)})\}=\tilde{p}_\ell^{(j)}.
\end{align*}
Equality (i) holds since $\sigma'$ keeps fix for $\ell\in\gU\setminus\{j\}$, thereby $x_{\sigma'(\ell)}=x_{\ell}$, and $\tilde{S}^{(j)}(x_\ell)$ is permutation invariant to any $\sigma'\in\Omega_j$. More specifically, by Lemma \ref{lemma:set_equal}, we have $\{{S}^{(j)}_{\sigma\cdot\sigma'}(x_\ell):\sigma\in\Omega_j\}=\{{S}^{(j)}_{\sigma}(x_\ell):\sigma\in\Omega_j\}$. Thus   $$\tilde{S}_{\sigma'}^{(j)}(x_\ell)=\text{Median}\{{S}^{(j)}_{\sigma\cdot\sigma'}(x_\ell):\sigma\in\Omega_j\}=\text{Median}\{{S}^{(j)}_{\sigma}(x_\ell):\sigma\in\Omega_j\}=\tilde{S}^{(j)}(x_\ell).$$
And equality (ii) is true by invoking Lemma \ref{lemma:set_equal} again as $\sum_{\sigma\in\Omega_j}\I\{\tilde{S}^{(j)}(x_{\ell})\leq S^{(j)}_{\sigma}(x_{\sigma(j)})\}=|\{S^{(j)}_{\sigma}(x_{\sigma(j)})\in [\tilde{S}^{(j)}(x_\ell),+\infty):\sigma\in\Omega_j\}|$ where $[\tilde{S}^{(j)}(x_\ell),+\infty)$ can be viewed as a fixed set.

\paragraph{(iii) Final FDR control}
Firstly, we can directly verify that $\mathcal{R}_j$ is measurable with respect to $\Psi_j$ conditional on $Y_j=0$, as it is determined by $\{\tilde{p}_\ell^{(j)}\}$ fully.

Recall that the final rejection set by conditional calibration can be formulated by
    $$\mathcal{R}=\left\{j\in\mathcal{R}^{\rm init}:\varepsilon_j\leq \frac{R^*}{|\gR_j|}\right\},$$
    where $R^* = |\mathcal{R}|$. Define $\varepsilon_{-j}=\{\varepsilon_{\ell}:~\ell \in \gU \setminus \{j\}\}$ as the collection of all $\varepsilon_{\ell}$ variables for $\ell \in \gU \setminus \{j\}$, and let $R^*_j=R(\varepsilon_j \leftarrow 0)$ denote the hypothetical total number of rejections obtained by fixing $\varepsilon_j=0$. Then, the FDR can be written as
    
\begin{align}\label{append_eq:FDR_pruning}
\FDR&=\sum_{j\in\gU}\E\left[\frac{\I\{j\in\gR^{\rm init}\}\I\{\varepsilon_j\leq \frac{R^*}{|\gR_j|}\}\I\{Y_j=0\}}{1\vee{R^*}}\right]\nonumber\\
    & \Eqmark{i}{=} \sum_{j\in\gU}\E\left[\frac{\I\{j\in\gR^{\rm init}\}\I\{\varepsilon_j\leq \frac{R_j^*}{|\gR_j|}\}\I\{Y_j=0\}}{1\vee{R_j^*}}\right] \nonumber\\
        & = \sum_{j\in\gU}\E\left[ \E\left[\frac{\I\{j\in\gR^{\rm init}\}\I\{\varepsilon_j\leq \frac{R_j^*}{|\gR_j|}\}\I\{Y_j=0\}}{1\vee{R_j^*}}\mid \varepsilon_{-j},\gD_0\cup\gD_1\cup\gD_u\right] \right]\nonumber\\
        &\Eqmark{ii}{\leq} \sum_{j\in\gU}\E\left[\frac{\I\{j\in\gR^{\rm init}\}\I\{Y_j=0\}}{1\vee{|\gR_j|}}\right]\,.
\end{align}
Equality (i) holds since the pruning procedure can be seen as a special case of the BH procedure. Specifically, it is equivalent to the BH procedure applied to $\{\varepsilon_j|\gR_j|/|\gR^{\rm init}|\}_{j\in\gR^{\rm init}}$ at level $1$. Therefore, by the property of the BH procedure, replacing a rejected  $\varepsilon_j$ with $0$ does not change the number of rejections, i.e., $R^*=R_j^*$ for $\varepsilon_j\leq R^*/|\gR_j|$.

Inequality (ii) holds because $\varepsilon_j$ is independent of $\varepsilon_{-j}$ given all available data $\gD_0\cup\gD_1\cup\gD_u$ by their independent generation, ensuring that $\varepsilon_j$ remains uniformly distributed. Furthermore, $\varepsilon_j$ is independent of $R_j^*/|\gR_j|$ given $\varepsilon_{-j}$ and $\gD_0\cup\gD_1\cup\gD_u$. Since $|\gR_j|$ is measurable with respect to $\gD_0\cup\gD_1\cup\gD_u$, and $\varepsilon_j$ has no influence on $R^*_j$ by the assignment ($\varepsilon_j\leftarrow 0$), the desired inequality holds.
    
Recall that $\mathcal{R}^{\rm init}=\left\{j\in\gU:{p}_j\leq\frac{\alpha |\gR_j|}{m}\right\}.$ We further analyse the FDR following \eqref{append_eq:FDR_pruning}:

\begin{align*}
\FDR&{\leq} \sum_{j\in\gU}\E\left[\frac{\I\{j\in\gR^{\rm init}\}\I\{Y_j=0\}}{1\vee{|\gR_j|}}\right]\\
        &=\sum_{j\in\gU}\E\left[\frac{\I\{{p}_j\leq\frac{\alpha |\gR_j|}{m}\}\I\{Y_j=0\}}{1\vee{|\gR_j|}}\right]\\
        &\Eqmark{i}{=}\sum_{j\in\gU}\E\left[\frac{\E\left[\I\{{p}_j\leq\frac{\alpha |\gR_j|}{m}\}\mid \Psi_j,Y_j=0\right]\I\{Y_j=0\}}{1\vee{|\gR_j|}}\right]\\
        &\Eqmark{ii}{\leq} \sum_{j\in\gU}\E\left[\frac{\alpha |\gR_j|}{m}\frac{\I\{Y_j=0\}}{1\vee{|\gR_j|}}\right]\\
&{\leq}\alpha\sum_{j\in\gU}\E\left[\frac{\I\{Y_j=0\}}{m}\right] =\alpha\E\left[\frac{|\mathcal{H}_0|}{|\gU|}\right]\\
\end{align*}
Equality (i) holds as $\gR_j$ is measurable with respect to $\Psi_j$ by the design of $\tilde{p}_{\ell}^{ (j)}$. Along with Theorem \ref{thm:FDR_control} (i) and the truth that $\gR_j$ is fixed given $\Psi_j$,
  %$$\Pr(\hat{p}_j\leq t\mid \gD_{\gC\cup\{j\}}^*,\gD_{\gU\backslash \{j\}} ,j\in\hgS_u,\theta_j=0 )\leq t,$$
  inequality (ii) holds correspondingly.

\subsection{Proof of Proposition \ref{pro:cali_sym}}
As we have verified that under Definition \ref{Assum:cali_sym}, the conformal p-value based on permutations in \eqref{eq:perm_pvalue} reduces to
$$ p_j=
\frac{1}{|\gC|+1}\sum_{i\in\gC\cup\{j\}}\I\{S^{(j)}(X_j)\leq S^{(j)}(X_{i})\},  $$
it suffices to prove the reduction of modified p-values $\{\tilde{p}_\ell^{(j)}\}_{\ell\in\gU}$ for any $j\in\gU$.

Recall the definition in \eqref{eq:modi_perm_pvalue}, we have for any $\ell\neq j$,
\begin{align*}
    \tilde{p}_\ell^{(j)}&=\frac{1}{|\Omega_j|}\sum_{\sigma\in\Omega_j}\I\{{S}^{(j)}(X_\ell)\leq S^{(j)}(X_{\sigma(j)})\}\\
    &=\frac{1}{|\Omega_j|}\sum_{i\in\gC\cup\{j\}}\sum_{\sigma\in\Omega_j,\sigma(j)=i}\I\{{S}^{(j)}(X_\ell)\leq S^{(j)}(X_{\sigma(j)})\}\\
    &=\frac{1}{|\gC|+1}\sum_{i\in\gC\cup\{j\}}\I\{{S}^{(j)}(X_\ell)\leq S^{(j)}(X_{i})\}
\end{align*}
The first equality holds due to Definition \ref{Assum:cali_sym}, making $S^{(j)}_\sigma\equiv S^{(j)}$ for any $\sigma\in\Omega_j$, thereby
$$\tilde{S}^{(j)}(X_\ell)=\text{Median}\{{S}^{(j)}_\sigma(X_\ell):\sigma\in\Omega_j\}={S}^{(j)}(X_\ell).$$

\subsection{Proof of Proposition \ref{pro:joint_sym}}\label{appen_sub_sec:proof_joint_sym}
It is obvious that under Definition \ref{Assum:joint_sym}, the conformal p-values in \eqref{eq:perm_pvalue} reduce to \eqref{eq:marg_cp}, which is a direct extension of Proposition \ref{pro:cali_sym}. Then it is left to prove the equivalence of our procedure applying conditional calibration with that of BH over the conformal p-values. Here we denote $S_k=S(X_k)$ for $k\in\gC\cup\gU$ for notational simplicity.

%Notice that during the conditional calibration procedure, if each $\gR_j=\gR^{\rm init}$, the output rejection set is $\gR=\gR^{\rm init}=\gR_j$. This indicates that if for each $j\in\gR^{\rm init}$, $\gR_j$ is equivalent to the rejection set $\gR^{\rm BH}$ by applying BH procedure at level $\alpha$ to original p-values $\{p_j\}_{j\in\gU}$, the output of conditional calibration is also equivalent to $\gR^{\rm BH}$.

Define $\gR^{\rm BH}$ as the rejection set by applying BH procedure at level $\alpha$ to original p-values $\{p_j\}_{j\in\gU}$.
Recall that $\gR_j$ is the rejection set by applying BH procedure at level $\alpha$ to modified p-values $\{\tilde{p}_\ell^{(j)}\}_{\ell\in\gU}$. 
The modified  p-value satisfies
\begin{equation}\label{appen_eq:connec_modify_p}
p_\ell=\tilde{p}_\ell^{(j)}+ \frac{\mathbbm{1}\{S_j<S_\ell\}}{|\gC|+1} \text{~~for~} j\neq \ell,
\end{equation}
Thereby $p_\ell\geq \tilde{p}_\ell^{(j)}$, meaning that $|\gR^{\rm BH}|\leq |\gR_j|$. So $p_j\leq \frac{\alpha |\gR^{\rm BH}|}{m}$ implies $ p_j\leq \frac{\alpha |\gR_j|}{m}$. From a similar discussion of Lemma D.6 in \citet{adadetect}, suppose that $j$ satisfies $ p_j\leq \frac{\alpha |\gR_j|}{m}$, then

\begin{align*}
    &\sum_{\ell\in\gU} \mathbbm{1}\left\{p_\ell\leq \frac{\alpha |\gR_j|}{|\gU|}\right\}\\
=&\sum_{\ell\in\gU} \mathbbm{1}\left\{p_\ell\leq \frac{\alpha |\gR_j|}{|\gU|}\right\}\mathbbm{1}\{p_\ell\leq p_j\}+\sum_{\ell\in\gU} \mathbbm{1}\left\{p_\ell\leq \frac{\alpha |\gR_j|}{|\gU|}\right\}\mathbbm{1}\{p_\ell> p_j\}\\
\Eqmark{i}{=}&\sum_{\ell\in\gU} \mathbbm{1}\left\{\tilde{p}_\ell^{(j)}\leq \frac{\alpha |\gR_j|}{|\gU|}\right\}\mathbbm{1}\{p_\ell\leq p_j\}+\sum_{\ell\in\gU} \mathbbm{1}\left\{\tilde{p}_\ell^{(j)}\leq \frac{\alpha |\gR_j|}{|\gU|}\right\}\mathbbm{1}\{p_\ell> p_j\}\\
=&\sum_{\ell\in\gU} \mathbbm{1}\left\{\tilde{p}_\ell^{(j)}\leq \frac{\alpha |\gR_j|}{|\gU|}\right\}\Eqmark{ii}{\geq} |\gR_j|.
\end{align*}
Here, equality (i) holds as $\tilde{p}_\ell^{(j)}\leq p_\ell\leq p_j\leq \frac{\alpha |\gR_j|}{|\gU|}$ when $p_\ell\geq p_j$ and $p_\ell=\tilde{p}_\ell^{(j)}$ when $p_\ell> p_j$. To see why, $p_\ell> p_j$ implies $S_\ell<S_j$. Through \eqref{appen_eq:connec_modify_p}, it directly follows $p_\ell=\tilde{p}_\ell^{(j)}$. Inequality (ii) is from the property of BH procedure.

Note that 
 $$|\mathcal{R}^{\rm BH}|=\max\left\{r:\left|\left\{\ell\in\gU:p_\ell\leq\frac{\alpha r}{m}\right\}\right|\geq r\right\}.$$
The above results indicate taking $r=|\gR_j|$ also satisfies $\left|\{\ell\in\gU:p_\ell\leq\frac{\alpha r}{|\gU|}\}\right|\geq r$, leading to $|\gR^{\rm BH}|\geq|\gR_j|$. Then, we have $|\gR^{\rm BH}|=|\gR_j|$. It shows
\begin{equation}\label{appen_eq:equi_rej_size}
    \mathbbm{1}\{p_j\leq \frac{\alpha |\gR^{\rm BH}|}{m}\}=\mathbbm{1}\{p_j\leq \frac{\alpha |\gR_j|}{m}\}.
\end{equation}
And \eqref{appen_eq:equi_rej_size} indicates that $\gR^{\rm init}=\sum_{j\in\gU}\mathbbm{1}\{p_j\leq \frac{\alpha |\gR_j|}{m}\}=\gR^{\rm BH}$. Note that for any $j\in\gR^{\rm init}$, $\gR_j=\gR^{\rm BH}=\gR^{\rm init}$. By the operation of conditional calibration, if for each $j\in\gR^{\rm init}$, $\gR_j=\gR^{\rm init}$, then the pruning procedure will be omitted and $\gR=\gR^{\rm init}$. This verifies our conclusion.

\subsection{Proof of Proposition \ref{pro:optimal}}

It suffices to prove that the likelihood ratio function $$s(x)=\frac{f_1(x)}{f_{\gL_0\cup\gU}(x)}$$ 
learned from distinguishing $\{X_j\}_{j\in\gL_1}$ and $\{X_j\}_{j\in\gL_0\cup\gU}$ is a strictly increasing transformation of $r(x)=(1-\pi)f_1(x)/(\pi f_0(x)+(1-\pi)f_1(x))$, where $f_{\gL_0\cup\gU}(x)$ denotes the average density of $\{X_i:i\in\gL_0\cup\gU\}$. 

To verify this, we have
\begin{align*}
    s(x)&=\frac{f_1(x)}{\frac{|\gH_0|+|\gL_0|}{|\gL_0|+|\gU|}f_0(x)+\frac{|\gH_1|}{|\gL_0|+|\gU|}f_1(x)}\\
    &=\left[{\frac{|\gH_1|}{|\gL_0|+|\gU|}+\frac{|\gH_0|+|\gL_0|}{|\gL_0|+|\gU|}\frac{f_0(x)}{f_1(x)}}\right]^{-1}\\
    &=\left[{\frac{|\gH_1|}{|\gL_0|+|\gU|}+\frac{|\gH_0|+|\gL_0|}{|\gL_0|+|\gU|}\frac{1-\pi}{\pi}\left(\frac{1}{r(x)}-1\right)}\right]^{-1}
\end{align*}
where the last equality is due to the transformation $$\frac{f_0(x)}{f_1(x)}=\frac{1-\pi}{\pi}\left(\frac{1}{r(x)}-1\right).$$
So $s(x)$ is a strictly increasing transformation of $r(x)$. Then the conclusions hold by directly applying Theorem 4.1 and Lemma 4.3 in \citet{adadetect}.

\subsection{Proof of Corollary \ref{cor:ECOT-ras}}

It suffices to verify that the final selected score function $S^{(j),k^*_j}$ is still symmetric to data in $\gC\cup\{j\}$.

Consider a permutation $\sigma\in\Omega_j$. The modified p-values constructed based on the datasets permuted by $\sigma$ would be
$$\frac{1}{|\gC|+1}\sum_{i\in\gC\cup\{j\}}\I\{S^{(j),k}_\sigma(X_{\sigma(\ell)})\leq S^{(j),k}_\sigma(X_{\sigma(i)})\}=\frac{1}{|\gC|+1}\sum_{i\in\gC\cup\{j\}}\I\{S^{(j),k}(X_{\ell})\leq S^{(j),k}(X_{\sigma(i)})\}=\tilde{p}_\ell^{(j),k}.$$
The first equality is from the calibration symmetric property of $S^{(j),k}$. The last equality holds by invoking Lemma \ref{lemma:set_equal}. Thus we verify that $\{\tilde{p}_\ell^{(j),k}\}_{\ell\in\gU}$ are permutation invariant to data in $\gC\cup\{j\}$ for any $k\in[K]$. Thereby the corresponding size of rejection set $R_{k}^{(j)}$ by applying BH procedure to them would be still permutation invariant to data in $\gC\cup\{j\}$.

It then follows that $k_j^*=\arg\max_{k\in [K]} R_{k}^{(j)}$ is also permutation invariant to data in $\gC\cup\{j\}$. This suggests that $S^{(j),k_j^*}$ satisfies Definition \ref{Assum:cali_sym} accordingly.

Thus by applying Proposition \ref{pro:cali_sym}, the FDR control is verified.

\subsection{Proof of Theorem \ref{thm:FDR_control_jin}}\label{appen_sub_sec:proof_thm:FDR_control_jin}

By Assumption \ref{Assum:data_full_exch} that $\big((X_i,Y_i):i\in \gC\cup\gU\big)$ are exchangeable, we have $\big((X_i,Y_i):i\in \gC\cup\{j\}\big)$ are exchangeable too.
Define the sets of all permutation over $\gC\cup\{j\}$ as $\Omega_j$. Therefore, for a sequence of realisations $\big((x_{k},y_k):k\in\gC\cup\{j\}\big)$ and any permutation $\sigma'\in\Omega_j$,
\begin{align}\label{eq_appen:prob_jin_perm}
 &\Pr\Big(\big((X_k,Y_k):k\in\gC\cup\{j\}\big)=\big((x_k,y_k):k\in\gC\cup\{j\}\big)\mid \Psi_j=\big(\gD^r,\{(x_{k},y_k):k\in\gC\cup\{j\}\}\big)\Big)\nonumber\\
 =&\Pr\Big(\big((X_k,Y_k):k\in\gC\cup\{j\}\big)=\big((x_{\sigma'(k)},y_{\sigma'(k)}):k\in\gC\cup\{j\}\big)\mid \Psi_j=\big(\gD^r,\{(x_{k},y_k):k\in\gC\cup\{j\}\}\big)\Big) 
\end{align}
The probability equals to $1/|\Omega_j|$ since for each $\sigma'\in\Omega_j$, the above probability is equally taken.

Denote $ Q_{t}(S_k:{k\in\gA})$ as the $(1-t)$-th quantile in the set $\{S_k:{k\in\gA}\}$. Then we have
\begin{align}\label{appen_eq:jin_pvalue_proof}
  &\Pr(p_j\leq t, Y_j=0\mid \Psi_j=\big(\gD^r,\{(x_{k},y_k):k\in\gC\cup\{j\}\}\big))\nonumber\\
  =&\E\left[\I\{S^{(j)}(X_j,\tilde{Y}_j)\geq Q_{t}(S_\sigma^{(j)}(X_{\sigma(j)},\tilde{Y}_{\sigma(j)}):{\sigma\in\Omega_j})\}\I\{Y_j=0\}\mid \Psi_j=\big(\gD^r,\{(x_{k},y_k):k\in\gC\cup\{j\}\}\big)\right] \nonumber\\
  \Eqmark{i}{\leq}&\E\left[\I\{S^{(j)}(X_j,Y_j)\geq Q_{t}(S_\sigma^{(j)}(X_{\sigma(j)},Y_{\sigma(j)}):{\sigma\in\Omega_j})\}\mid \Psi_j=\big(\gD^r,\{(x_{k},y_k):k\in\gC\cup\{j\}\}\big)\right]\nonumber\\
\Eqmark{ii}{=}&\sum_{\sigma'\in\Omega_j}\Big[\Pr\left(\big((X_{k},Y_k):k\in\gC\cup\{j\}\big)=\big((x_{\sigma'(k)},y_{\sigma'(k)}):k\in\gC\cup\{j\})\mid \Psi_j=\big(\gD^r,\{(x_{k},y_k):k\in\gC\cup\{j\}\}\big)\right)\nonumber\\
&\quad\times\I\{S_{\sigma'}^{(j)}(x_{\sigma'(j)},y_{\sigma'(j)})\geq Q_{t}(S_{\sigma\cdot\sigma'}^{(j)}(x_{\sigma\cdot\sigma'(j)},y_{\sigma\cdot\sigma'(j)}):{\sigma\in\Omega_j})\}\Big]\nonumber\\
  \Eqmark{iii}{=}&\frac{1}{|\Omega_j|}\sum_{\sigma'\in\Omega_j} \I\{S_{\sigma'}^{(j)}(x_{\sigma'(j)},y_{\sigma'(j)})\geq Q_{t}(S_{\sigma'}^{(j)}(x_{\sigma'(j)},y_{\sigma'(j)}):\sigma'\in\Omega_j)\}\leq t.
\end{align}
The (i) holds since on the event $Y_j=0$, the score function satisfies $S^{(j)}(X_j,\tilde{Y}_j)=S^{(j)}(X_j,0)= S^{(j)}(X_j,Y_j)$. By the definition of $\{\tilde{Y}_k\}_{k\in\gC\cup\gU}$, we have $\tilde{Y}_k=Y_k$ for all $k\in\gC\cup\{j\}$ under the event $Y_j=0$. Equality (ii) comes from the fact that given $\{(x_{k},y_k):k\in\gC\cup\{j\}\}$ and $\gD^r$, the only randomness of $(S^{(j)}_\sigma(x_{\sigma(j)},y_{\sigma(j)}):\sigma\in\Omega_j)$ is on the order of $\big((x_{k},y_k):k\in\gC\cup\{j\}\big)$. As for equality (iii), the first part is directly from \eqref{eq_appen:prob_jin_perm}. To analyse the second part, it suffices to prove that the sets $\{S_{\sigma'}^{(j)}(x_{\sigma'(j)},y_{\sigma'(j)}):\sigma'\in\Omega_j\}$ is equivalent to $\{S_{\sigma\cdot\sigma'}^{(j)}(x_{\sigma\cdot\sigma'(j)},y_{\sigma\cdot\sigma'(j)}):\sigma\in\Omega_j\}$. Noticing that $S_{\sigma'}^{(j)}(x_{\sigma'(j)},y_{\sigma'(j)})$ is a fixed function of $\sigma'$, by Lemma \ref{lemma:set_equal}, we can directly verify this analog to the proof in Section \ref{appen_sub_sec:proof_of_FDR_control}.

Next, define ${\gR}^*_j$ as the rejection set by applying the BH procedure at level $\alpha$ to the oracle modified conformal p-values $\{\tilde{p}_\ell^{*(j)}\}_{\ell\in\gU}$ as
\begin{equation}\label{eq:modi_perm_pvalue_full_oracle}
     \tilde{p}_\ell^{*(j)}=\frac{1}{|\Omega_j|}\sum_{\sigma\in\Omega_j}\I\{\tilde{S}^{(j)}(X_\ell,0)\leq S^{(j)}_\sigma(X_{\sigma(j)},{Y}_{\sigma(j)})\}\quad \ell\neq j\quad\text{and}\quad \tilde{p}_j^{*(j)}=0.   
    \end{equation}
Under the event $Y_j=0$, we have 
\begin{align*}
\tilde{p}_\ell^{(j)}&=\frac{1}{|\Omega_j|}\sum_{\sigma\in\Omega_j}\I\{\tilde{S}^{(j)}(X_\ell,\tilde{Y}_\ell)\leq S^{(j)}_\sigma(X_{\sigma(j)},\tilde{Y}_{\sigma(j)})\}    \\
&=\frac{1}{|\Omega_j|}\sum_{\sigma\in\Omega_j}\I\{\tilde{S}^{(j)}(X_\ell,0)\leq S^{(j)}_\sigma(X_{\sigma(j)},{Y}_{\sigma(j)})\} .
\end{align*}
Therefore, we also have $|{\gR}^*_j|=|\gR_j|$. 

Moreover, we can verify that ${\gR}^*_j$ is permutation invariant to $\{(X_i,Y_i):i\in\gC\cup\{j\}\}$. This is a direct extension of the proof strategy in Section \ref{appen_sub_sec:proof_of_FDR_control} (ii).

Based on the analysis of conditional calibration, we have
\begin{align*}
\FDR&\Eqmark{i}{\leq} \sum_{j\in\gU}\E\left[\frac{\I\{j\in\gR^{\rm init}\}\I\{Y_j=0\}}{1\vee{|\gR_j|}}\right]\\
        &=\sum_{j\in\gU}\E\left[\frac{\I\{{p}_j\leq\frac{\alpha |\gR_j|}{m}\}\I\{Y_j=0\}}{1\vee{|\gR_j|}}\right]\\
        &\Eqmark{ii}{=}\sum_{j\in\gU}\E\left[\frac{\E\left[\I\{{p}_j\leq\frac{\alpha |\gR^*_j|}{m}\}\I\{Y_j=0\}\mid \Psi_j\right]}{1\vee{|\gR^*_j|}}\right]\\
        &\Eqmark{iii}{\leq} \sum_{j\in\gU}\E\left[\frac{\alpha |\gR^*_j|}{m}\frac{1}{1\vee{|\gR^*_j|}}\right]{\leq}\alpha.
\end{align*}
The (i) holds by the property of conditional calibration, which is proved in \eqref{append_eq:FDR_pruning}. Equality (ii) comes from the property that under the event $Y_j=0$, $|\gR_j^*|=|\gR_j|$ by its definition. The last (iii) is directly from the conclusion in \eqref{appen_eq:jin_pvalue_proof}.

\subsection{Proof of Theorem \ref{thm:FDR_control_jin_weighted}}
We will check the super-uniformity of weighted permutation p-value under Assumption \ref{Assum:covariate_shift} based on a similar discussion in Section \ref{appen_sub_sec:proof_thm:FDR_control_jin}. 
For notational simplicity, write
\[
Z_k=(X_k,Y_k),\qquad z_k=(x_k,y_k),\qquad 
\gD^r=(Z_k:k\in (\gU\setminus\{j\})\cup (\gL\setminus \gC)).
\]
Let $\Omega_j$ be the set of all permutations on $\gC\cup\{j\}$.
Conditional on
\[
\Psi_j=\big(\gD^r,\{z_k:k\in\gC\cup\{j\}\}\big),
\]
the only randomness in the ordered tuple $(Z_k:k\in\gC\cup\{j\})$ is which element of the unordered set
$\{z_k:k\in\gC\cup\{j\}\}$ is assigned to the distinguished index $j$.
By Assumption~\ref{Assum:covariate_shift}, for any $\sigma'\in\Omega_j$,
\begin{align}\label{eq_appen:prob_jin_perm_weighted}
&\Pr\Big((Z_k:k\in\gC\cup\{j\})=(z_{\sigma'(k)}:k\in\gC\cup\{j\})
\;\Big|\;
\Psi_j=\big(\gD^r,\{z_k:k\in\gC\cup\{j\}\}\big)\Big) \nonumber\\
&\qquad = \frac{Q_X(x_{\sigma'(j)})P_{Y\mid X}(y_{\sigma'(j)}\mid x_{\sigma'(j)})
\prod_{k\in\gC}
P_X(x_{\sigma'(k)})P_{Y\mid X}(y_{\sigma'(k)}\mid x_{\sigma'(k)})}{\sum_{\sigma\in\Omega_j}Q_X(x_{\sigma(j)})P_{Y\mid X}(y_{\sigma(j)}\mid x_{\sigma(j)})
\prod_{k\in\gC}
P_X(x_{\sigma(k)})P_{Y\mid X}(y_{\sigma(k)}\mid x_{\sigma(k)})}
 \nonumber\\
 &\qquad = \frac{w(x_{\sigma'(j)})
\prod_{k\in\gC\cup\{j\}}
P_X(x_{\sigma'(k)})P_{Y\mid X}(y_{\sigma'(k)}\mid x_{\sigma'(k)})}{\sum_{\sigma\in\Omega_j}w(x_{\sigma(j)})
\prod_{k\in\gC\cup\{j\}}
P_X(x_{\sigma(k)})P_{Y\mid X}(y_{\sigma(k)}\mid x_{\sigma(k)})}
 \nonumber\\
&\qquad 
= \frac{w(x_{\sigma'(j)})}{\sum_{\sigma\in\Omega_j}w(x_{\sigma(j)})}.
\end{align}
The last equality holds since $\prod_{k\in\gC\cup\{j\}}
P_X(x_{\sigma(k)})P_{Y\mid X}(y_{\sigma(k)}\mid x_{\sigma(k)})$ is invariant for any $\sigma\in\Omega_j$.

For a weighted collection $\{(v_\sigma,a_\sigma):\sigma\in\Omega_j\}$ with $a_\sigma\ge 0$,
define the upper weighted $t$-quantile by
\[
Q_t^w(v_\sigma:\sigma\in\Omega_j)
:=
\inf\left\{q\in\mathbb R:
\frac{\sum_{\sigma\in\Omega_j}a_\sigma\I\{v_\sigma\ge q\}}
{\sum_{\sigma\in\Omega_j}a_\sigma}
\le t
\right\}.
\]
In our setting, we take
$a_\sigma = w(x_{\sigma(j)})$ and $v_\sigma = S_\sigma^{(j)}(x_{\sigma(j)},y_{\sigma(j)}).$ 
Then we have
\begin{align}\label{appen_eq:jin_pvalue_proof}
  &\Pr(p_j\leq t, Y_j=0\mid \Psi_j=\big(\gD^r,\{z_k:k\in\gC\cup\{j\}\}\big))\nonumber\\
  =&\E\left[\I\{S^{(j)}(X_j,\tilde{Y}_j)\geq Q^w_{t}(S_\sigma^{(j)}(X_{\sigma(j)},\tilde{Y}_{\sigma(j)}):{\sigma\in\Omega_j})\}\I\{Y_j=0\}\mid \Psi_j=\big(\gD^r,\{z_k:k\in\gC\cup\{j\}\}\big)\right] \nonumber\\
  {\leq}&\E\left[\I\{S^{(j)}(X_j,Y_j)\geq Q^w_{t}(S_\sigma^{(j)}(X_{\sigma(j)},Y_{\sigma(j)}):{\sigma\in\Omega_j})\}\mid \Psi_j=\big(\gD^r,\{z_k:k\in\gC\cup\{j\}\}\big)\right]\nonumber\\
{=}&\sum_{\sigma'\in\Omega_j}\Big[\Pr\left(\big(Z_k:k\in\gC\cup\{j\}\big)=\big(z_{\sigma'(k)}:k\in\gC\cup\{j\}\big)\mid \Psi_j=\big(\gD^r,\{z_k:k\in\gC\cup\{j\}\}\big)\right)\nonumber\\
&\quad\times\I\{S_{\sigma'}^{(j)}(x_{\sigma'(j)},y_{\sigma'(j)})\geq Q^w_{t}(S_{\sigma\cdot\sigma'}^{(j)}(x_{\sigma\cdot\sigma'(j)},y_{\sigma\cdot\sigma'(j)}):{\sigma\in\Omega_j})\}\Big]\nonumber\\
  \Eqmark{i}{=}&\frac{1}{\sum_{\sigma\in\Omega_j}w(X_{\sigma(j)})}\sum_{\sigma'\in\Omega_j} w(X_{\sigma'(j)})\I\{S_{\sigma'}^{(j)}(x_{\sigma'(j)},y_{\sigma'(j)})\geq Q^w_{t}(S_{\sigma}^{(j)}(x_{\sigma(j)},y_{\sigma(j)}):\sigma\in\Omega_j)\}\leq t.\nonumber
\end{align}
 As for equality (i), the first part is directly from \eqref{eq_appen:prob_jin_perm_weighted}. 
To justify the second part of (iii), it suffices to show that the weighted collection
$\Big\{\big(w(x_{\sigma\cdot\sigma'(j)}),\,S_{\sigma\cdot\sigma'}^{(j)}(x_{\sigma\cdot\sigma'(j)},y_{\sigma\cdot\sigma'(j)})\big):\sigma\in\Omega_j\Big\}$
coincides with 
$\Big\{\big(w(x_{\sigma(j)}),\,S_{\sigma}^{(j)}(x_{\sigma(j)},y_{\sigma(j)})\big):\sigma\in\Omega_j\Big\}.$ 
This follows because $\{\sigma\cdot\sigma':\sigma\in\Omega_j\}=\Omega_j$ by the bijection property of permutations.
Hence the corresponding upper weighted quantile is unchanged. The final inequality follows directly from the definition of the upper weighted $t$-quantile.

Next, define ${\gR}^*_j$ as the rejection number by applying the BH procedure at level $\alpha$ to the oracle modified conformal p-values $\{\tilde{p}_\ell^{*(j)}\}_{\ell\in\gU}$ as
\begin{equation}\label{eq:modi_perm_pvalue_full_oracle_weighted}
     \tilde{p}_\ell^{*(j)}=\frac{\sum_{\sigma\in\Omega_j}w(X_{\sigma(j)})\I\{\tilde{S}^{(j)}(X_\ell,0)\leq S^{(j)}_\sigma(X_{\sigma(j)},{Y}_{\sigma(j)})\}}{\sum_{\sigma\in\Omega_j}w(X_{\sigma(j)})}\quad \ell\neq j\quad\text{and}\quad \tilde{p}_j^{*(j)}=0.   \nonumber
    \end{equation}
Under the event $Y_j=0$, we have 
\begin{align*}
\tilde{p}_\ell^{(j)}&=\frac{\sum_{\sigma\in\Omega_j}w(X_{\sigma(j)})\I\{\tilde{S}^{(j)}(X_\ell,\tilde{Y}_\ell)\leq S^{(j)}_\sigma(X_{\sigma(j)},\tilde{Y}_{\sigma(j)})\}}{\sum_{\sigma\in\Omega_j}w(X_{\sigma(j)})}    \\
&=\frac{\sum_{\sigma\in\Omega_j}w(X_{\sigma(j)})\I\{\tilde{S}^{(j)}(X_\ell,0)\leq S^{(j)}_\sigma(X_{\sigma(j)},{Y}_{\sigma(j)})\}}{\sum_{\sigma\in\Omega_j}w(X_{\sigma(j)})} .
\end{align*}
Therefore, we also have $|{\gR}^*_j|=|\gR_j|$. Moreover, we can verify that ${\gR}^*_j$ is permutation invariant to $\{(X_i,Y_i):i\in\gC\cup\{j\}\}$. This is a direct extension of the proof strategy in Section \ref{appen_sub_sec:proof_of_FDR_control} (ii). Then based on the same proof in Section \ref{appen_sub_sec:proof_thm:FDR_control_jin}, the FDR control is verified.

\subsection{Proof of Proposition \ref{pro:storey_est}}
By our construction, we can directly evaluate that the auxiliary p-values $\{\tilde{p}^{(j),{\rm join}}_\ell\}_{\ell\in\gU\setminus\{j\}}$ are measurable with respect to $\Psi_j$. Therefore, applying conditional calibration procedure over $\{\hat{\pi}_j p_j\}$ yields

\begin{align*}
\FDR&{\leq} \sum_{j\in\gU}\E\left[\frac{\I\{j\in\gR^{\rm init}\}\I\{Y_j=0\}}{1\vee{|\gR_j|}}\right]\\
        &=\sum_{j\in\gU}\E\left[\frac{\I\{\hat{\pi}_j{p}_j\leq\frac{\alpha |\gR_j|}{m}\}\I\{Y_j=0\}}{1\vee{|\gR_j|}}\right]\\
        &=\sum_{j\in\gU}\E\left[\frac{\E\left[\I\{{p}_j\leq\frac{\alpha |\gR_j|}{m\hat{\pi}_j}\}\mid \Psi_j,Y_j=0\right]\I\{Y_j=0\}}{1\vee{|\gR_j|}}\right]\\
        &{\leq} \sum_{j\in\gU}\E\left[\frac{\alpha |\gR_j|}{m\hat{\pi}_j}\frac{\I\{Y_j=0\}}{1\vee{|\gR_j|}}\right]\\
&{\leq}\alpha\sum_{j\in\gU}\E\left[\frac{\I\{Y_j=0\}}{m\hat{\pi}_j}\right].
\end{align*}
Now we check the property of $\E\left[{1}/{\hat{\pi}_j}\right]$ following the strategies in \citet{lee2025full}.

For notational convenience, we denote $S^{\rm join}_i=S^{\rm join}(X_i)$. 
By the design of Storey's estimator, we have
$$\frac{1}{\hat{\pi}_j}=\frac{ m(1-\lambda)}{1+\sum_{\ell\in\gU\setminus\{j\}}\I\{\tilde{p}^{(j),{\rm join}}_\ell\geq\lambda\} }=\frac{m}{|\gC|+1}\frac{\sum_{i\in\gC\cup\{j\}}\I\{S^{\rm join}_i\leq Q_{1-\lambda}(S^{\rm join}_k:k\in\gC\cup\{j\})\} }{ 1+\sum_{\ell\in\gU\setminus\{j\}}\I\{S^{\rm join}_\ell\leq Q_{1-\lambda}(S^{\rm join}_k:k\in\gC\cup\{j\})\}}.$$
Then it has
\begin{align}\label{appen_eq:storey_analy}
    &\frac{\I\{S^{\rm join}_i\leq Q_{1-\lambda}(S^{\rm join}_k:k\in\gC\cup\{j\})\} }{ 1+\sum_{\ell\in\gU\setminus\{j\}}\I\{S^{\rm join}_\ell\leq Q_{1-\lambda}(S^{\rm join}_k:k\in\gC\cup\{j\})\}}\nonumber\\
    \leq&\frac{\I\{S^{\rm join}_i\leq Q_{1-\lambda}(S^{\rm join}_k:k\in\gC\cup\{j\})\} }{ 1+\sum_{\ell\in\gH_0\setminus\{j\}}\I\{S^{\rm join}_\ell\leq Q_{1-\lambda}(S^{\rm join}_k:k\in\gC\cup\{j\})\}} \nonumber\\
    =&\frac{\I\{S^{\rm join}_i\leq Q_{1-\lambda}(S^{\rm join}_k:k\in\gC\cup\{j\})\} }{ 1\vee(\I\{S^{\rm join}_i\leq Q_{1-\lambda}(S^{\rm join}_k:k\in\gC\cup\{j\})\}+\sum_{\ell\in\gH_0\setminus\{j\}}\I\{S^{\rm join}_\ell\leq Q_{1-\lambda}(S^{\rm join}_k:k\in\gC\cup\{j\})\})}\\
    =& \frac{\I\{S^{\rm join}_i\leq q(\lambda)\} }{ 1\vee(\sum_{\ell\in\gH_0\setminus\{j\}\cup\{i\}}\I\{S^{\rm join}_\ell\leq q(\lambda)\})},\nonumber
\end{align}
where $q(\lambda)$ is a value determined by $\lambda$ and $\{S^{\rm join}_k:k\in\gH_0\setminus\{j\}\cup\{i\}\}$, since \eqref{appen_eq:storey_analy} changes only when $Q(S^{\rm join}_k:k\in\gC\cup\{j\})$ takes values at $\{S^{\rm join}_k:k\in\gH_0\setminus\{j\}\cup\{i\}\}$.

Then, denote the set of all permutations of indices $\gH_0\setminus\{j\}\cup\{i\}$  as $\bar{\Omega}_{i,j}$. As $\{S^{\rm join}_\ell\}_{\ell\in\gC\cup\gH_0}$ are exchangeable, for a realisation $\{S^{\rm join}_\ell:\ell\in \gH_0\setminus\{j\}\cup\{i\}\}=\{s_\ell:\ell\in \gH_0\setminus\{j\}\cup\{i\}\}$, we have
\begin{align*}
   &\E\left[ \frac{\I\{S^{\rm join}_i\leq q(\lambda)\} }{ 1\vee(\sum_{\ell\in\gH_0\setminus\{j\}\cup\{i\}}\I\{S^{\rm join}_\ell\leq q(\lambda)\})}\mid \{S^{\rm join}_\ell:\ell\in \gH_0\setminus\{j\}\cup\{i\}\}=\{s_\ell:\ell\in \gH_0\setminus\{j\}\cup\{i\}\},Y_j=0\right]\\
   =&\frac{1}{|\gH_0|!}\sum_{\bar{\sigma}\in\bar{\Omega}_{i,j}}\frac{\I\{s_{\bar{\sigma}(i)}\leq q(\lambda)\} }{ 1\vee(\sum_{\ell\in\gH_0\setminus\{j\}\cup\{i\}}\I\{s_{\bar{\sigma}(\ell)}\leq q(\lambda)\})}\\
   =&\frac{1}{|\gH_0|}\sum_{k\in\gH_0\setminus\{j\}\cup\{i\}}\frac{\I\{s_{k}\leq q(\lambda)\} }{ 1\vee(\sum_{\ell\in\gH_0\setminus\{j\}\cup\{i\}}\I\{s_{\ell}\leq q(\lambda)\})}\leq \frac{1}{|\gH_0|}.
\end{align*}
The first equality holds as $q(\lambda)$ is fully determined by 
$\{s_\ell:\ell\in \gH_0\setminus\{j\}\cup\{i\}\}$. 

Thus 
\begin{align*}
   \sum_{j\in\gU} \E\left[\frac{\I\{Y_j=0\}}{m\hat{\pi}_j}\right]&\leq\frac{1}{|\gC|+1}\sum_{j\in\gU} \E\left[\I\{Y_j=0\}\sum_{i\in\gC\cup\{j\}}\frac{\I\{S^{\rm join}_i\leq q(\lambda)\} }{ 1\vee(\sum_{\ell\in\gH_0\setminus\{j\}\cup\{i\}}\I\{S^{\rm join}_\ell\leq q(\lambda)\})}\right]\\
   &\leq \frac{1}{|\gC|+1}\sum_{j\in\gU} \E\left[\I\{Y_j=0\}\sum_{i\in\gC\cup\{j\}}\frac{1}{|\gH_0|}\right]\\
   &=\E\left[\frac{\sum_{j\in\gU}\I\{Y_j=0\}}{|\gH_0|}\right]=1.
\end{align*}

\subsection{Proof of Proposition \ref{pro:cali_est}}

Note that $\gC=\gC_0$ by definition. We can rewrite $\hat{\pi}p_j$ as
$$ \hat{\pi}p_j=\frac{1 + |\{i\in \gC: S^{(j)}(X_j) \leq S^{(j)}(X_i)\}|}{1+|\gC_0\cup\gC_1|}=\frac{1 + |\{i\in \gC_0\cup\gC_1: S^{(j)}(X_j,0) \leq S^{(j)}(X_i,Y_i)\}|}{1+|\gC_0\cup\gC_1|},$$
where $S^{(j)}(X_k,1)=-\infty$ and $S^{(j)}(X_k,0)=S^{(j)}(X_k)$ otherwise for any $k\in\gC_0\cup\gC_1\cup\gU$. Through the lens of \citet{jin2023selection}, we can view  $\hat{\pi}p_j$ as a special p-value constructed by $\{S^{(j)}(X_i,Y_i)\}_{i\in\gC_0\cup\gC_1}$ defined in \eqref{eq:perm_pvalue_full} instead of $\{{S}^{j}(X_i)\}_{i\in\gC_0}$. Specifically, the score function $S^{(j)}$ is symmetric to $\gC_0\cup\gC_1\cup\{j\}$ by its definition. Then following Theorem \ref{thm:FDR_control_jin} with the simplification of the full permutation strategy, the FDR control is direct.

\subsection{Proof of Proposition \ref{pro:jack_type}}
Firstly, we verify under Definition \ref{Assum:jack_type}, the p-values in \eqref{eq:perm_pvalue} are reduced to \eqref{eq:jack_pvalue}. 

Denote $\sigma(i,j)$ as the permutation that only swaps the position of $i$ and $j$. 
For any permutation $\sigma\in\Omega_j$ such that $\sigma(j)=i$, we can evaluate $S^{(j)}_\sigma=S^{(j)}_{\sigma(i,j)}$. For simplicity, we assume that $S^{(j)}$ is symmetric to $\{X_k:k\in\gC\cup\gU\setminus\{j\}\}$. 
Because $\sigma$ satisfying $\sigma(j)=i$ only changes the order of $\gC\cup\gU\setminus\{j\}$ and $S^{(j)}$ is symmetric to $\{X_k:k\in\gC\cup\gU\setminus\{j\}\}$, where the order in $\gC\cup\gU\setminus\{j\}$ does not change the output.

Therefore, we have
\begin{align*}
  p_j&=\frac{1}{(|\gC|+1)!}\sum_{i\in\gC\cup\{j\}}\sum_{\sigma\in\Omega_j,\sigma(j)=i}\I\{S^{(j)}(X_j)\leq S^{(j)}_\sigma(X_{\sigma(j)})\}\nonumber\\
  &=\frac{1}{(|\gC|+1)!}\sum_{i\in\gC\cup\{j\}}\sum_{\sigma\in\Omega_j,\sigma(j)=i}\I\{S^{(j)}(X_j)\leq S^{(j)}_{\sigma(i,j)}(X_{i})\}\nonumber\\
    &=\frac{1}{(|\gC|+1)!}\sum_{i\in\gC\cup\{j\}}|\gC|!\I\{S^{(j)}(X_j)\leq S^{(j)}_{\sigma(i,j)}(X_{i})\}\nonumber\\
  &=
\frac{1}{|\gC|+1}\sum_{i\in\gC\cup\{j\}}\I\{S^{(j)}(X_j)\leq S^{(i)}(X_{i})\}.  \nonumber
\end{align*}
Next, we can also evaluate the modified p-values for $\ell\neq j$ are reduced to
\begin{align*}
    \tilde{p}_\ell^{(j)}&=\frac{1}{|\Omega_j|}\sum_{\sigma\in\Omega_j}\I\{{S}^{(\ell)}(X_\ell)\leq S_\sigma^{(j)}(X_{\sigma(j)})\}\\
    &=\frac{1}{|\Omega_j|}\sum_{i\in\gC\cup\{j\}}\sum_{\sigma\in\Omega_j,\sigma(j)=i}\I\{{S}^{(\ell)}(X_\ell)\leq S_\sigma^{(j)}(X_{\sigma(j)})\}\\
    &=\frac{1}{|\gC|+1}\sum_{i\in\gC\cup\{j\}}\I\{{S}^{(\ell)}(X_\ell)\leq S^{(i)}(X_{i})\}.
\end{align*}

Denote $S_k=S^{(k)}(X_k)$ for $k\in\gC\cup\gU$. Then following the same proof strategy in Section \ref{appen_sub_sec:proof_joint_sym}, we can show that applying BH procedure over these p-values is equivalent to applying conditional calibration at the same level. If one only assumes the null-wise symmetry in Definition~\ref{Assum:jack_type},
the same argument applies after restricting attention to null indices and replacing
$\gU$ by $\gH_0$ in the permutation.  Thus, the proof is completed.

\end{document}